# Main Manuscript for

Stability of Nucleic Acid Bases in Concentrated Sulfuric Acid: Implications for the Habitability of Venus' Clouds.

Sara Seager[1,2,3,4,#,*] Janusz J. Petkowski[1,5,#], Maxwell D. Seager[4,6], John H. Grimes Jr.[7], Zachary Zinsli[8], Heidi R. Vollmer-Snarr[8], Mohamed K. Abd El-Rahman[8], David S. Wishart[9,10], Brian L. Lee[9], Vasuk Gautam[9], Lauren Herrington[1], William Bains[1,11,12], Charles Darrow[4]

[1] Department of Earth, Atmospheric and Planetary Sciences, Massachusetts Institute of Technology, Cambridge, MA, USA
[2] Department of Physics, Massachusetts Institute of Technology, Cambridge, MA, USA
[3] Department of Aeronautical and Astronautical Engineering, Massachusetts Institute of Technology, Cambridge, MA, USA
[4] Nanoplanet Consulting, Concord, MA, USA
[5] JJ Scientific, Warsaw, Mazowieckie, Poland
[6] Department of Chemistry and Biochemistry, Worcester Polytechnic Institute, Worcester, MA, USA
[7] Department of Chemistry, Massachusetts Institute of Technology, Cambridge, MA, USA
[8] Department of Chemistry and Chemical Biology, Harvard University, Cambridge, MA, USA
[9] Department of Biological Sciences, University of Alberta, Edmonton, Alberta, Canada
[10] Department of Computing Science, Department of Laboratory Medicine and Pathology, Faculty of Pharmacy and Pharmaceutical studies, University of Alberta, Edmonton, Alberta, Canada
[11] School of Physics and Astronomy, Cardiff University, 4 The Parade, Cardiff CF24 3AA, UK
[12] Rufus Scientific, Melbourn, Royston, Herts, UK
# contributed equally to this work

*Corresponding author: Sara Seager
**Email:** seager@mit.edu

**This PDF file includes:** Main Text; Figures 1 to 9; Tables 1 to 3.

## Abstract

What constitutes a habitable planet is a frontier to be explored and requires pushing the boundaries of our terracentric viewpoint for what we deem to be a habitable environment. Despite Venus' 700 K surface temperature being too hot for any plausible solvent and most organic covalent chemistry, Venus' cloud-filled atmosphere layers at 48-60 km above the surface hold the main requirements for life: suitable temperatures for covalent bonds, an energy source (sunlight), and a liquid solvent. Yet the Venus clouds are widely thought to be incapable of supporting life because the droplets are composed of concentrated liquid sulfuric acid—an aggressive solvent that is assumed to rapidly destroy most biochemicals of life on Earth. Recent work, however, demonstrates that a rich organic chemistry can evolve from simple precursor molecules seeded into concentrated sulfuric acid, a result that is corroborated by domain knowledge in industry that such chemistry leads to complex molecules, including aromatics. We aim to expand the set of molecules known to be stable in concentrated sulfuric acid. Here we show that nucleic acid bases adenine, cytosine, guanine, thymine, and uracil, as well as 2,6-diaminopurine and the "core" nucleic acid bases purine and pyrimidine, are stable in sulfuric acid in the Venus cloud temperature and sulfuric acid concentration range, using UV spectroscopy and combinations of 1D and 2D $^1$H $^{13}$C $^{15}$N NMR spectroscopy. The stability of nucleic acid bases in concentrated sulfuric acid advances the idea that chemistry to support life may exist in the Venus cloud particle environment.

## Significance Statement

The search for signs of life beyond Earth is a motivator in modern-day planetary exploration. While our "sister" planet Venus has a surface too hot for life, scientists have speculated that the much cooler atmosphere at 48-60 km above the surface might host life in Venus' perpetual cloud cover, as Earth's clouds do. The Venus clouds, however, are composed of concentrated sulfuric acid—an aggressive chemical that destroys most of Earth life's biochemicals and are thought to be sterile to life of any kind. Here we show that key molecules needed for life (nucleic acid bases) are stable in concentrated sulfuric acid, advancing the notion that the Venus atmosphere environment may be able to support complex chemicals needed for life.

**Main Text**

**1. Introduction: Background and motivation**

The search for signs of life is a key motivator in modern-day planetary exploration. Our "sister" planet Venus has long been relegated to the uninhabitable category because its surface at 735 K is too hot for any plausible solvent and for most organic covalent chemistry. Yet for over five decades a small but growing group of scientists have continued to support the speculative idea that microbial-type life might permanently occupy the temperate cloud layers of Venus (1–14). The cloud layers are permanent, continuous across the planet, and vertically extensive (48 to 60 km altitude above the surface). The cloud layers have the main requirements for life (15): suitable temperatures for covalent bonds; a liquid environment; and an energy source. So the Venus cloud layers could, in principle, host systems with the key characteristics of life, including their ability to undergo Darwinian evolution, even if the specifics of their chemistry are very different from those of terrestrial life.

The concept of life in the clouds of Venus is both motivated and supported by the fact that Earth has an aerial biosphere. A diverse set of microbes are transported up from the Earth surface (16–19) and remain aloft for on average 3 to 7 days (20) before being rained out. Earth's clouds are both transient and fragmented—a challenging ecological niche for permanent habitation. Venus, in contrast, has a permanent and continuous cover from vertically extensive clouds. Venus microbial type life residing inside cloud particles could remain aloft indefinitely via a life cycle of droplet growth, sedimentation, sporulation to a dormant hibernation in a haze layer beneath the clouds, followed by upward mixing via gravity waves and cloud nucleation (6).

The Venus cloud particles are made of concentrated sulfuric acid as inferred by Pioneer Venus in situ measurements of the backscattered polarized radiation (21). Technically, the measurements yield the particle refractive index and particle size assuming spherical particles. These measurements confirmed earlier polarization observations from Earth-based telescopes (22). The concentration of sulfuric acid in the Venus cloud particles has not been directly measured, but inferred from Pioneer Venus measurements of gases by the mass spectrometer. The gases evolved from cloud particles that clogged the inlet and are consistent with cloud droplets composed of 85% w/w $H_2SO_4$ and 15% w/w $H_2O$ (23). Based on models, it is likely that the sulfuric acid concentration of the cloud particles varies with altitude, in the cloud tops reaching 79% w/w while in the lower clouds the sulfuric acid concentration could reach 98% w/w (24).

Two potential show-stoppers exist for the survival of life in the clouds of Venus. First, there is nearly no water available, as the atmosphere is extremely dry and the cloud sulfuric acid droplets have very low water activity. Any water in the sulfuric acid droplets is locked away in strong hydrogen bonds to sulfuric acid. Second, the concentrated sulfuric acid is an aggressive solvent and destroys most biochemicals (25). Indeed, concentrated sulfuric acid is orders of magnitude more acidic than the most acidic environments on Earth that host acid-adapted microorganisms (6). The general accepted viewpoint is that concentrated sulfuric acid is sterile for any interesting chemistry and for any life (26).

We promote the concept that the concentrated sulfuric acid droplets may support a rich organic chemistry of a kind that might be able to support life different from Earth life. Spacek and Benner (27–29) have shown that a rich organic chemistry in sulfuric acid evolves from a simple organic seed molecule such as formaldehyde (and propose that such organic chemistry may be the origin of the perplexing and unidentified material in Venus' atmosphere that absorbs 50% of all incident ultraviolet (UV) sunlight (30), the so-called "unknown UV absorber" (31)). Even gas phase CO and $CO_2$ can act as seed molecules, themselves originating from atmospheric photochemical processes. The experiments by Spacek and Benner use a precursor molecule containing only C, H and O elements and yet even such limited elemental composition can yield complex organic molecules. The Venus clouds may have a much more diverse organic chemistry than these laboratory models, because the clouds also contain molecules that have N, S, or even P atoms (32).

The idea of a rich organic chemistry in concentrated sulfuric acid has long been known outside of planetary science. The oil refinement industry, for example, uses concentrated sulfuric acid to process crude oil. As a byproduct, "red oil" waste product is generated which includes a diversity of organic compounds including aromatic ringed molecules dissolved in concentrated sulfuric acid (33–35).

Complex organic chemistry is of course not life, but there is no life without it. We can consider the potential stability of complex organic chemistry in an environment as a prerequisite to habitability. Thus, we are motivated to study the stability of complex molecules that may make up a biochemistry of non-Earth-like life in concentrated sulfuric acid to explore whether sulfuric acid is inherently uninhabitable to all possible life, rather than just to terrestrial life. While the Venus cloud droplets likely contain dissolved substances that might influence reactivity, such as atmospheric gases and metal ions, pure concentrated sulfuric acid is a foundational starting point for investigating stability of chemicals needed for life to function.

Specifically, we are motivated to study the stability of the components that could make up an information polymer, one that would be very different from DNA or RNA on Earth due to the overall instability of DNA or RNA in concentrated sulfuric acid. We focus on nucleic acid bases and related molecules (Figure 1) as a starting point and consider sulfuric acid concentrations and temperatures found in the Venus cloud layers. Studies from decades ago explored not only stability and chemistry of aromatics in concentrated sulfuric acid (e.g., (36–38)) but also, curiously, some isolated experiments on protonation of a few of the nucleic acid bases in acidic conditions (39–42).

We describe our experimental results followed by contextual discussion, then present the methods.

## 2. Results

We find that nucleic acid bases adenine, cytosine, guanine, thymine, and uracil are stable in sulfuric acid ranging from 81 to 98% concentration by weight (the rest being water) at room temperature (18 to 21 deg ºC). We also find the stability under the same conditions of the core structure of the nucleic acid bases purine and pyrimidine. We additionally tested the adenine-like compound 2,6-diaminopurine, used as a genetic base substitute for adenine by specialized viruses (43–46), and found it stable as well.

By stable, we mean no detectable reactivity and degradation of the tested compound after incubation in concentrated sulfuric acid up to two weeks. We first analyzed the compounds after 18 to 24 hours in concentrated sulfuric acid and next repeated one of the analyses after a two-week incubation period for an additional stability test. By stable we also mean long-term stability of the heavy atom bonding topology in the bases, i.e. the bonding of non-hydrogen atoms to each other, and not rapid, reversible exchange of protons between the bases and the solvent.

We begin with evidence that the aromatic compound ring structure is not broken, using data from both UV spectroscopy (Section 2.1, Figure 2) and $^{13}$C nuclear magnetic resonance spectroscopy (NMR; Section 2.2, Figures 3 and 4). In order to demonstrate stability of a given molecule in concentrated sulfuric acid, we elucidate the molecular structure using a combination of $^1$H NMR, $^{13}$C NMR, $^{15}$N NMR, and/or 2D Heteronuclear Multiple-Quantum Correlation spectroscopy (HMQC) and Heteronuclear Multiple Bond Correlation spectroscopy (HMBC) ($^1$H-$^{13}$C and $^1$H-$^{15}$N) NMR (Section 3.3 and Figures 5-6).

### 2.1 Ultraviolet Spectroscopy

For an initial investigation of the compounds' stability in concentrated sulfuric acid, we used UV spectroscopy (190 to 300 nm; Figure 2). The two absorption peaks at ~180 nm (in the range 160 to 210 nm) and at ~260 nm wavelengths or longer are characteristic of nucleic acid bases and are due to the presence of conjugated π bonds in the aromatic rings (e.g. (47)). In other words, absorption is caused when π or nonbonding (nb) electrons jumping from $\pi \to \pi^*$ or nb $\to \pi^*$ molecular orbitals and so these absorptions are dependent on the presence of conjugated π systems. The wavelength of the two peaks for each nucleic acid base may slightly differ in concentrated sulfuric acid as compared to other solvents (e.g., (42, 48) and Table S9) due to the change in pH, which would cause a change to the conjugation on the nucleic acid bases (including likely protonation), thus leading to the wavelength shifts.

For each compound, the maximum wavelength and shape of each of the two major peaks did not change after about 24 hours in concentrated sulfuric acid (Figure 2), indicating stability of each compound, i.e., that the aromatic ring remained stable and intact. We do note a slight increase in absorbance for the sample after 24 hours due to increased dissolution of the tested compound over time. We emphasize that the shape of the curve stays the same over time, meaning that the rings themselves remain stable as the compound dissolves. We note that we do not detect any features beyond 300 nm (apart from a very low-amplitude signal from contamination around 320 nm), because the electron excitation is confined to shorter wavelengths.

Because we do not see any dramatically different absorption characteristics, we can further rule out a change in the substituent groups (apart from likely protonation). Such a substitution would preserve the ring structures but would alter the shape of the spectrum. We can also rule out any reaction to link two rings which would produce a multi-ring conjugated molecule, which would still have conjugated rings but be expected to have different absorption characteristics.

## 2.2 $^{13}$C NMR

To further confirm the stability for each compound and validate the compound's structure we used $^{13}$C NMR spectroscopy (Figure 3). We mixed 10 to 40 mg of each compound in 500 to 600 µl of a combination of $D_2SO_4$ and $D_2O$, ranging from 81% to 98% $D_2SO_4$ by weight, with 10% of DMSO-$d_6$ added as a chemical shift reference and took a $^{13}$C NMR spectrum 30 to 48 hours after sample preparation. We used $D_2O$ instead of $H_2O$ for all NMR experiments, so that we could use the same sample and minimize the intensity of the hydrogen peaks in any $^1$H NMR experiment. To confirm the stability of the bases over time, the $^{13}$C NMR measurements for all compounds in 98% w/w D$_2$SO$_4$ were additionally measured after two weeks later and the peak positions remain unchanged (Figures 4, S12-S14).

For each compound, the number of carbon peaks and their chemical shift position in the $^{13}$C NMR spectrum show that the aromatic ring structure and number of carbon atoms of the original compound is preserved over extended time periods in concentrated sulfuric acid (Figure 3 and Figure 4). The carbon peak assignments shown in Figure 3 and Figure 4 are described in Section 3.3 and in the SI. We made the assignment for a given compound in 98% w/w sulfuric acid and since the peaks in lower concentration of sulfuric acid are the same, we conclude that the bases in lower acid concentrations (i.e., 81-94% w/w) are also stable.

### 2.3 Molecular Structure Determination by NMR in 98% Concentrated Sulfuric Acid

To confirm that the molecular structure of nucleic acid bases does not change in concentrated sulfuric acid we use results from a series of 1D and 2D NMR experiments of nucleic acid bases dissolved in 98% D$_2$SO$_4$ with 2% D$_2$O (by weight). In this section we choose two exemplar nucleic acid bases to describe fully, purine and pyrimidine. Details of the assignments of the remaining six compounds that we investigated are presented in the SI.

Our aim in this section is to confirm the known chemical structure. A key point is that while the approximate chemical shifts of $^1$H and $^{13}$C are predictable from the proposed structure, their exact positions will vary with the solvent, and hence will not be known in a rarely studied solvent such as concentrated sulfuric acid. To identify NMR spectral peaks, we rely on general rules for chemical shifts for $^1$H, $^{13}$C, and $^{15}$N as well as known chemical shifts in solvents other than concentrated sulfuric acid (Tables S7 and S8), and combined with 2D NMR.

### 2.3.1 Purines in Concentrated Sulfuric Acid

For purine, in the 1D $^{13}$C NMR spectrum (Figure 3 and Figure 5A), we find five peaks for carbon that correspond to the five carbons in the purine ring. Each of the five peaks are found in the region of the NMR spectra associated with aromatic compounds.

We assign the carbon peaks by comparison with literature data (Table S7) and by our 2D NMR experiments (Figure 5D, E). We can assign C5 because it is the most magnetically shielded atom in the ring structure with a chemical shift distinctly upfield from the other four carbon peaks (Table S7). To assign C2, C6, and C8 we use our 2D NMR spectra where we correlate the positions of H and C within the ring. Purine has three protonated carbons, at C2, C6, and C8. Our 2D $^1$H-$^{13}$C HMQC shows three signals that correspond to C2 (at 149.39 ppm) attached to H2 (at 9.01 ppm), C6 (at 140.66 ppm) attached to H6 (at 9.20 ppm), and the C8 carbon peak at 150.00 ppm attached to H8 at 9.20 ppm. The two peaks corresponding to H6 and H8 hydrogens (at 9.20 ppm) on the 1D $^1$H NMR spectrum are overlapping (Figure 5B).

We assign C4 on the basis of $^1$H-$^{13}$C HMBC data (Figure 5E). As expected, the $^1$H-$^{13}$C HMBC correlates the chemical shift of carbon C4 at 151.04 ppm to the nearby H2, at 9.01 ppm, and H6, H8 at 9.20 ppm (all separated from C4 by three chemical bonds) (Figure 5E).

The $^1$H-$^{13}$C HMBC spectra also further confirms the assignments of C2, C5, C6, and C8. The C6 (at 140.66 ppm) correlates with the chemical shift of H2, at 9.02 ppm. C6 and H2 are at the right separation of three chemical bonds, supporting the assignment of 140.66 ppm and 9.02 ppm chemical shifts to C6 and H2 respectively. Further supporting these assignments and the integrity of the imidazole ring, the $^1$H-$^{13}$C HMBC correlates the distinct $^{13}$C chemical shift of C5 to the nearby H8 which in turn is correlated with C4. Note that there is also a weak signal on the $^1$H-$^{13}$C HMBC spectra (coordinates: 9.01 ppm, 120.62 ppm) (Figure 5E). The signal corresponds to the carbon atom C5 at the separation of four bonds from the H2 hydrogen.

Finally, we use 1D $^{15}$N NMR to confirm the presence of the N atoms in the purine ring (Figure 5C). We see four peaks corresponding to four nitrogen atoms of the purine ring and make assignments of N1 and N3 to 185.99 ppm and 262.08 ppm respectively based on the $^1$H-$^{15}$N HMBC experiment (Figure 5F). We assign N7 and N9 to 158.11 ppm and 163.02 ppm peaks respectively, based on the known N7 and N9 chemical shifts of purine in concentrated sulfuric acid (40), where N7 is more magnetically shielded (shifted towards lower ppm) in acidic conditions than N9 (Figure 7, Table S7).

Taken together the NMR data confirms that the purine ring structure remains intact in 98% w/w $D_2SO_4$ in $D_2O$.

We now turn to evidence of protonation of purine in concentrated sulfuric acid. We summarize pioneering work by Schumacher and Günther (40) and show that our NMR $^{15}$N chemical shifts agree in acidic solvents, thus demonstrating protonation of N atoms in the purine ring (Figure S15). Schumacher and Günther show that the purine ring is protonated in 90% w/w concentrated sulfuric acid by analyzing the chemical shift change of $^{15}$N NMR peaks of purine in different solvents ranging from strongly basic to strongly acidic (Figure 7). Purine $^{15}$N NMR spectra in DMSO-$d_6$ and $H_2O$ can be considered well-known standards. In the basic aqueous solution of 5% NaOH, the N7 and N9 atom spectral peaks are shifted downfield as compared to their peaks in $H_2O$ and DMSO-$d_6$ solutions indicating deprotonation of N atoms. With increasing solvent acidity, N atoms sequentially get protonated, causing dramatic upfield spectral peak migration. In some cases (e.g., H$_2$O and DMSO-d$_6$) two tautomeric structures exist in equilibrium due to fast proton exchange; the relative abundance controls the spectral peak positions. For example, in H$_2$O, the N7 and N9 spectral peaks are relatively close to each other, indicating a similar chemical environment, including a similar time-averaged protonation state, i.e., a relatively equal abundance of an N7 and N9 protonated tautomeric structures (40). In comparison, in DMSO vs. H$_2$O, the purine tautomeric structure with protonated N9 is more abundant than that with protonated N7 (49). Note that the N3 atom is not protonated even in 90% H$_2$SO$_4$ (40), as shown by the insignificant migration of its spectral peak. However, N3 does get protonated in superacids (41). Our data at 98% D$_2$SO$_4$ match the high acidity solvents 90% H$_2$SO$_4$ and FSO$_3$H from (40), demonstrating protonation of the N1, N7, and N9 nitrogen atoms in the purine molecule (Figure 7).

For other purines, adenine, guanine, and diaminopurine, we demonstrate stability in concentrated sulfuric acid as follows. In the 1D $^{13}$C NMR spectra (Figure 3), we find five peaks for carbon that correspond to the five carbons in the adenine, guanine, and diaminopurine ring. Each of the five peaks are found in the region of the NMR spectra associated with aromatic compounds. To further confirm the integrity and stability of the adenine, guanine, and diaminopurine nucleic acid bases in concentrated sulfuric acid we employ $^1$H and $^{15}$N NMR, combined with HMQC and HMBC 2D NMR (Figures S1, S3, and S5). The results of the $^{15}$N NMR show the correct number of nitrogen atoms in the adenine, guanine, and diaminopurine rings. The 2D NMR spectra further confirms the $^{13}$C, $^{15}$N, and $^1$H peak assignments. Taken together the results show that adenine, guanine, and diaminopurine are stable in concentrated sulfuric acid for a prolonged time. For detailed justification and description of the results, please see the SI.

We now turn to assessment of the protonation state of adenine, guanine, and diaminopurine. Apart from purine itself, the protonation of other purines in concentrated sulfuric acid has not been previously studied. Comparison of the $^{13}$C NMR chemical shift changes in different solvents suggests that guanine, adenine, and diaminopurine are protonated in concentrated sulfuric acid (Table S1, Table S3, Table S5). Protonation of nitrogen atoms results in greater shielding of the nearby carbon atoms and the upfield shift (towards lower ppm) of peaks corresponding to carbon atoms directly adjacent to the nitrogen atoms that are protonated in the acidic solvent. This effect is uniform among all tested purines, where the upfield shift of $^{13}$C NMR carbon peaks in concentrated sulfuric acid is particularly pronounced for C2, C4, and C6 peaks (Table S1, Table S3, Table S5). All nitrogen atoms of adenine, guanine, and diaminopurine appear to be protonated in concentrated sulfuric acid, while the protonation of the oxygen atom in the carbonyl group of guanine is a possibility.

## 2.3.2 Pyrimidines in Concentrated Sulfuric Acid

We now turn to our second example, pyrimidine, again as in the purine case aiming to confirm the known structure to demonstrate stability in concentrated sulfuric acid. Also as for purine, while the chemical shifts are known for pyrimidine, they are different in concentrated sulfuric acid as compared to other solvents (Table S8).

For pyrimidine, the 1D $^{13}$C NMR spectrum (Figure 3 and Figure 6A) shows three peaks for carbon that correspond to the four carbons in the pyrimidine ring. Since carbons C4 and C6 are distributed symmetrically within the pyrimidine ring they present as a single NMR peak. All peaks are found in the region of the NMR spectra associated with aromatic compounds.

We assign the carbon peaks by comparison with literature data (Table S8) and by our 2D NMR experiments (Figure 6D, E). We can assign C5 because it is the most magnetically shielded atom in the ring structure with a chemical shift distinctly upfield from the other two carbon peaks (Table S8). To assign C2, C4, and C6 we use our 2D NMR spectra where we correlate the positions of H and C within the ring (Figure 6D, E). Pyrimidine has four carbons with directly attached hydrogen atoms. Our 2D $^{1}$H-$^{13}$C HMQC shows three distinct signals that we use to assign C2, C4, C6, and C5 (Figure 6D). The first signal corresponds to the C2 carbon peak at 149.85 ppm and H2 (bonded to C2) at 9.68 ppm. We assign the peak at 158.18 ppm to symmetric carbon atoms C4 and C6. The 2D $^{1}$H-$^{13}$C HMQC also confirms the assignment of C5 at 127.77 ppm.

We further confirm our carbon atoms peak assignments on the basis of $^{1}$H-$^{13}$C HMBC data (Figure 6E).

Finally, we use 1D $^{15}$N NMR to confirm the presence of the symmetric N atoms in the pyrimidine ring (Figure 6C). We detect a single peak at 201.21 ppm corresponding to two symmetrical nitrogen atoms, N1 and N3, of the pyrimidine ring. The $^{1}$H-$^{15}$N HMBC experiment (Figure 6F) further confirms this assignment.

Taken together the NMR data confirms that the pyrimidine ring structure remains intact in 98% w/w $D_2SO_4$ in $D_2O$.

We now describe evidence of protonation of pyrimidine in concentrated sulfuric acid. Work by Wagner and von Philipsborn (39) shows that pyrimidine is protonated in strong acid solvents, as compared to less acidic solvents. While Wagner and von Philipsborn do not use $^{15}$N NMR, they instead analyze $^{1}$H NMR data to demonstrate protonation in fluorosulfuric acid (FSO$_3$H). The effects of the two acids, concentrated sulfuric acid and fluorosulfuric acid, are shown to be the same as demonstrated by purine showing the same chemical shift changes and protonation state in fluorosulfuric acid as in sulfuric acid (40) (Figure 7).

In our $^{13}$C NMR spectra we see a significant spectral peak upfield shift of C2 as compared to C2 spectral peaks in lower acidity solvents such as DMSO-d$_6$ and D$_2$O (Table S8). Protonation of the adjacent ring N atom results in shielding of the nearby C2 nucleus, causing the upfield shift. For pyrimidine, we see shifts in $^{13}$C peaks as the concentration of acid increases from 81% to 98% (Figure 4). For the same reasons, carbon C5 becomes significantly deshielded by the protonation of N atoms in the pyrimidine ring.

For other pyrimidines, cytosine, uracil, and thymine, we demonstrate stability in concentrated sulfuric acid as follows. In the 1D $^{13}$C NMR spectra (Figure 3 and Figure 6A), we find four peaks for carbon that correspond to the four carbons in the pyrimidine ring. Each of the four peaks are found in the region of the NMR spectra associated with aromatic compounds. To further confirm the integrity and stability of the cytosine, uracil and thymine nucleic acid bases in concentrated sulfuric acid we employ $^1$H and $^{15}$N NMR, combined with 2D HMQC and HMBC NMR. The results of the $^{15}$N NMR show the correct number of nitrogen atoms in the cytosine, uracil, and thymine rings (Figures S2, S4, S6). The 2D NMR further confirms the $^{13}$C, $^{15}$N, and $^1$H peak assignments. Taken together the results show that cytosine, uracil and thymine are stable in concentrated sulfuric acid for a prolonged time. For detailed justification and the description of the results please see SI.

We now turn to assessment of the protonation state of other pyrimidines cytosine, uracil, and thymine. For cytosine, uracil, and thymine, the N atoms and the carbonyl group O atoms are likely protonated in concentrated sulfuric acid (Figure S15). The $^{13}$C NMR carbon chemical shifts of cytosine, uracil, and thymine in concentrated sulfuric acid reported by Benoit and Frechette agree with ours (Table S2, S4, S6) and are consistent with protonation of nitrogen atoms and oxygens of the carbonyl groups (42). The protonation of cytosine, uracil, and thymine in strong acid is further supported by early studies on protonation of pyrimidines in FSO$_3$H (39).

In our NMR analysis we have not fully addressed protonation of the nucleic acid bases, likely in the high acid solvent concentrated sulfuric acid. In this paper we are concerned with the long-term stability of the heavy atom bonding topology in the bases, i.e., the bonding of non-hydrogen atoms to each other, and not with rapid, reversible exchange of protons between the nucleic acid bases and the solvent. Thus, while protonation of the ring N atoms is formally a chemical change, it will at least not destabilize the structure of the bases as conventionally drawn (i.e. without explicit hydrogens), and may well stabilize it to certain reactivity (e.g. once protonated, the bases will react only as electrophiles).

In this assessment of stability, we follow convention, as illustrated by the common understanding that carboxylic acids and amines are stable to pH difference; even though reducing the pH causes the molecules to lose or gain a proton respectively, this change does not alter how the non-hydrogen atoms are bonded to each other, and losing or gaining a proton is readily reversed by reversing the pH change. The stability of DNA bases shown in this paper is in contrast to the instability of other species in concentrated sulfuric acid, such as some aliphatic carbonyl compounds, where protonation of the oxygen leads to rapid and irreversible rearrangement of the carbon skeleton (50).

## 3. Discussion

For life to exist, complex organic chemistry has to be capable of forming an informational biopolymer. On Earth those polymers are DNA and RNA. The complementarity of the nucleic acid bases is the key to the DNA and RNA structure and to its informational function. The stability of nucleic acid bases in concentrated sulfuric acid therefore goes far beyond the finding of a new category of molecules stable in this aggressive solvent—and towards the design of an informational biopolymer that employs the canonical or other, modified nucleic acid bases (e.g., if protonation of the nitrogen atoms in the ring prevents their pairing). While it is already well known that DNA's phosphate ester backbone and ribose linker are unstable in concentrated sulfuric acid, we can aim to find a stable substitute, with similar structural characteristics, size, flexibility, shape, etc. By designing a stable informational biopolymer in concentrated sulfuric acid we can support the idea that concentrated sulfuric acid can support some kind of life different from Earth's, because such a polymer can in principle carry information and support genetics and Darwinian evolution.

One key gap in understanding a path from sulfuric acid-stable bases to a sulfuric acid-stable informational polymer is on the likely protonation of the nucleic acid bases. Because concentrated sulfuric acid is such a strong acid it may be that every atom that could serve as a hydrogen bond acceptor is protonated. Protonation will change the nucleic acid base's physical shape and hydrogen bonding potential, and hence is relevant for formation of macromolecular structures such as a double helix. Specifically, protonation of the N atoms in the ring would interfere with the canonical Watson-Crick base pairing between the bases (Figure S16). Understanding the protonation of N atoms in the canonical DNA/RNA nucleic acid bases at different concentrations of sulfuric acid is essential for progress. As relevant to a sulfuric acid-stable informational polymer, we emphasize that alternative nucleic acid bases (e.g. (51)), such as 4-methylbenzimidazole and 2,4-difluorotoluene, are known to form functional nucleic base pairs (52) that will not be subjected to N protonation in concentrated sulfuric acid and hence the N protonation will not affect their potential base pair complementarity. Whether such nucleic acid bases can form pairs in solution in concentrated sulfuric acid is a subject for future work.

We do not know if the origin of life in concentrated sulfuric acid is possible, but such a possibility cannot be excluded a priori. Life could use concentrated sulfuric acid as a solvent instead of water and could have originated in the cloud droplets in liquid concentrated sulfuric acid. This scenario does not require Venus to be canonically habitable in the past, i.e., with liquid surface water, and is relevant because some researchers think Venus was always too warm for surface liquid water oceans to condense and may have looked like it does today for most of its geological history (53). In this scenario, the Venus atmosphere could still support the strictly aerial concentrated sulfuric acid-based life.

Another hypothetical scenario for how life might have come to be in the Venusian clouds involves life changing its solvent in the course of evolution (i.e., undergo solvent replacement). In this scenario Venus had early water oceans, that later evaporated (54, 55), and life originated in water on the early Venus billions of years ago. As water became very scarce, life would have adapted its biochemistry to the concentrated sulfuric acid environment, eventually substituting water for concentrated sulfuric acid. The solvent replacement might appear as an impossible adaptation as most of Earth's life biochemicals that are stable in water are not stable in concentrated sulfuric acid (25). We do not however know for sure if such gradual adaptation to another solvent is possible or impossible, if the change would have been gradual enough (over millions, if not billions of years). It may have been, considering the evolutionary pressure would have been constant and strong. However, we definitely have no example of such a mechanism and adaptation on Earth.

For completeness we also mention a third possibility for how life could be in the Venusian clouds, in which the Venus cloud-based life originated in water on the early Venus billions of years ago and as water became very scarce, life would have adapted its biochemistry to the concentrated sulfuric acid environment but never to the point of using it as a solvent. Such a scenario involves "neutralization" of sulfuric acid droplets with biologically produced ammonia (4, 5) and other specialized adaptations to the hostile sulfuric acid environment (56). Many unexplained atmospheric gases detected in the clouds of Venus and other cloud properties are consistent with this hypothesis (32). We also emphasize that if life exist in the clouds of Venus it has to be very different than life on Earth, with biochemical and evolutionary adaptations that have no precedent here on Earth.

All life needs a liquid solvent and concentrated sulfuric acid is one of only three liquids known to exist on rocky solar system planets and moons. Liquid water in present on the surface of the Earth and in the subsurface of several icy moons (e.g., Europa and Enceladus). Liquid hydrocarbons have been identified on Saturn's moon Titan. Concentrated sulfuric acid as a planetary solvent is not only relevant to Venus, but also for exoplanets; while we know rocky exoplanets are common, we do not know if Venus-like planets are more prevalent than Earth-like exoplanets. Ballesteros et al. (57) postulate that concentrated sulfuric acid is one of the most common liquids in the Galaxy, although the stability of concentrated sulfuric acid as a surface liquid is unknown. Further studies of concentrated sulfuric acid organic chemistry can therefore move the search for habitable worlds forward.

Venus is right "next door" and the cloud particles can be directly probed by space missions. NASA's Pioneer Venus *in situ* missions in 1978, and several Soviet descent craft, established the particle size distribution in the clouds, but did not unambiguously determine the composition of all types of cloud particles or search for organic chemicals. Today NASA and ESA have plans to send missions to Venus at the end of this decade, though none of the three planned missions will probe the cloud particles directly (58–60).

Our findings that show that complex organic chemistry, including DNA nucleic acid bases, can be stable in concentrated sulfuric acid and motivates us to design missions that directly probe the cloud particles for the presence of organic material. As a first mission in our "Morning Star Missions to Venus" program, our Rocket Lab Mission to Venus is planned for launch in January 2025 (61). The Rocket Lab mission will deliver a probe containing one instrument, the autofluorescence nephelometer, to search for autofluorescence (and backscattered polarized radiation) in the Venus cloud droplets, indicative of organic molecules, especially aromatic ring systems (62). More sophisticated future missions (63–65), can then identify organics if they are present (66). Ultimately a sample return from the Venus atmosphere may be needed to robustly identify life, if present (67).

## 4. Materials and Methods

### 4.1 Materials

We purchased chemical compounds from Millipore-Sigma, with pyrimidine ordered from Thermo Scientific. The compounds were used without further purification. The catalog numbers and purities are as follows: adenine A8626-5G ≥99%; cytosine C3506-1G ≥99%; 2,6-diaminopurine 247847-1G 98%; guanine G11950-10G 98%; purine P55805-1G 98%; pyrimidine 289-95-2 99%; thymine T0376-10G ≥99%; and uracil U0750-25G ≥99%. We used $H_2SO_4$ from Sigma Aldrich (used for UV spectroscopy) and $D_2SO_4$ from ACROS Organics (sulfuric acid-d2 for NMR, 98 wt.% in $D_2O$, 99.5+ atom % D) and $D_2O$ (deuteration degree min 99.9%) from MagniSolv.

### 4.2 Ultraviolet and Visible Wavelength Spectroscopy

We used ultraviolet and visible (UV-Vis) wavelength spectrometry from 190-600 nm for our initial investigation of compound stability in 98% w/w $H_2SO_4$ in $H_2O$. We prepared our samples by mixing each compound into a small stock solution and diluted the solution until the measurement reached a desired absorbance value near $A = 1$, where, from the Beer-Lambert law $A = \varepsilon L c$, where $A$ is the absorbance (dimensionless), $\varepsilon$ is the molar absorption coefficient in units of $M^{-1}$ $cm^{-1}$ (on order of 6,000 to 13,000 in water (68)), $L$ is pathlength in cm (the cuvette width was 1 cm), and $c$ is concentration in molarity. We stored the solution in a sealed glass vial and after about 24 hours put the solution back into a quartz cuvette for remeasurement. Our UV-Vis spectrometer is a HELIOS OMEGA, 244001, v8.01 with a scan rate of 1200 nm/min and data interval of 1 nm. No data analysis was required other than the baseline correction from an initially measured sample of 98% w/w $H_2SO_4$ with 2% $H_2O$.

The concentration of each compound in 98% w/w $H_2SO_4$ in $H_2O$ is given in Table S9, and comes from the mass (or volume in the case of pyrimidine) in 3.2 mL of $H_2SO_4$. The concentrations are low, less than one mM, because the compounds are strong UV absorbers. The original data for all UV-Vis experiments are available for download as Supplementary Dataset 1.

### 4.3 1D and 2D NMR Spectroscopy

We prepared our NMR samples by dissolving 10 to 80 mg of the aromatic compounds into 500 to 600 μL of solvent $D_2SO_4$ in $D_2O$ in glass vials. We added DMSO-$d_6$, used as a chemical shift reference compound, to a final concentration of 10% by volume. We used 10 to 40 mg of compounds for the 1D $^1H$ and $^{13}C$ NMR. We used 80 mg for the $^{15}N$

NMR due to low sensitivity (except for adenine for which we used 50 mg) and also used 80 mg for the 2D NMR (though using the lower 30 to 40 mg concentrations did not change the results). For 2,6-diaminopurine we used only 10 mg for the 6-19% $D_2O$ solutions due to limited solubility. We heated sealed glass vials in a hot water bath (~80 ºC) when this was needed to dissolve the compounds. Out of all the compounds analyzed only 2,6-diaminopurine partially precipitated back out on cooling to room temperature. We stored the sealed glass vials for 12 to 48 hours before transferring the solution to 5 mm NMR tubes. After NMR measurements we stored the solutions in the NMR tubes, where the storage room temperature varied from about 18 to 24 ºC.

To acquire NMR data, we used a Bruker Avance III-HD 400 MHz spectrometer equipped with a Prodigy liquid nitrogen cryoprobe (BBO) at 25 °C. We acquired 1D $^1H$, $^{13}C$, $^{15}N$, 2D $^1H$-$^{13}C$ HMBC and HMQC, and $^1H$-$^{15}N$ HMBC NMR spectra to confirm the structures and hence stability of the compounds in 98% w/w $D_2SO_4$ in $D_2O$. For our solutions of 94%, 88%, and 81% $D_2SO_4$ in $D_2O$ (by weight) we acquired 1D $^{13}C$ NMR spectra. In all cases we locked on DMSO-$d_6$ for consistency, where we found the DMSO-$d_6$ peak to be at 33.44 ppm +/- 0.02 ppm in our 98% w/w $D_2SO_4$ solutions. For solutions of different acidity, the DMSO-$d_6$ varied by about +/- 0.1 ppm. We note that for uracil, cytosine and pyrimidine in 81% w/w sulfuric acid the DMSO-$d_6$ was significantly shifted to about 26.70 ppm. For Figures 3 and 4 we set the DMSO-$d_6$ reference to 33.44 ppm in all cases.

We used MNova software (Mestrelab Research) to process and analyze the NMR data (69). The original data for all NMR experiments are available for download as Supplementary Dataset 2.

## Acknowledgements
We thank the MIT Department of Chemistry Instrumentation Facility NMR Consultant Bruce Adams and Director Walter Massefski. We thank Adam Jost for experimental assistance with the mass microbalance. We thank Jingcheng Huang for help with the compilation of the spectroscopic data from the literature. We thank Steven Benner and Jan Spacek for useful discussion. This work was partially funded by MIT and Nanoplanet Consulting LLC.

## References

1. H. Morowitz, C. Sagan, Life in the clouds of venus? *Nature* **215**, 1259–1260 (1967).
2. D. H. Grinspoon, *Venus revealed: a new look below the clouds of our mysterious*

*twin planet* (1997).
3. M. R. Patel, J. P. Mason, T. A. Nordheim, L. R. Dartnell, Constraints on a potential aerial biosphere on Venus: II. Ultraviolet radiation. *Icarus*, 114796 (2021).
4. R. Mogul, S. S. Limaye, Y. J. Lee, M. Pasillas, Potential for Phototrophy in Venus' Clouds. *Astrobiology* **21**, 1237–1249 (2021).
5. W. Bains, J. J. Petkowski, P. B. Rimmer, S. Seager, Production of Ammonia Makes Venusian Clouds Habitable and Explains Observed Cloud-Level Chemical Anomalies. *Proc. Natl. Acad. Sci.* **118** (2021).
6. S. Seager, *et al.*, The Venusian Lower Atmosphere Haze as a Depot for Desiccated Microbial Life: a Proposed Life Cycle for Persistence of the Venusian Aerial Biosphere. *Astrobiology* **21**, 1206–1223 (2021).
7. C. S. Cockell, Life on venus. *Planet. Space Sci.* **47**, 1487–1501 (1999).
8. D. Schulze-Makuch, L. N. Irwin, Reassessing the possibility of life on venus: Proposal for an astrobiology mission. *Astrobiology* **2**, 197–202 (2002).
9. D. Schulze-Makuch, L. N. Irwin, The prospect of alien life in exotic forms on other worlds. *Naturwissenschaften* **93**, 155–172 (2006).
10. D. Schulze-Makuch, D. H. Grinspoon, O. Abbas, L. N. Irwin, M. A. Bullock, A sulfur-based survival strategy for putative phototrophic life in the Venusian atmosphere. *Astrobiology* **4**, 11–18 (2004).
11. D. H. Grinspoon, M. A. Bullock, Astrobiology and Venus exploration. *Geophys. Monogr. Geophys. Union* **176**, 191 (2007).
12. S. S. Limaye, *et al.*, Venus' Spectral Signatures and the Potential for Life in the Clouds. *Astrobiology* **18**, 1181–1198 (2018).
13. O. R. Kotsyurbenko, *et al.*, Exobiology of the Venusian Clouds: New Insights into Habitability through Terrestrial Models and Methods of Detection. *Astrobiology* (2021).
14. L. R. Dartnell, *et al.*, Constraints on a potential aerial biosphere on Venus: I. Cosmic rays. *Icarus* **257**, 396–405 (2015).
15. J. Baross, *et al.*, *The limits of organic life in planetary systems* (National Academies Press, 2007).
16. M. Vaïtilingom, *et al.*, Long-term features of cloud microbiology at the puy de Dôme (France). *Atmos. Environ.* **56**, 88–100 (2012).
17. P. Amato, *et al.*, Active microorganisms thrive among extremely diverse communities in cloud water. *PLoS One* **12**, e0182869 (2017).
18. P. Amato, *et al.*, Metatranscriptomic exploration of microbial functioning in clouds. *Sci. Rep.* **9** (2019).
19. N. C. Bryan, B. C. Christner, T. G. Guzik, D. J. Granger, M. F. Stewart, Abundance and survival of microbial aerosols in the troposphere and stratosphere. *ISME J.*, 1–11 (2019).
20. S. M. Burrows, *et al.*, Bacteria in the global atmosphere–Part 2: Modeling of emissions and transport between different ecosystems. *Atmos. Chem. Phys.* **9**, 9281–9297 (2009).
21. R. Knollenberg, *et al.*, The clouds of Venus: A synthesis report. *J. Geophys. Res. Sp. Phys.* **85**, 8059–8081 (1980).
22. J. E. Hansen, J. W. Hovenier, Interpretation of the polarization of Venus. *J.*

*Atmos. Sci.* **31**, 1137–1160 (1974).
23. J. H. Hoffman, V. I. Oyama, U. Von Zahn, Measurements of the Venus lower atmosphere composition: A comparison of results. *J. Geophys. Res. Sp. Phys.* **85**, 7871–7881 (1980).
24. V. A. Krasnopolsky, Vertical profiles of H2O, H2SO4, and sulfuric acid concentration at 45–75 km on Venus. *Icarus* **252**, 327–333 (2015).
25. W. Bains, J. J. Petkowski, Z. Zhan, S. Seager, Evaluating Alternatives to Water as Solvents for Life: The Example of Sulfuric Acid. *Life* **11**, 400 (2021).
26. J. E. Hallsworth, *et al.*, Water activity in Venus's uninhabitable clouds and other planetary atmospheres. *Nat. Astron.*, 1–11 (2021).
27. J. Spacek, Organic Carbon Cycle in the Atmosphere of Venus. *arXiv Prepr. arXiv2108.02286* (2021).
28. S. A. Benner, J. Spacek, The Limits to Organic Life in the Solar System: From Cold Titan to Hot Venus. *LPI Contrib.* **2629**, 4003 (2021).
29. J. Spacek, S. A. Benner, The organic carbon cycle in the atmosphere of venus and evolving red oil. *LPI Contrib.* **2629**, 4052 (2021).
30. J. Spacek, *et al.*, Organics produced in the clouds of Venus resemble the spectrum of the unknown absorber. *LPI Contrib.* **2807**, 8060 (2023).
31. D. V Titov, N. I. Ignatiev, K. McGouldrick, V. Wilquet, C. F. Wilson, Clouds and hazes of Venus. *Space Sci. Rev.* **214**, 1–61 (2018).
32. J. J. Petkowski, *et al.*, Astrobiological Potential of Venus Atmosphere Chemical Anomalies and Other Unexplained Cloud Properties. *Astrobiology* **in press** (2023).
33. S. Miron, R. J. Lee, Molecular Structure of Conjunct Polymers. *J. Chem. Eng. Data* **8**, 150–160 (1963).
34. L. F. Albright, L. Houle, A. M. Sumutka, R. E. Eckert, Alkylation of isobutane with butenes: effect of sulfuric acid compositions. *Ind. Eng. Chem. Process Des. Dev.* **11**, 446–450 (1972).
35. Q. Huang, G. Zhao, S. Zhang, F. Yang, Improved catalytic lifetime of H2SO4 for isobutane alkylation with trace amount of ionic liquids buffer. *Ind. Eng. Chem. Res.* **54**, 1464–1469 (2015).
36. H. Cerfontain, A. Telder, The solubility of toluene and benzene in concentrated aqueous sulfuric acid; implications to the kinetics of aromatic sulfonation. *Recl. des Trav. Chim. des Pays‑Bas* **84**, 545–550 (1965).
37. H. Cerfontain, F. L. J. Sixma, L. Vollbracht, Aromatic sulphonation IX heterogeneous sulphonation of toluene with aqueous sulphuric acid. *Recl. des Trav. Chim. des Pays‑Bas* **83**, 226–232 (1964).
38. H. Cerfontain, Solubility of aromatic hydrocarbons in aqueous sulfuric acid. *Recl. des Trav. Chim. des Pays‑Bas* **84**, 491–502 (1965).
39. R. Wagner, W. von Philipsborn, Protonierung von Amino-und Hydroxypyrimidinen NMR-Spektren und Strukturen der Mono-und Dikationen. *Helv. Chim. Acta* **53**, 299–320 (1970).
40. M. Schumacher, H. Günther, Beiträge zur 15N-NMR-Spektroskopie Protonierung und Tautomerie in Purinen: Purin und 7-und 9-Methylpurin. *Chem. Ber.* **116**, 2001–2014 (1983).
41. R. Wagner, W. von Philipsborn, Protonierung von Purin, Adenin und Guanin

NMR.-Spektren und Strukturen der Mono-, Di-und Tri-Kationen. *Helv. Chim. Acta* **54**, 1543–1558 (1971).
42. R. L. Benoit, M. Frechette, 1H and 13C nuclear magnetic resonance and ultraviolet studies of the protonation of cytosine, uracil, thymine, and related compounds. *Can. J. Chem.* **64**, 2348–2352 (1986).
43. Y. Zhou, *et al.*, A widespread pathway for substitution of adenine by diaminopurine in phage genomes. *Science (80-. ).* **372**, 512–516 (2021).
44. M. D. Kirnos, I. Y. Khudyakov, N. I. Alexandrushkina, B. F. Vanyushin, 2-Aminoadenine is an adenine substituting for a base in S-2L cyanophage DNA. *Nature* **270**, 369–370 (1977).
45. V. Pezo, *et al.*, Noncanonical DNA polymerization by aminoadenine-based siphoviruses. *Science (80-. ).* **372**, 520–524 (2021).
46. D. Sleiman, *et al.*, A third purine biosynthetic pathway encoded by aminoadenine-based viral DNA genomes. *Science (80-. ).* **372**, 516–520 (2021).
47. M. S. H. Akash, K. Rehman, "Ultraviolet-visible (UV-VIS) spectroscopy" in *Essentials of Pharmaceutical Analysis*, (Springer, 2020), pp. 29–56.
48. L. F. Cavalieri, A. Bendich, J. F. Tinker, G. B. Brown, Ultraviolet absorption spectra of purines, pyrimidines and triazolopyrimidines. *J. Am. Chem. Soc.* **70**, 3875–3880 (1948).
49. M. Schumacher, H. Guenther, Carbon-13-proton spin-spin coupling. 9. Purine. *J. Am. Chem. Soc.* **104**, 4167–4173 (1982).
50. W. Bains, J. J. Petkowski, S. Seager, A Data Resource for Sulfuric Acid Reactivity of Organic Chemicals. *Data* **6**, 24 (2021).
51. K. Dhami, *et al.*, Systematic exploration of a class of hydrophobic unnatural base pairs yields multiple new candidates for the expansion of the genetic alphabet. *Nucleic Acids Res.* **42**, 10235–10244 (2014).
52. J. C. Morales, E. T. Kool, Efficient replication between non-hydrogen-bonded nucleoside shape analogs. *Nat. Struct. Biol.* **5**, 950–954 (1998).
53. M. Turbet, *et al.*, Day–night cloud asymmetry prevents early oceans on Venus but not on Earth. *Nature* **598**, 276–280 (2021).
54. M. J. Way, *et al.*, Was Venus the first habitable world of our solar system? *Geophys. Res. Lett.* **43**, 8376–8383 (2016).
55. M. J. Way, A. D. Del Genio, Venusian Habitable Climate Scenarios: Modeling Venus Through Time and Applications to Slowly Rotating Venus-Like Exoplanets. *J. Geophys. Res. Planets* **125**, e2019JE006276 (2020).
56. W. Bains, J. J. Petkowski, S. Seager, Venus' atmospheric chemistry and cloud characteristics are compatible with Venusian life. *Astrobiology* **in press** (2023).
57. F. J. Ballesteros, A. Fernandez-Soto, V. J. Martínez, Diving into Exoplanets: Are Water Seas the Most Common? *Astrobiology* (2019).
58. J. B. Garvin, *et al.*, Revealing the Mysteries of Venus: The DAVINCI Mission. *Planet. Sci. J.* **3**, 117 (2022).
59. A. Freeman, *et al.*, Veritas: A discovery-class Venus surface geology and geophysics mission (2016).
60. M. R. R. de Oliveira, P. J. S. Gil, R. Ghail, A novel orbiter mission concept for venus with the EnVision proposal. *Acta Astronaut.* **148**, 260–267 (2018).
61. R. French, *et al.*, Rocket Lab Mission to Venus. *Aerospace* **9**, 445 (2022).

62. D. Baumgardner, *et al.*, Deducing the Composition of Venus Cloud Particles with the Autofluorescence Nephelometer (AFN). *Aerospace* **9**, 492 (2022).
63. S. Seager, *et al.*, Venus Life Finder Habitability Mission: Motivation, Science Objectives, and Instrumentation. *Aerospace* **9**, 733 (2022).
64. R. Agrawal, *et al.*, Mission Architecture to Characterize Habitability of Venus Cloud Layers via an Aerial Platform. *Aerospace* **9**, 359 (2022).
65. W. P. Buchanan, *et al.*, Aerial Platform Design Options for a Life-Finding Mission at Venus. *Aerospace* **9**, 363 (2022).
66. N. F. W. Ligterink, *et al.*, The ORIGIN Space Instrument for Detecting Biosignatures and Habitability Indicators on a Venus Life Finder Mission. *Aerospace* **9**, 312 (2022).
67. S. Seager, *et al.*, Venus Life Finder Missions Motivation and Summary. *Aerospace* **9**, 385 (2022).
68. M. Taniguchi, J. S. Lindsey, Database of absorption and fluorescence spectra of> 300 common compounds for use in photochem CAD. *Photochem. Photobiol.* **94**, 290–327 (2018).
69. M. R. Willcott, MestRe Nova. *J. Am. Chem. Soc.* **131**, 13180 (2009).
70. T. Saito, *et al.*, Spectral database for organic compounds (sdbs). *Natl. Inst. Adv. Ind. Sci. Technol.* (2006).
71. M. T. Chenon, R. J. Pugmire, D. M. Grant, R. P. Panzica, L. B. Townsend, Carbon-13 magnetic resonance. XXVI. Quantitative determination of the tautomeric populations of certain purines. *J. Am. Chem. Soc.* **97**, 4636–4642 (1975).
72. M. C. Thorpe, W. C. Coburn Jr, J. A. Montgomery, The 13C nuclear magnetic resonance spectra of some 2-, 6-, and 2, 6-substituted purines. *J. Magn. Reson.* **15**, 98–112 (1974).
73. A. Unciti-Broceta, M. J. Pineda de las Infantas, M. A. Gallo, A. Espinosa, Reduction of Different Electron-Poor N-Heteroarylhydrazines in Strong Basic Conditions. *Chem. Eur. J.* **13**, 1754–1762 (2007).
74. M. Česnek, *et al.*, Synthesis and properties of 2-guanidinopurines. *Collect. Czechoslov. Chem. Commun.* **71**, 1303–1319 (2006).
75. H. J. Schneider, *et al.*, Complexation of nucleosides, nucleotides, and analogs in an azoniacyclophane. Van der Waals and electrostatic binding increments and NMR shielding effects. *J. Am. Chem. Soc.* **114**, 7704–7708 (1992).
76. T. H. Graham, W. Liu, D.-M. Shen, A Method for the Reductive Scission of Heterocyclic Thioethers. *Org. Lett.* **13**, 6232–6235 (2011).
77. W. C. J. Coburn, M. C. Thorpe, J. A. Montgomery, K. Hewson, Correlation of the Proton Magnetic Resonance Chemical Shifts of Substituted Purines with Reactivity Parameters. I. 2,6-Disubstituted Purines. *J. Org. Chem.* **30**, 1110–1113 (1965).
78. W. C. J. Coburn, M. C. Thorpe, J. A. Montgomery, K. Hewson, Correlation of the Proton Magnetic Resonance Chemical Shifts of Substituted Purines with Reactivity Parameters. II. 6-Substituted Purines. *J. Org. Chem.* **30**, 1114–1117 (1965).
79. R. Marek, V. Sklenar, NMR Studies of Purines. *Annu. reports NMR Spectrosc.* **54**, 201–242 (2005).

80. R. K. Harris, E. D. Becker, S. M. Cabral de Menezes, R. Goodfellow, P. Granger, NMR Nomenclature: Nuclear Spin Properties and Conventions for Chemical Shifts. IUPAC Recommendations 2001. *Solid State Nucl. Magn. Reson.* **22**, 458–483 (2002).
81. K. Goel, S. Bera, M. Singh, D. Mondal, Synthesis of dual functional pyrimidinium ionic liquids as reaction media and antimicrobial agents. *RSC Adv.* **6**, 106806–106820 (2016).
82. J. Clark, G. Hitiris, Covalent hydration of 5-substituted pyrimidines. *Spectrochim. Acta Part A Mol. Spectrosc.* **40**, 75–79 (1984).
83. R. J. Pugmire, D. M. Grant, Carbon-13 magnetic resonance. X. Six-membered nitrogen heterocycles and their cations. *J. Am. Chem. Soc.* **90**, 697–706 (1968).
84. J. Riand, M. T. Chenon, N. Lumbroso-Bader, Etude par rmn du carbone-13 des effets de substituants dans le noyau de la pyrimidine. *Tetrahedron Lett.* **15**, 3123–3126 (1974).
85. A. Y. Denisov, V. I. Mamatyuk, O. P. Shkurko, Additivity of 13C-1H and 1H-1H spin-spin coupling constants in six-membered aromatic nitrogen-containing heterocycles. *Chem. Heterocycl. Compd.* **21**, 821–825 (1985).
86. K. J. Sheehy, L. M. Bateman, N. T. Flosbach, M. Breugst, P. A. Byrne, Identification of N-or O-Alkylation of Aromatic Nitrogen Heterocycles and N-Oxides Using 1H–15N HMBC NMR Spectroscopy. *European J. Org. Chem.* **2020**, 3270–3281 (2020).
87. A. Dokalik, H. Kalchhauser, W. Mikenda, G. Schweng, NMR spectra of nitrogen-containing compounds. Correlations between experimental and GIAO calculated data. *Magn. Reson. Chem.* **37**, 895–902 (1999).

**Figure Legends**

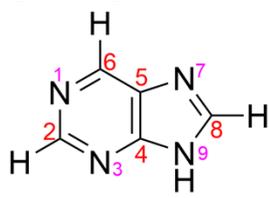
Purine

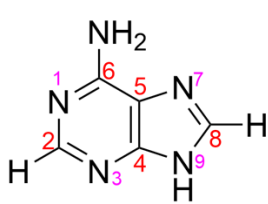
Adenine

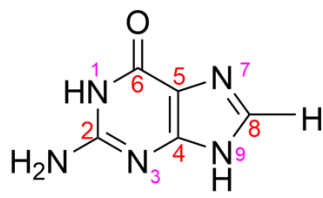
Guanine

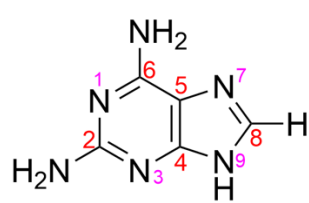
Diaminopurine

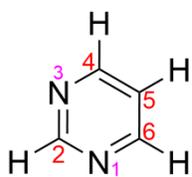
Pyrimidine

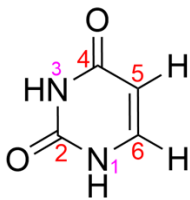
Uracil

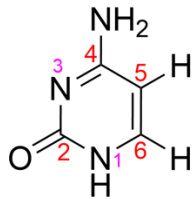
Cytosine

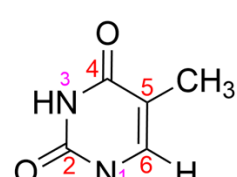
Thymine

**Figure 1.** The nucleic acid bases and related molecules studied in concentrated sulfuric acid in this work. Red numbers indicate carbon atoms and pink numbers indicate nitrogen atoms in the ring.

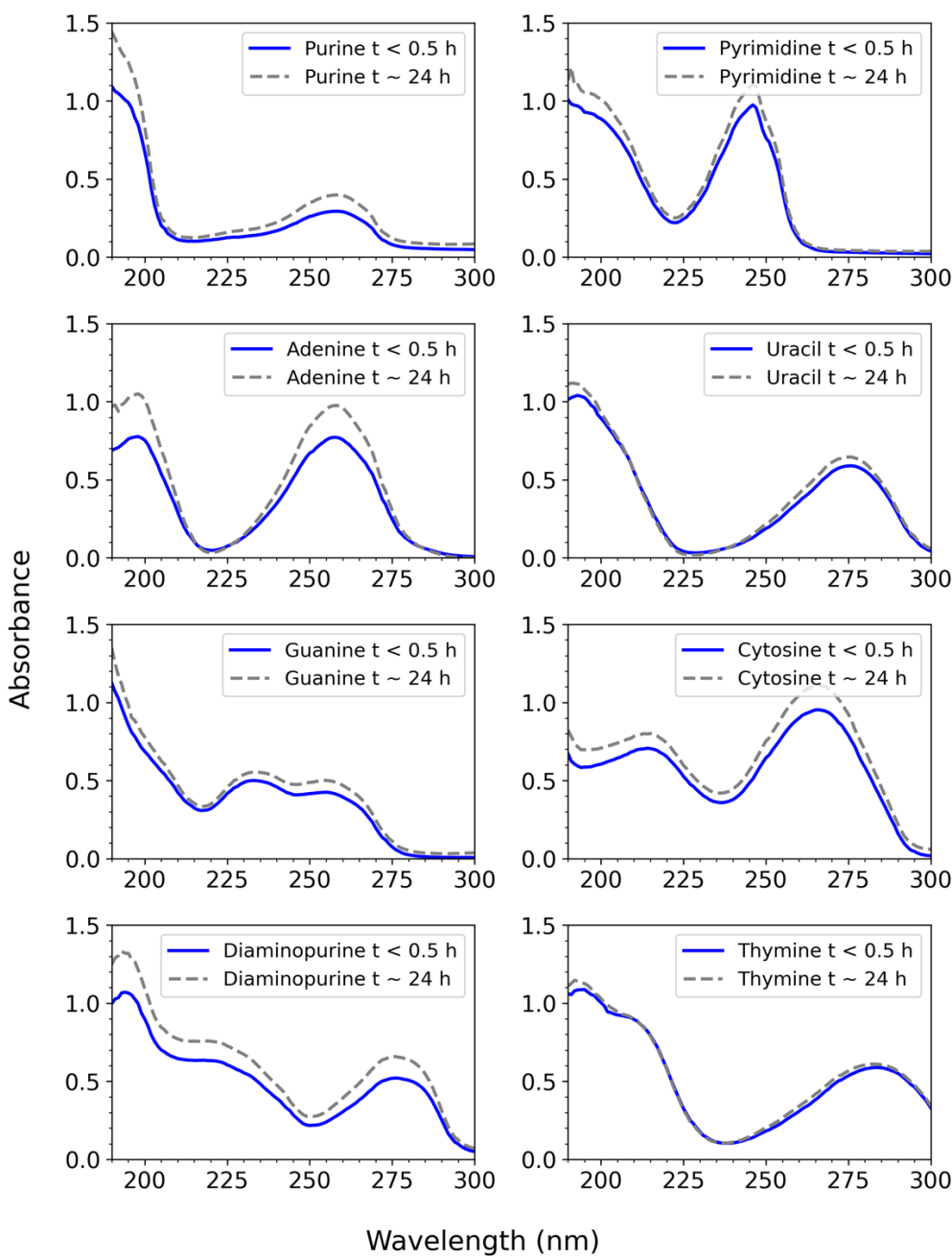

**Figure 2.** Ultraviolet spectroscopy of the eight compounds studied in 98% w/w sulfuric acid. The absorbance, defined as $A = \varepsilon L c$, where $A$ is absorbance (dimensionless), $\varepsilon$ is the molar absorption coefficient (in units of $M^{-1}$ $cm^{-1}$), $L$ is pathlength (in cm), and $c$ is concentration (in units of M), as a function of wavelength. Each compound shows two characteristic UV peaks, due to π-π conjugated bonds. The blue line shows the UV spectrum measured within about 15 to 20 minutes after mixing of the compound in 98% w/w $H_2SO_4$ in $H_2O$ and the grey dashed line is the same compound measured after about 24 hours. While some compounds have a higher absorbance due to more dissolution over the 24 hours, the same peak wavelength maximum and peak shape demonstrates stability of each compound in 98% w/w sulfuric acid.

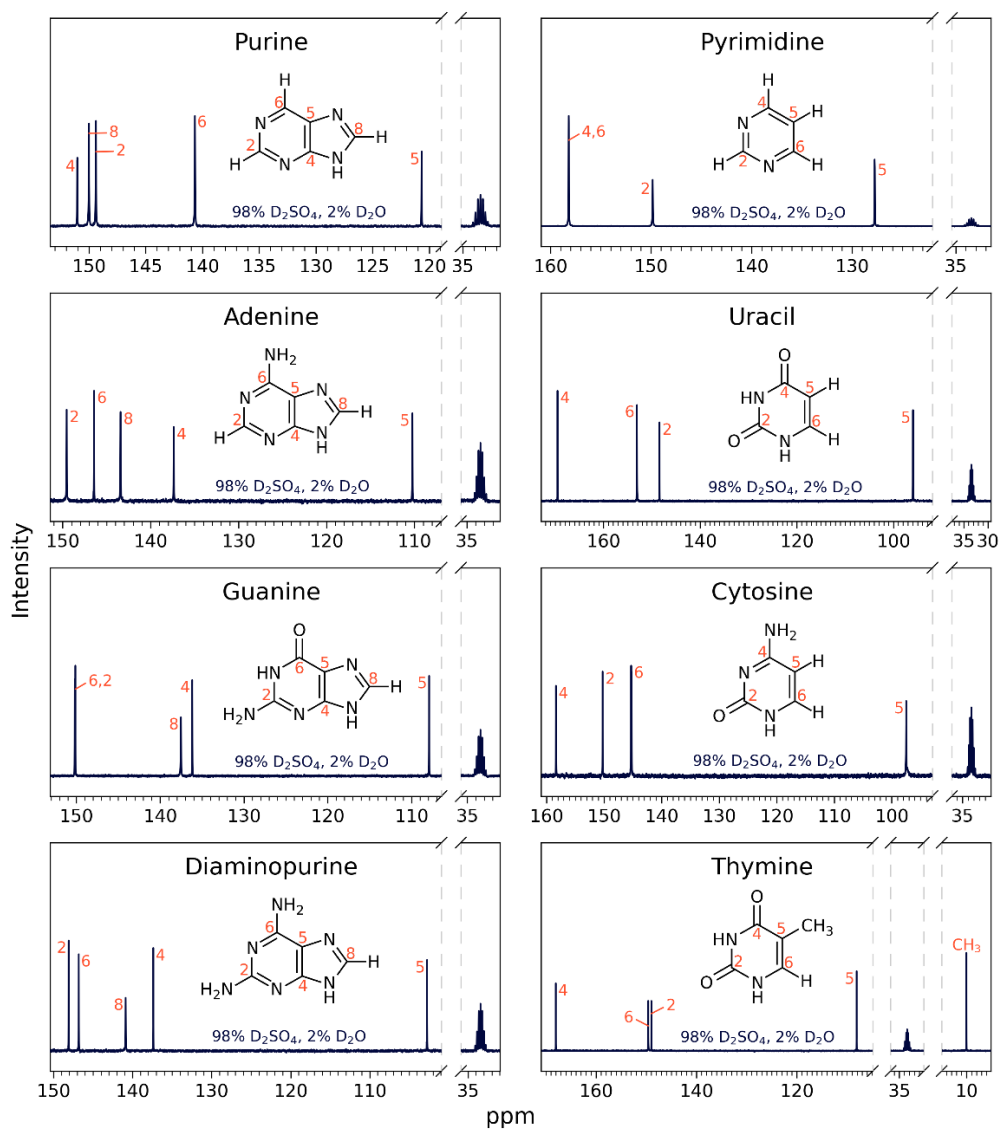

**Figure 3.** $^{13}$C NMR spectra for eight nucleic acid bases: purine, adenine, guanine, diaminopurine, pyrimidine, uracil, cytosine, and thymine in 98% D$_2$SO$_4$/2% D$_2$O (by weight) with DMSO-d$_6$ as a reference, at room temperature. The labeled NMR carbon peaks match the number of carbon atoms in the known molecular structure for a given compound. All peaks are consistent with the molecules being stable and the structure not being affected by the concentrated sulfuric acid solvent. For a description of peak assignments, see Section 3.3., Figures 5 and 6, and SI Figures S1-S11.

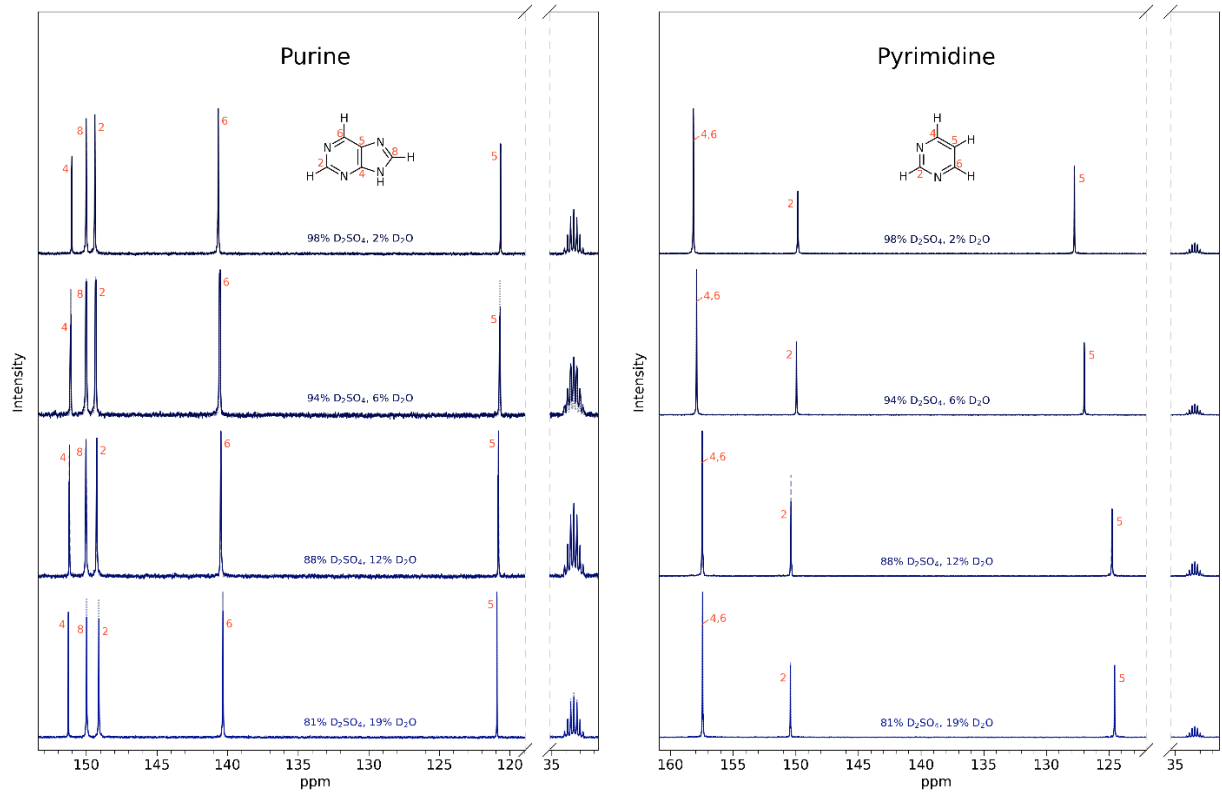

**Figure 4.** Purine (left) and pyrimidine (right) are stable for two weeks in a range of sulfuric acid concentrations found in the Venus clouds. We have incubated 30-40 mg of each base in 81-98% w/w $D_2SO_4$ for two weeks. After the two-week incubation, we measured 1D $^{13}C$ NMR spectra (solid line spectra), at each of the tested acid concentrations, and compared them to the original 1D $^{13}C$ NMR spectra collected after ~30-48 h (dashed line spectra and Figure S11). The two-week spectra and the ~30-48 h spectra look virtually identical for all tested concentrations, confirming long-term stability of the compounds in concentrated sulfuric acid solvent. From top to bottom are different concentrations (by weight) of sulfuric acid in water: 98% $D_2SO_4$/2% $D_2O$; 94% $D_2SO_4$/6% $D_2O$; 88% $D_2SO_4$/12% $D_2O$; 81% $D_2SO_4$/19% $D_2O$ with DMSO-$d_6$ as a reference and at room temperature. All peaks are consistent with the molecules being stable and the structure not being affected by the concentrated sulfuric acid solvent. For two-week stability of other compounds see Figures S12-S14.

**Figure 5**. NMR spectra for purine in concentrated sulfuric acid (98% $D_2SO_4$ and 2% $D_2O$, by weight, with reference DMSO-$d_6$) at room temperature. The NMR experiments confirm the stability of purine in concentrated sulfuric acid. A) 1D $^{13}$C NMR. B) 1D $^1$H NMR. The solvent peak is suppressed for clarity. C) 1D $^{15}$N NMR. D) The 2D $^1$H-$^{13}$C HMQC NMR shows direct bonding between H and C atoms in the purine ring structure. E) The 2D $^1$H-$^{13}$C HMBC NMR shows signals that correspond to hydrogen and carbon

atoms separated from each other by the distance of 2-4 chemical bonds in the purine ring structure (blue arrows). The HMBC "one-bond artifacts" are marked with an asterisk (*). F) The 2D $^1$H-$^{15}$N HMBC NMR shows 2 bond distances between hydrogen and nitrogen atoms (blue arrows).

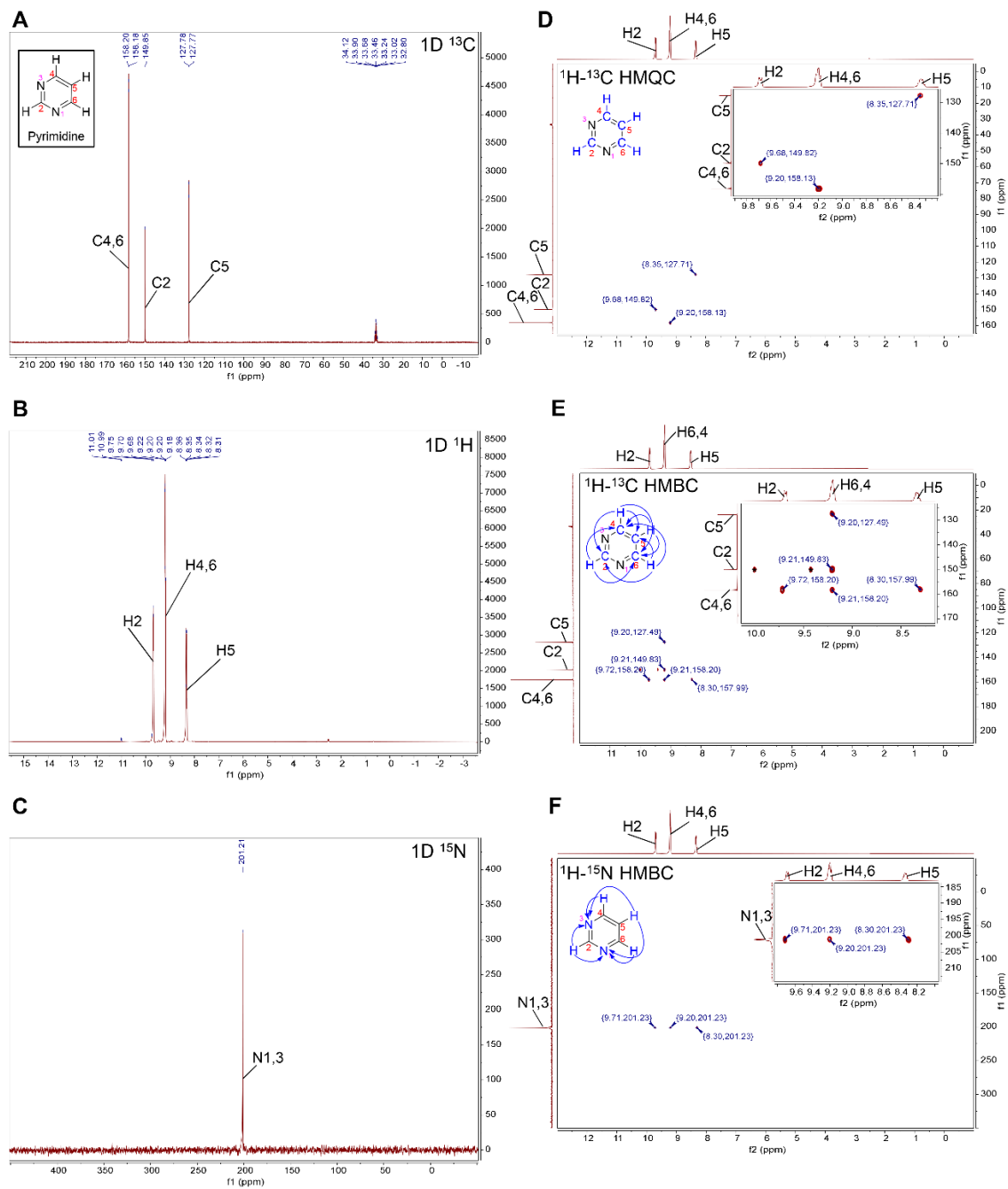

**Figure 6**. NMR spectra for pyrimidine in concentrated sulfuric acid (98% $D_2SO_4$ and 2% $D_2O$, by weight, with reference DMSO-$d_6$) at room temperature. The NMR experiments confirm the stability of pyrimidine in concentrated sulfuric acid. A) 1D $^{13}$C NMR. B) 1D $^1$H NMR. The solvent peak is suppressed for clarity. C) 1D $^{15}$N NMR. D) The 2D $^1$H-$^{13}$C HMQC NMR shows direct bonding between H and C atoms in the pyrimidine ring structure. E) The 2D $^1$H-$^{13}$C HMBC NMR shows signals that correspond to hydrogen and carbon atoms separated from each other by the distance of 2 or 3 chemical bonds in the pyrimidine ring structure (blue arrows). The HMBC "one-bond artifacts" are marked with an asterisk (*). F) The 2D $^1$H-$^{15}$N

HMBC NMR shows 2 or 3 bond distances between hydrogens attached to carbon and nitrogen atoms (blue arrows).

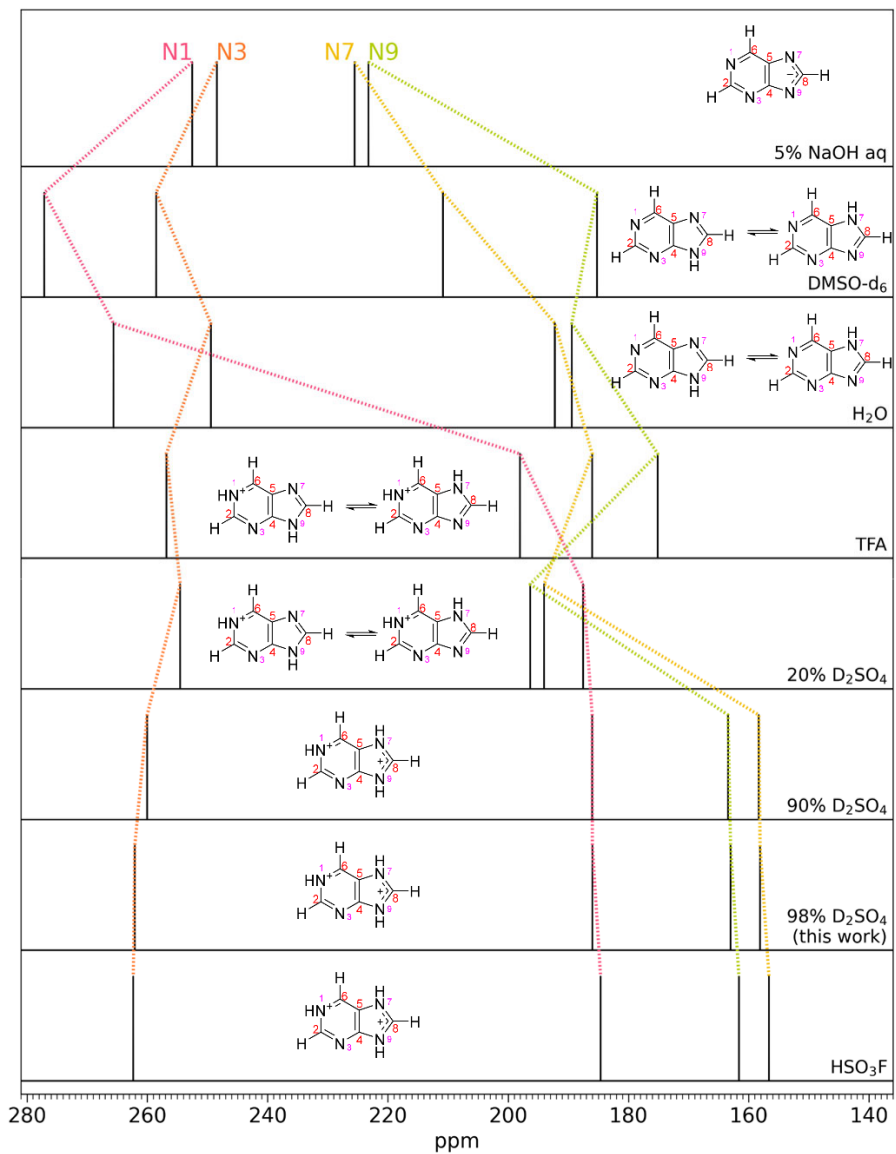

**Figure 7.** Purine $^{15}$N NMR spectral peaks for different solvent acidities. The ordering from top to bottom is in increasing acidity of the solvent. The change in chemical shifts of different N atoms indicate different protonation states of the N atoms; each of the four peaks correspond to a different N atom as indicated. The second and third row shows purine in DMSO-d$_6$ and H$_2$O which can be considered well-known standards. Top: in the basic aqueous solution of 5% NaOH, all of the N atoms are deprotonated (i.e., magnetically deshielded) and the N atom spectral peaks are shifted downfield as compared to the H$_2$O and DMSO solutions. With increasing solvent acidity, N atoms sequentially get protonated, causing dramatic upfield spectral peak migration as added protons provide more magnetic shielding. In some cases (e.g., DMSO-d$_6$ and H$_2$O) two tautomeric structures exist in equilibrium due to fast proton exchange; the relative abundance controls the spectral peak positions. See text for more details. Data and figure adapted from (40). Our data at 98% D$_2$SO$_4$ match the high acidity solvents 90% H$_2$SO$_4$ and FSO$_3$H from (40), demonstrating protonation of the N1, N7, and N9 nitrogen atoms in the purine molecule. TFA is trifluoroacetic acid.

# Supplementary Information for
Stability of Nucleic Acid Bases in Concentrated Sulfuric Acid: Implications for the Habitability of Venus' Clouds

Sara Seager[1,2,3,4,#,*] Janusz J. Petkowski[1,5,#], Maxwell D. Seager[4,6], John H. Grimes Jr.[7], Zachary Zinsli[8], Heidi Vollmer-Snarr[8], Mohamed K. Abd El-Rahman[8], David S. Wishart[9,10], Brian L. Lee[9], Vasuk Gautam[9], Lauren Herrington[1], William Bains[1,11,12], Charles Darrow[4]

[1] Department of Earth, Atmospheric and Planetary Sciences, Massachusetts Institute of Technology, Cambridge, MA, USA
[2] Department of Physics, Massachusetts Institute of Technology, Cambridge, MA, USA
[3] Department of Aeronautical and Astronautical Engineering, Massachusetts Institute of Technology, Cambridge, MA, USA
[4] Nanoplanet Consulting, Concord, MA, USA
[5] JJ Scientific, Warsaw, Mazowieckie, Poland
[6] Department of Chemistry and Biochemistry, Worcester Polytechnic Institute, Worcester, MA, USA
[7] Department of Chemistry, Massachusetts Institute of Technology, Cambridge, MA, USA
[8] Department of Chemistry and Chemical Biology, Harvard University, Cambridge, MA, USA
[9] Department of Biological Sciences, University of Alberta, Edmonton, Alberta, Canada
[10] Department of Computing Science, Department of Laboratory Medicine and Pathology, Faculty of Pharmacy and Pharmaceutical studies, University of Alberta, Edmonton, Alberta, Canada
[11] School of Physics and Astronomy, Cardiff University, 4 The Parade, Cardiff CF24 3AA, UK
[12] Rufus Scientific, Melbourn, Royston, Herts, UK

# contributed equally to this work

*Correspondence: Sara Seager
Email: seager@mit.edu

**This PDF file includes:**

    Supplementary text
    Figures S1 to S16
    Tables S1 to S9
    Legends for Datasets S1 to S2
    SI References

**Other supplementary materials for this manuscript include the following:**

    Datasets S1 to S2

**Supplementary Information Text**

**1. Molecular Structure Determination of Guanine by NMR in 98% w/w Concentrated Sulfuric Acid.** For guanine we find five $^{13}$C NMR peaks for carbon that correspond to the five carbons in the purine ring. Each of the five peaks are found in the region of the NMR spectra associated with aromatic compounds.

We assign the carbon peaks by comparison with literature data (Table S1) and by our 2D NMR experiments (Figure S1). We can assign C5 because it is the most magnetically shielded atom in the ring structure with a chemical shift distinctly upfield from the other four carbon peaks (Table S1). The C5 chemical shift also agrees with the literature values (including in solvents DMSO-$d_6$ and $D_2O$) (1, 2). To assign C4 and C8 we use our 2D NMR where we correlate the positions of H and C within the ring. Guanine has only one protonated carbon, at C8. Our 2D $^1$H-$^{13}$C HMQC shows a signal that corresponds to the C8 carbon peak at 137.53 ppm and H at 8.29 ppm. Further supporting these assignments and the integrity of the imidazole ring, the $^1$H-$^{13}$C HMBC correlates the distinct $^{13}$C chemical shift of C5 to the nearby H8 which in turn is correlated with C4 (see Figure S1E).

This leaves C2 and C6, which are especially difficult to assign due to their similar magnetically deshielded environment, which results in very similar, almost overlapping chemical shifts. C6 is attached to the carbonyl group that withdraws electrons from the ring resulting in deshielding of the C6 atom. C2 is deshielded by the electron withdrawing amino group. So while the two peaks that have the highest chemical shift values in the 1D $^{13}$C NMR spectra belong to carbons that are attached to the most electronegative atoms, C2 and C6, it is difficult to determine which is which. Since the peak at 150.13 ppm has higher intensity than the peak at 150.19 ppm we tentatively assign the 150.13 ppm peak as carbon C2. The higher peak intensity is due to the Nuclear Overhauser Effect (NOE) that is produced by the proton decoupling carried out during the experiment. The NOE is proportional to the distance ($r^{-6}$) and C2 is closer to the $NH_2$ protons. The general trend in chemical shifts of C2 and C6 for DMSO-$d_6$ and $D_2O$ reported in the literature (Table S1), where C2 is more shielded than C6, also supports this assignment. Note that the C2 and C6 can also be distinguished thanks to the very weak signal on 2D $^1$H-$^{13}$C HMBC spectra (coordinates: 8.29 ppm, 150.26 ppm; not shown for clarity). The signal corresponds to carbon C6, at 150.19 ppm, at the distance of four bonds from the H8 hydrogen.

To further confirm the structure, we turn to $^1$H (Figure S1B). For guanine the $^1$H NMR spectral peak for the H attached to carbon is highly consistent with the peaks in other solvents to about 0.5 ppm (Table S1) and we use this consistency to assign the H attached to C8. The strong peak at 8.29 ppm corresponds to H8 of the purine ring (consistent with the literature data on the $^1$H NMR of guanine hydrochloride in $D_2O$ (2)).

The other peaks in our $^1$H NMR spectrum are due to N-H hydrogens. We assign the peaks on the basis of comparison to literature data to other purines in the solvents DMSO-$d_6$ and $D_2O$ (Tables S1-8). The broad peak at 7.05 ppm corresponds to the hydrogen atoms of the intact amino group of the guanine ring, consistent with literature chemical shifts for other purines. N9 and N7 are chemically similar; we tentatively assign H9 at 12.21 ppm and H7 at 11.73 ppm.

Finally, we use 1D $^{15}$N NMR to show that the N atoms in the ring and in the amino group remain intact in concentrated sulfuric acid. The amino group contains the most magnetically shielded N and we therefore can assign the peak at 85.92 ppm to N atom attached to C2. We see four other peaks corresponding to nitrogen atoms of the guanine ring and make assignments (Table S1) based on other purine data available in the literature (see Table S5 and Table S7) and the $^1$H-$^{15}$N HMBC experiment (Figure S1F). Together with the results of $^1$H-$^{13}$C HMBC discussed above, the $^1$H-$^{15}$N HMBC experiment further confirms the integrity of the imidazole ring and allows for tentative assignments of N9 and N7 atoms to signals at 159.34 ppm and 154.76 ppm respectively.  We assign N7 and N9 peaks based on the known N7 and N9 chemical shifts of purine, where N7 is more magnetically shielded (shifted towards lower ppm) in acidic conditions than N9 (Figure 7).

We now turn to the discussion of the protonation state of the guanine ring in concentrated sulfuric acid. The upfield shift (towards lower ppm) in acidic solvent of $^{13}$C NMR peaks corresponding to carbon atoms directly adjacent to the nitrogen atoms, as compared to DMSO-d$_6$ and D$_2$O (Table S1), could indicate protonation of nitrogen atoms in the guanine ring. As the ring nitrogen atoms get protonated in acidic conditions the neighboring carbon atoms become more shielded which results in the upfield shift of $^{13}$C NMR carbon peaks. The upfield shift is particularly pronounced for C2, C4, and C6 $^{13}$C NMR peaks, which could indicate the protonation of N1, N3, N9 as well as possibly the carbonyl group oxygen of the carbon C6 in concentrated sulfuric acid. This conclusion is supported by the early work of Wagner and von Philipsborn (3) who claim that nitrogen atoms N1, N3 as well as N7 and N9 of guanine are protonated in fluorosulfuric acid (HSO$_3$F), a strong acid, analogous to H$_2$SO$_4$. The fact that HSO$_3$F and H$_2$SO$_4$ are chemically closely related and that the protonation state of purine is the same in both acids (4) further suggests that protonation of guanine in conc. H$_2$SO$_4$ will be analogous to the protonation in HSO$_3$F reported by Wagner and von Philipsborn (3) (Figure 7 and Figure S15).

**2. Molecular Structure Determination of Cytosine by NMR in 98% w/w Concentrated Sulfuric Acid.** For cytosine we see four peaks for carbon that correspond to the four carbons in the pyrimidine ring; they are found in the region of the NMR spectra associated with aromatic compounds.

We assign the carbon peaks by comparison with literature data (Table S2) and by our 2D NMR experiments (Figure S2). We can assign C5 because it is the most magnetically shielded atom in the ring structure with a chemical shift distinctly upfield from the other three carbon peaks (Table S8). The C5 chemical shift largely agrees with the literature values (including DMSO and D$_2$O) (1, 2). To assign C2, C4 and C6 we use our 2D NMR where we correlate the positions of H and C within the ring (Figure S2). Cytosine has only two carbons with directly attached hydrogen, and this is for C5 and C6. Our 2D $^1$H-$^{13}$C HMQC shows a signal that corresponds to the C6 carbon peak at 145.28 ppm and H (attached to C6) at 7.17 ppm. The 2D $^1$H-$^{13}$C HMQC also further confirms the assignment of C5 at 97.48 ppm. We assign C2 and C4, and confirm the assignments of C5 and C6, on the basis of $^1$H-$^{13}$C HMBC data (see Figure S2E).

To further confirm the structure, we turn to $^1$H NMR (Figure S2B). For cytosine the $^1$H NMR spectral peak for the H attached to carbons are highly consistent with the peaks in other solvents (Table S2) and we use this consistency to assign the H attached to C5 and C6. The peak at 7.17 ppm corresponds to H6 of the pyrimidine ring, while the second peak at 5.88 ppm corresponds to H5 (the chemical shifts for hydrogen atoms are highly consistent with the literature data on the $^1$H NMR of cytosine in D$_2$O (2)). No peaks corresponding to hydrogens attached to nitrogen have been detected. Such non-detection is expected as hydrogen atoms exchange very rapidly between the N atom and the solvent in the acidic solution. Similarly, if the carbonyl oxygens were protonated in concentrated sulfuric acid those hydrogen atoms would not be detected.

Finally, we use 1D $^{15}$N NMR to show that the N atoms, especially the amino group remain present in the cytosine structure in concentrated sulfuric acid (Table S2). The amino group contains the most shielded N and we therefore can assign the peak at 102.59 ppm to the N attached to C4. We see an additional two nitrogen atoms of the pyrimidine ring and tentatively assign N1 and N3, to 136.44 ppm and 138.61 ppm based on the $^1$H-$^{15}$N HMBC data (Figure S2F). We can further confirm the assignment of the amino group to the 102.59 ppm peak on the basis of the very weak signal on the 2D $^1$H-$^{15}$N HMBC spectra (coordinates: 5.86 ppm, 102.29 ppm; not shown for clarity). The signal corresponds to the nitrogen atom of the amino group at the distance of three bonds from the H5 hydrogen.

We now turn to the discussion of the protonation state of the cytosine ring in concentrated sulfuric acid. The upfield shift (towards lower ppm) of C2 and C4 $^{13}$C NMR peaks in concentrated sulfuric acid, as compared to DMSO-d$_6$ and D$_2$O (Table S2), suggests protonation of neighboring nitrogen atom N3 and possibly also carbonyl oxygen. The $^{13}$C NMR chemical shifts of cytosine in concentrated sulfuric acid reported by Benoit and Frechette are consistent with protonation of cytosine (5) and agree with ours, supporting the protonation of nitrogen N3 and carbonyl oxygen atoms in 98% w/w D$_2$SO$_4$ (Table S2). The protonation of cytosine in strong acid is also supported by early studies on protonation of pyrimidines in FSO$_3$H (6).

We note that the slight shifts of the carbon peaks on the $^{13}$C NMR, as the acid concentration changes from 81% to 98% (Figure S7) are likely due to different relative amounts of double protonated (2H) vs single protonated (1H) species in solution as the acidity of the solution changes. Different tautomeric structures of cytosine, including protonation of carbonyl group could also contribute to this effect (6).

**3. Molecular Structure Determination of 2,6-Diaminopurine by NMR in 98% w/w Concentrated Sulfuric Acid.** For 2,6-diaminopurine, in 1D $^{13}$C NMR spectrum (Figure S3A), we find five peaks for carbon that correspond to the five carbons in the diaminopurine ring. Each of the five peaks are found in the region of the NMR spectra associated with aromatic compounds.

There is very scarce NMR data for 2,6-diaminopurine available in the literature (Table S3). We therefore assign the carbon peaks by comparison of the diaminopurine spectra with the spectra of the closely-related molecules, adenine (Figure S5, Table S5) and guanine (Figure 5, Table S7) and by our 2D NMR experiments (Figure S3D, E). As for the other compounds we can assign C5 because it is the most magnetically shielded atom in the ring structure with a chemical shift distinctly upfield from the other four carbon peaks. To assign C4 and C8 we use our 2D NMR spectra where we correlate the positions of H and C within the ring. Diaminopurine has only one protonated carbon, at C8, therefore, the single strong peak at 8.53 ppm in the 1D $^1$H NMR corresponds to the hydrogen atom H8 of the diaminopurine ring (Figure S3B). Our 2D $^1$H-$^{13}$C HMQC correctly shows a signal that corresponds to the C8 carbon peak at 140.87 ppm and H (bonded to C8) at 8.53 ppm (Figure S3C). We assign C4 to the 137.38 ppm peak on the basis of our $^1$H-$^{13}$C HMBC data (Figure S3E). Further supporting these assignments and the integrity of the imidazole ring, the HMBC correlates the distinct $^{13}$C chemical shift of C5 to the nearby H8 which in turn is correlated with C4.

This leaves C2 and C6 which are especially difficult to assign due to their similarly magnetically deshielded environment which results in very similar chemical shifts. Nevertheless, the presence of a very weak signal in the $^1$H-$^{13}$C HMBC spectra (coordinates: 8.52 ppm, 146.82 ppm; not shown for clarity) allows us to distinguish carbons C2 and C6 from each other. The very weak signal in the $^1$H-$^{13}$C HMBC spectra corresponds to carbon C6, at 146.80 ppm, at the separation of four bonds from the H8 hydrogen.

Finally, we use 1D $^{15}$N NMR to show that the diaminopurine structure contains all six expected N atoms. The 1D $^{15}$N spectra (Figure S3C) shows six peaks, as expected, corresponding to six nitrogen atoms, four Ns of the aromatic rings and two Ns belonging to the amino groups. The $^1$H-$^{15}$N HMBC experiment (Figure S1F) helps in the identification of N7 and N9. We can tentatively assign N7 and N9 to 156.96 ppm and 160.12 ppm respectively, on the basis of the similarities with purine (Table S7). We assign N7 and N9 peaks based on the known N7 and N9 chemical shifts of purine, where N7 is more magnetically shielded (shifted towards lower ppm) in acidic conditions than N9 (Figure 7). However, based on the data at hand, we cannot unambiguously assign N1, N3 and the two remaining amino groups.

We now turn to the discussion of the protonation state of the diaminopurine ring in concentrated sulfuric acid. There are no studies on protonation of diaminopurine in acidic conditions. The upfield shift (towards lower ppm) of C2, C4, C5, and C6 $^{13}$C NMR peaks in concentrated sulfuric acid, as compared to DMSO-d$_6$ (Table S3), suggests protonation of neighboring nitrogen atoms. As the ring nitrogen atoms get protonated in acidic conditions the neighboring carbon atoms become more shielded which results in the upfield shift of $^{13}$C NMR carbon peaks. Assuming that the protonation state of diaminopurine is similar to adenine we can hypothesize that N1, N3 as well as N7 and N9 of diaminopurine are protonated in 98% $D_2SO_4$ (Figure S15).

Taken together the NMR data confirms that the 2,6-diaminopurine ring structure remains intact in 98% w/w $D_2SO_4$ in $D_2O$.

**4. Molecular Structure Determination of Thymine by NMR in 98% w/w Concentrated Sulfuric Acid.** For thymine, in the 1D $^{13}$C NMR spectrum (Figure S4A), we see four peaks for carbon that correspond to the four carbons in the pyrimidine ring; they are found in the region of the NMR spectra associated with aromatic compounds. We also see one distinctive peak corresponding to thymine's methyl group.

We assign the carbon peaks by comparison with literature data (Table S4) and by our 2D NMR experiments (Figure S4D, E). We can assign C5 because it is the most magnetically shielded atom in the ring structure with a chemical shift distinctly upfield from the other three carbon peaks (Table S4). We assign the distinct peak at 9.99 ppm to the methyl group carbon ($CH_3$), following the established literature values (Table S4). To assign C2, C4, and C6 we use our 2D NMR where we correlate the positions of H and C within the ring (Figure S4D, E). Thymine has only one carbon in the pyrimidine ring with a directly attached hydrogen, and this is for C6. Our 2D $^1$H-$^{13}$C HMQC shows a signal that corresponds to the C6 carbon peak at 149.59 ppm and H6 (attached to C6) at 7.43 ppm (Figure S4D). We assign C2 and C4, and confirm the assignments of C5 and C6, on the basis of $^1$H-$^{13}$C HMBC data (Figure S4E).

To further confirm the structure, we turn to $^1$H (Figure S4B). For thymine the $^1$H NMR spectral peak for the H attached to carbons are highly consistent with literature values of peaks in DMSO-$d_6$ (Table S4) and we use this consistency to assign the Hs attached to the carbon of the methyl group and carbon C6. No peaks corresponding to hydrogens attached to nitrogen have been detected. Such non-detection is not surprising as hydrogen atoms are expected to exchange very rapidly between the N atom and the solvent in the acidic solution. Similarly, if the carbonyl oxygens were protonated in concentrated sulfuric acid those hydrogen atoms would likely also not be detected.

Finally, we use 1D $^{15}$N NMR to show the two N atoms of the thymine ring (Figure S4C). We assign N1 and N3 to 146.64 ppm and 154.42 ppm respectively based on the $^1$H-$^{15}$N HMBC data (Figure S4F).

We now turn to the discussion of the protonation state of the thymine ring in concentrated sulfuric acid. The $^{13}$C NMR chemical shift changes of thymine in different solvents are similar to the chemical shift changes reported for uracil (Table S4). The $^{13}$C NMR chemical shifts of thymine in concentrated sulfuric acid reported by Benoit and Frechette (5) have been interpreted as signs of protonation of carbonyl oxygen atoms. The results of Benoit and Frechette (5) generally agree with ours. The change in chemical shifts of C2, C4, C5 and C6 carbon peaks between acidic and neutral media could indicate protonation of the neighboring carbonyl oxygen atoms in 98% w/w $D_2SO_4$ (Table S4 and Figure S15). The downfield shift of the C2 peak as the concentration of acid increases (Figure S8) could indicate the protonation of the carbonyl oxygen O2. The downfield shift of C6 peak in the acidic medium (Table S4) is consistent with previously reported chemical shifts of thymine in concentrated sulfuric acid (5).
The protonation of thymine carbonyl groups in strong acid is also supported by early studies on protonation of pyrimidines in $FSO_3H$ (6).

Taken together the NMR data confirms that the thymine ring structure remains intact in intact in 98% w/w $D_2SO_4$ in $D_2O$.

**5. Molecular Structure Determination of Adenine by NMR in 98% w/w Concentrated Sulfuric Acid.** For Adenine, in the 1D $^{13}$C NMR spectrum (Figure S5A), we find five peaks for carbon that correspond to the five carbons in the adenine ring. Each of the five peaks are found in the region of the NMR spectra associated with aromatic compounds.

We assign the carbon peaks by comparison with literature data (Table S5) and by our 2D NMR experiments (Figure S5D, E). We can assign C5 because it is the most magnetically shielded atom in the ring structure with a chemical shift distinctly upfield from the other four carbon peaks (Table S5). To assign C2, C4, C6, and C8 we use our 2D NMR where we correlate the positions of H and C within the ring. Adenine has two protonated carbons, at C2 and C8. Our 2D $^1$H-$^{13}$C HMQC data show two signals that correspond to C2 (peak at 149.55 ppm) and C8 peak (at 143.42 ppm) attached to H2 (at 8.83 ppm) and H8 (at 8.86 ppm) respectively (Figure S5D). We assign C4 and C6 to 137.37 ppm and 146.43 ppm respectively on the basis of $^1$H-$^{13}$C HMBC data (Figure S5E). The assignment of C5 also supported by the $^1$H-$^{13}$C HMBC NMR. Furthermore, the $^1$H-$^{13}$C HMBC data are consistent with carbon atom assignments derived from the $^1$H-$^{13}$C HMQC data.

Finally, we use 1D $^{15}$N NMR to show that the adenine structure contains all five expected N atoms. The 1D $^{15}$N spectra (Figure S5C) shows five peaks corresponding to five nitrogen atoms, four Ns of the aromatic ring and one N belonging to the amino group. The $^1$H-$^{15}$N HMBC experiment (Figure S5F) allows for the assignment of the peaks at 161.23 ppm and 160.89 ppm to N7 and N9 respectively. The amino group contains the most magnetically shielded N and we therefore can assign the peak at 113.44 ppm to N atom attached to C6. Based on the data at hand we cannot distinguish N1, N3 from each other.

We now turn to the discussion of the protonation state of the adenine ring in concentrated sulfuric acid. The general upfield shift (towards lower ppm) of $^{13}$C NMR carbon peaks in concentrated sulfuric acid, as compared to DMSO-$d_6$ and $D_2O$ (Table S5), is consistent with protonation of neighboring nitrogen atoms. As the ring nitrogen atoms get protonated in acidic conditions the neighboring carbon atoms become more shielded which results in the upfield shift of $^{13}$C NMR carbon peaks.

To the authors' knowledge direct measurements of the protonation state of N atoms in adenine at different concentrations of sulfuric acid have never been published. However, the assessment of the protonation state in a closely related fluorosulfuric acid ($HSO_3F$) has been attempted for adenine, guanine and purine (3). Since the protonation state of purine is the same in both acids (Figure 7)(3, 4) we can hypothesize that the state of protonation of adenine in conc. $H_2SO_4$ will be analogous to the state of protonation in $HSO_3F$. Wagner and von Philipsborn postulate that N1, N3 as well as N7 and N9 of adenine are protonated in $HSO_3F$ (3), we can therefore assume that the same nitrogen atoms get protonated in 98% w/w $D_2SO_4$ (Figure S15).

Taken together the NMR data confirms that the adenine ring structure remains intact in 98% w/w $D_2SO_4$ in $D_2O$.

**6. Molecular Structure Determination of Uracil by NMR in 98% w/w Concentrated Sulfuric Acid.** For uracil, in the 1D $^{13}$C NMR spectrum (Figure S6A), we see four peaks for carbon that correspond to the four carbons in the pyrimidine ring; they are found in the region of the NMR spectra associated with aromatic compounds.

We assign the carbon peaks by comparison with literature data (Table S6) and by our 2D NMR experiments (Figure S6D, E). We can assign C5 because it is the most magnetically shielded atom in the ring structure with a chemical shift distinctly upfield from the other three carbon peaks (Table S6). To assign C2, C4, and C6 we use our 2D NMR data where we correlate the positions of H and C within the ring (Figure S6D, E). Uracil has only two carbons with directly attached hydrogens, and this is for C5 and C6. Our 2D $^1$H-$^{13}$C HMQC shows a signal that corresponds to the C6 carbon peak at 153.07 ppm and H (attached to C6) at 7.61 ppm. The 2D $^1$H-$^{13}$C HMQC also further confirms the assignment of C5 at 95.93 ppm (Figure S6D). We assign C2 and C4, and confirm the assignments of C5 and C6, on the basis of $^1$H-$^{13}$C HMBC data (Figure S6E).

To further confirm the structure, we turn to $^1$H (Figure S6B). For uracil the $^1$H NMR spectral peak for the Hs attached to carbons are highly consistent with the peaks of uracil in DMSO-d$_6$ found in the literature (Table S6) and we use this consistency to assign the Hs attached to C5 and C6. No peaks corresponding to hydrogens attached to nitrogen have been detected. Such non-detection is not surprising as hydrogen atoms are expected to exchange very rapidly between the N atom and the solvent in the acidic solution. Similarly, if the carbonyl oxygens were protonated in concentrated sulfuric acid those hydrogen atoms would likely also not be detected.

Finally, we use 1D $^{15}$N NMR to show the two N atoms of the uracil ring (Figure S6C). We assign N1 and N3 to 149.50 ppm and 154.75 ppm respectively based on the $^1$H-$^{15}$N HMBC data (Figure S6F).

We now turn to the discussion of the protonation state of the uracil ring in concentrated sulfuric acid. Uracil is protonated in acidic conditions. The $^{13}$C NMR chemical shifts of uracil in concentrated sulfuric acid reported by Benoit and Frechette (5) have been interpreted as signs of protonation of carbonyl oxygen atoms. The results of Benoit and Frechette (5) generally agree with ours. The change in chemical shifts of C2, C4, C5 and C6 carbon peaks between acidic and neutral media could indicate protonation of the neighboring carbonyl oxygen atoms in 98% w/w D$_2$SO$_4$ (Table S6 and Figure S15). The protonation of uracil carbonyl groups in strong acid is also supported by early studies on protonation of pyrimidines in FSO$_3$H (6).

Taken together the NMR data confirms that the uracil ring structure remains intact in 98% w/w D$_2$SO$_4$ in D$_2$O.

**7. Stability of the Nucleic Acid Bases in 98% w/w Concentrated Sulfuric Acid After Two Week Incubation.** To confirm the long-term stability of all eight nucleic acid bases in concentrated sulfuric acid we have incubated 10 to 40 mg of each base in 81-98% w/w D$_2$SO$_4$ in D$_2$O for two weeks, stored in the NMR tubes with room temperature varying from about 18 to 24 °C. After the two-week incubation we acquired 1D $^{13}$C NMR spectra of each base, at each of the tested acid concentrations, and compared them to the original 1D $^{13}$C NMR spectra collected after ~30-48 h. The two-week spectra and the ~30-48 h spectra look virtually identical for all eight tested bases, at all tested concentrations, confirming long-term stability of the nucleic acid bases in concentrated sulfuric acid solvent (Figure 4 and Figures S12-S14).

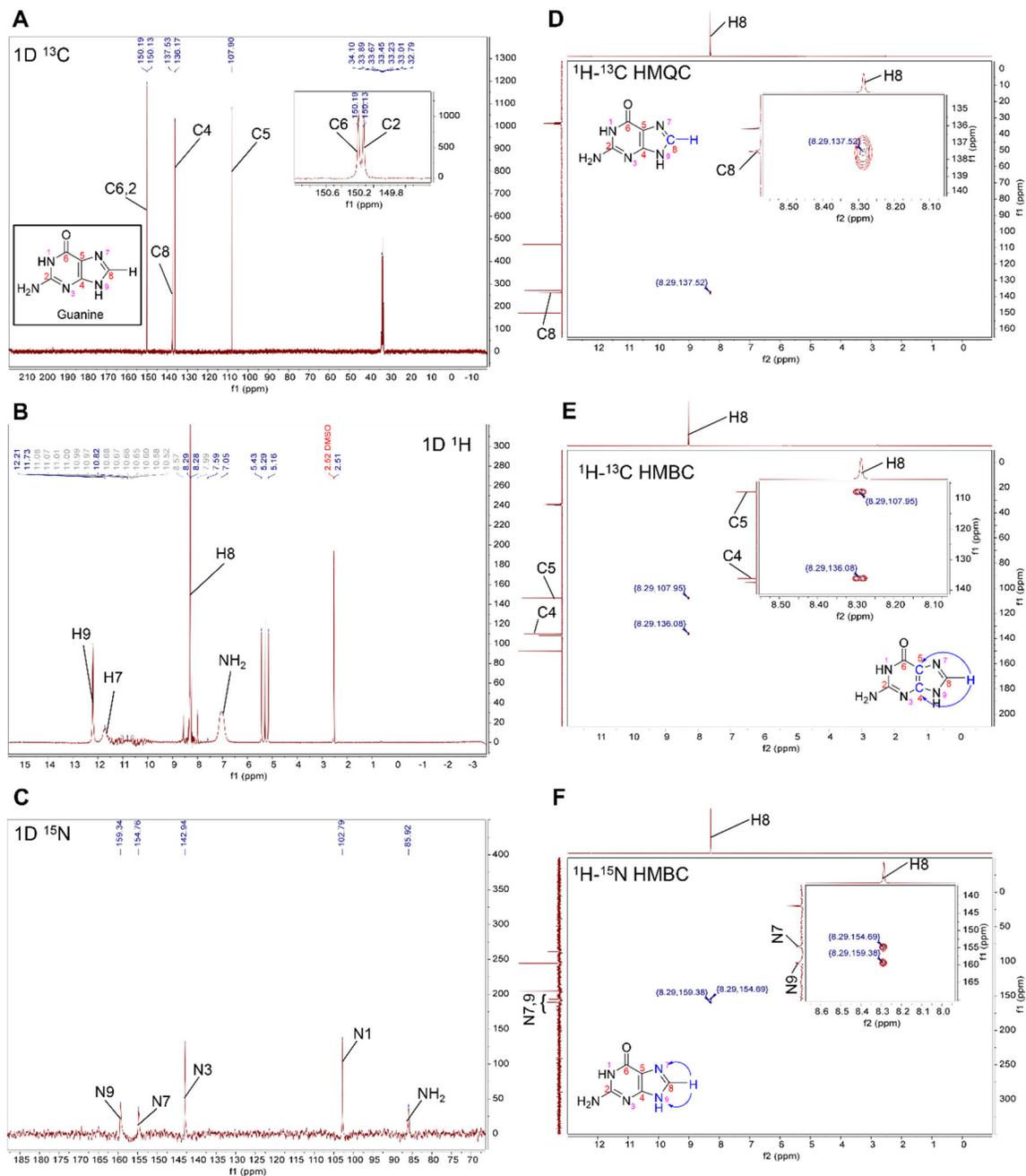

**Fig. S1.** NMR spectra for guanine in concentrated sulfuric acid (98% $D_2SO_4$ and 2% $D_2O$ (by weight) with reference DMSO-$d_6$) at room temperature. The NMR experiments confirm the stability of guanine in concentrated sulfuric acid. **A)** 1D $^{13}C$ NMR shows five peaks corresponding to five carbons in the guanine ring. DMSO-$d_6$ reference peak shown at 33.45 ppm. **B)** 1D $^1H$ NMR shows peaks corresponding to hydrogen atoms in the guanine ring, including hydrogens belonging to the C2 amino group. The solvent peak is suppressed for clarity. **C)** 1D $^{15}N$ NMR further reaffirms the integrity of the guanine ring in the concentrated sulfuric acid by showing four peaks of the nitrogen atoms of the aromatic ring and one nitrogen belonging to the intact amino group attached to carbon C2. **D)** The 2D $^1H$-$^{13}C$ HMQC NMR shows direct bonding between H and C atoms in the guanine ring structure. As expected it shows only one signal, at the intersection of hydrogen atom H8 (f2: $^1H$ tracer spectra) and carbon atom C8 (f1: $^{13}C$ tracer spectra). This confirms the identity of 137.53 ppm peak in 1D $^{13}C$ NMR spectra as carbon C8. **E)** The 2D $^1H$-$^{13}C$ HMBC NMR shows signals that correspond to hydrogen and carbon atoms

separated from each other by the distance of 3 chemical bonds in the guanine ring structure (blue arrows). As expected for guanine, the spectrum shows only two signals at the expected positions. The two signals correspond to a distance of 3 chemical bonds between H8 and C4 and C5 in the guanine structure. The correct distances between atoms derived from the 2D $^1$H-$^{13}$C HMBC NMR confirm the identity of 107.90 ppm peak and the 136.17 ppm peak in 1D $^{13}$C NMR spectra as C5 and C4 respectively. **F)** The 2D $^1$H-$^{15}$N HMBC NMR shows bond distances between hydrogen atoms attached to carbons and nitrogen atoms (blue arrows). The spectrum shows two signals at the expected positions. The two signals correspond to a distance of 2 chemical bonds between H8 and N7 or N9. The relationships between atoms derived from the 2D NMR, taken together with the 1D NMR data, further confirm peak assignments of the carbon, hydrogen and nitrogen 1D NMR spectra and support the hypothesis that the guanine ring remains unchanged and is stable in 98% w/w concentrated sulfuric acid. The NMR experiments confirm the integrity of guanine in concentrated sulfuric acid.

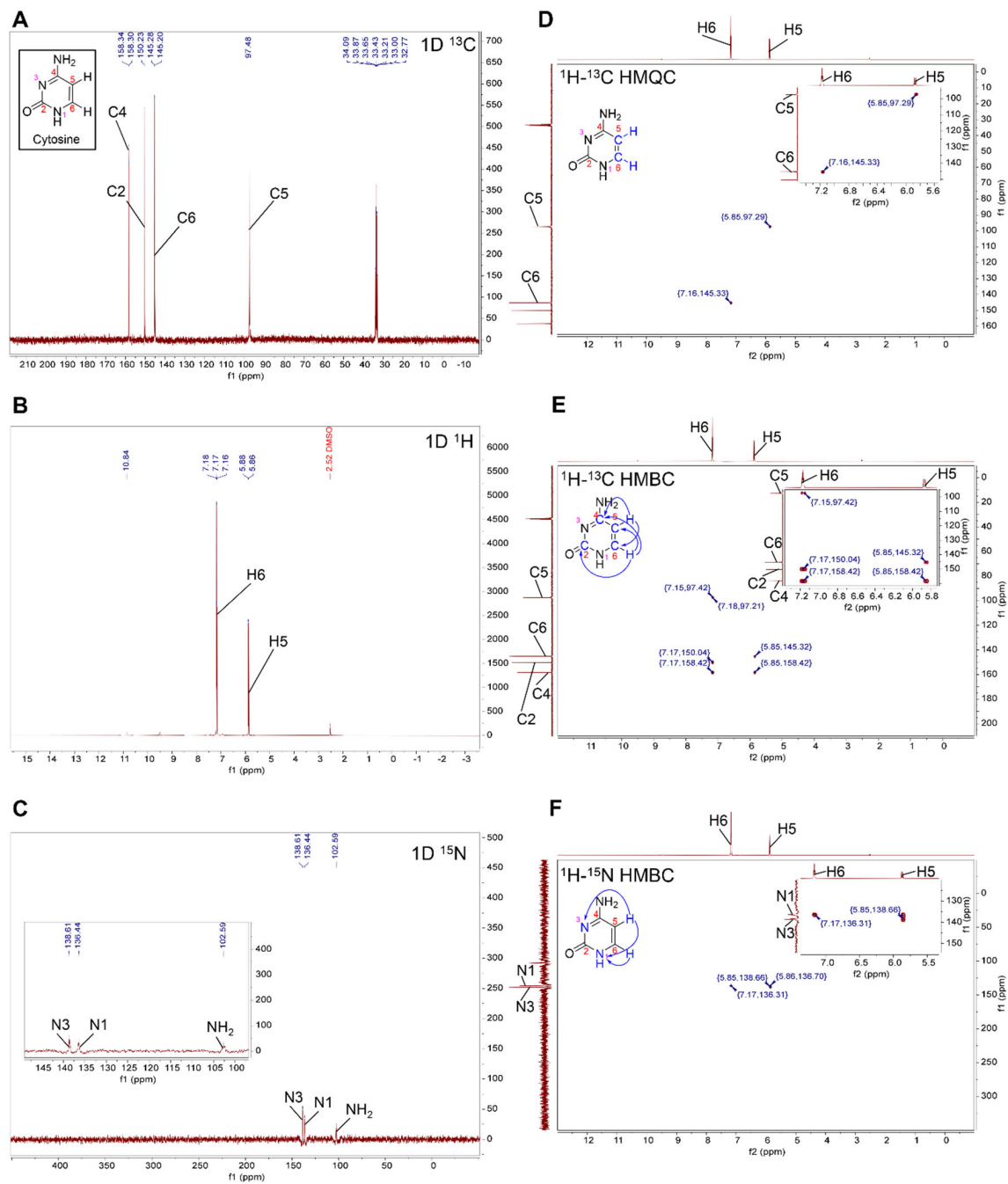

**Fig. S2.** NMR spectra for cytosine in concentrated sulfuric acid (98% D$_2$SO$_4$ and 2% D$_2$O (by weight) with reference DMSO-d$_6$) at room temperature. The NMR experiments confirm the stability of cytosine in concentrated sulfuric acid. **A)** 1D $^{13}$C NMR shows four peaks corresponding to four carbons in the cytosine ring. The DMSO-d$_6$ reference peak appears at 33.43 ppm. **B)** 1D $^1$H NMR shows peaks corresponding to hydrogen atoms attached to carbons in the cytosine ring. The Hs associated with Ns have not been detected. The solvent peak is suppressed for clarity. **C)** 1D $^{15}$N NMR further reaffirms the integrity of the cytosine ring in concentrated sulfuric acid by showing two peaks corresponding to the nitrogen atoms of the aromatic ring and one nitrogen belonging to the intact amino group attached to carbon C4. **D)** The 2D $^1$H-$^{13}$C HMQC NMR shows direct bonding between H and C atoms in the cytosine ring structure. As expected it shows two signals, at the intersection of hydrogen atoms H5 and H6 (f2: $^1$H trace spectra) and carbon atoms

C5 and C6 (f1: $^{13}$C trace spectra). The experiment confirms the identity of 97.48 ppm and 145.28 ppm peaks in 1D $^{13}$C NMR spectra as carbons C5 and C6 respectively. **E)** The 2D $^{1}$H-$^{13}$C HMBC NMR shows signals that correspond to hydrogen and carbon atoms separated from each other by the distance of 2 or 3 chemical bonds in the cytosine ring structure (blue arrows). The correct distances between atoms derived from the 2D $^{1}$H-$^{13}$C HMBC NMR confirm the identity of carbon peaks in 1D $^{13}$C NMR spectra. **F)** The 2D $^{1}$H-$^{15}$N HMBC NMR shows 2 or 3 bond distances between hydrogens attached to carbon and nitrogen atoms (blue arrows). The spectrum shows three signals at the expected positions. The three signals correspond to a distance of 2 or 3 chemical bonds between hydrogens H5 and H6 and nitrogens N1 and N3 in the cytosine structure. The correct distances between atoms derived from the 2D $^{1}$H-$^{15}$N HMBC NMR confirm the identity of nitrogen peaks in the 1D $^{15}$N NMR spectra. The relationships between atoms derived from the 2D NMR, taken together with the 1D NMR data, further confirm peak assignments of the carbon, hydrogen and nitrogen 1D NMR spectra and support the hypothesis that the cytosine ring remains unchanged and is stable in 98% w/w concentrated sulfuric acid. The NMR experiments confirm the integrity of cytosine in concentrated sulfuric acid.

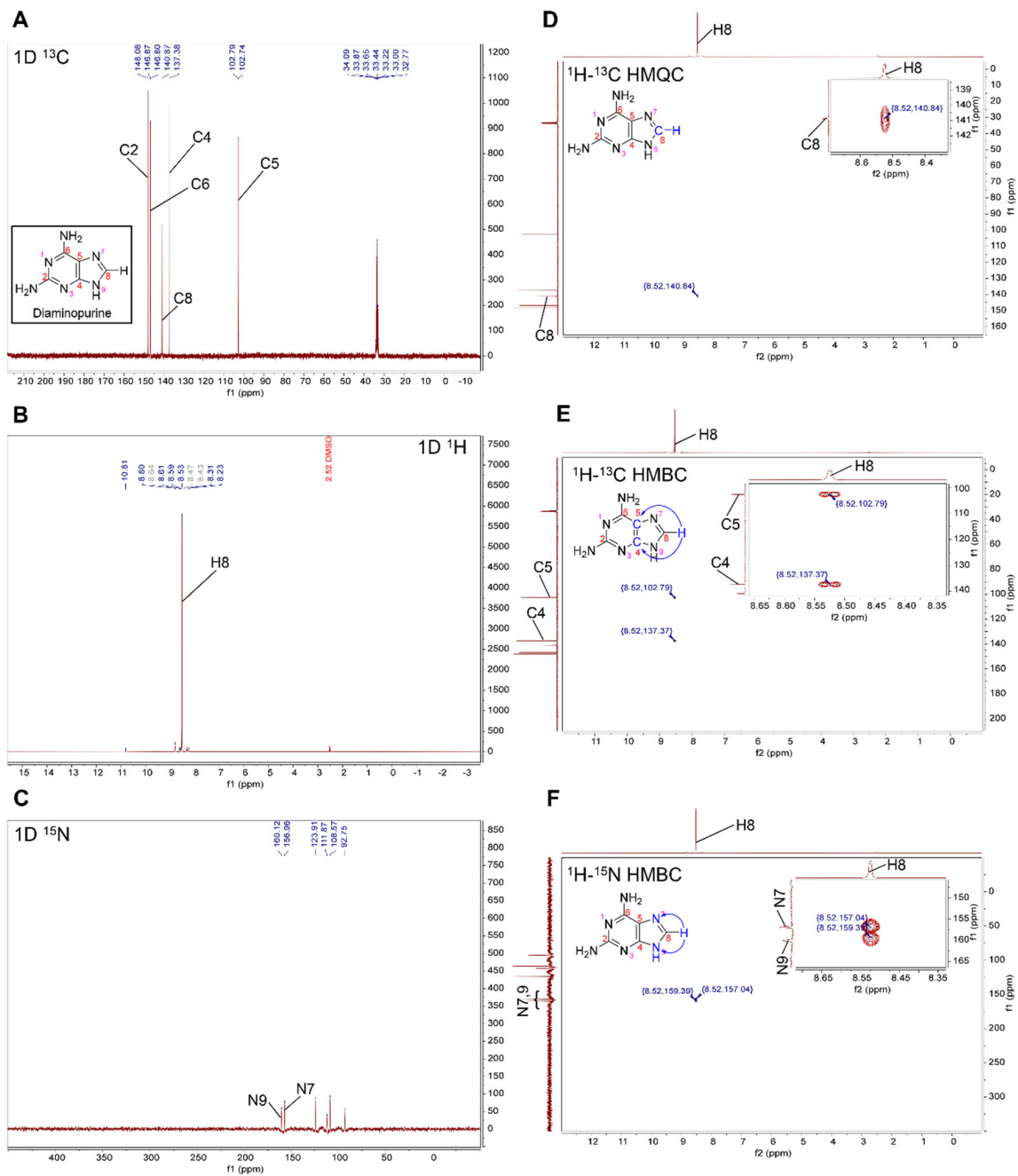

**Fig. S3.** NMR spectra for 2,6-diaminopurine concentrated sulfuric acid (98% D$_2$SO$_4$ and 2% D$_2$O, by weight, with reference DMSO-d$_6$) at room temperature. The NMR experiments confirm the stability of 2,6-diaminopurine in concentrated sulfuric acid. **A**) 1D $^{13}$C NMR shows five peaks corresponding to five carbons in the 2,6-diaminopurine ring. The DMSO-d$_6$ reference peak shown at 33.44 ppm. **B**) 1D $^1$H NMR shows a single peak corresponding to the H8 hydrogen atom. The solvent peak is suppressed for clarity. **C**) 1D $^{15}$N NMR further reaffirms the integrity of the 2,6-diaminopurine ring in the concentrated sulfuric acid by showing four peaks of the nitrogen atoms of the aromatic ring and two nitrogen atoms belonging to the intact amino groups attached to carbons C2 and C6. **D**) The 2D $^1$H-$^{13}$C HMQC NMR shows direct bonding between H and C atoms in the 2,6-diaminopurine ring structure. As expected it shows only one signal, at the intersection of hydrogen atom H8 (f2: $^1$H tracer spectra) and carbon atom C8 (f1: $^{13}$C tracer spectra). This confirms the identity of 140.87 ppm peak in 1D $^{13}$C NMR spectra as carbon C8. **E**)

The 2D $^1$H-$^{13}$C HMBC NMR shows signals that correspond to hydrogen and carbon atoms separated from each other by the distance of 3 chemical bonds in the 2,6-diaminopurine ring structure (blue arrows). As expected for 2,6-diaminopurine, the spectrum shows only two signals at the expected positions. The two signals correspond to a distance of 3 chemical bonds between H8 and C4 and C5 in the 2,6-diaminopurine structure. The correct distances between atoms derived from the 2D $^1$H-$^{13}$C HMBC NMR confirm the identity of the 102.74 ppm peak and the 137.38 ppm peak in 1D $^{13}$C NMR spectra as C5 and C4 respectively. **F**) The 2D $^1$H-$^{15}$N HMBC NMR shows bond distances between hydrogen and nitrogen atoms (blue arrows). The spectrum shows two signals at the expected positions. The two signals correspond to a distance of two chemical bonds between H8 and N7 or N9. The relationships between atoms derived from the 2D NMR, taken together with the 1D NMR data, further confirm peak assignments of the carbon, hydrogen and nitrogen 1D NMR spectra and support the hypothesis that the 2,6-diaminopurine ring remains unchanged and is stable in 98% w/w concentrated sulfuric acid.

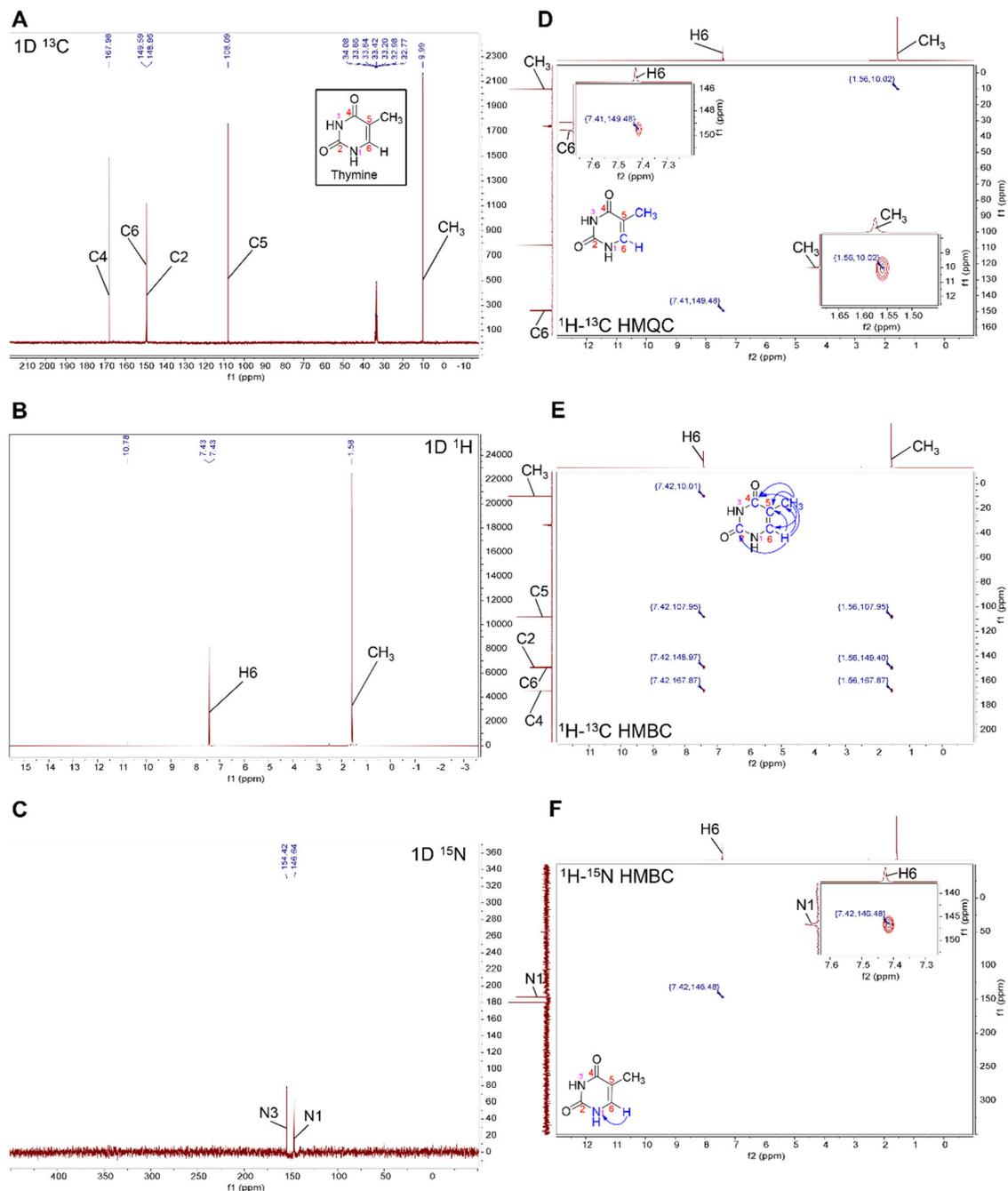

**Fig. S4.** NMR spectra for thymine in concentrated sulfuric acid (98% $D_2SO_4$ and 2% $D_2O$, by weight, with reference DMSO-$d_6$) at room temperature. The NMR experiments confirm the stability of thymine in concentrated sulfuric acid. **A**) 1D $^{13}C$ NMR shows four peaks corresponding to four carbons in the thymine ring. The DMSO-$d_6$ reference peak shown at 33.42 ppm. **B**) 1D $^1H$ NMR shows two peaks corresponding to H6 in the thymine ring and the hydrogens of the methyl group. The Hs associated with Ns have not been detected. The solvent peak is suppressed for clarity. **C**) 1D $^{15}N$ NMR further reaffirms the integrity of the thymine ring in concentrated sulfuric acid by showing two peaks corresponding to the nitrogen atoms of the aromatic ring. **D**) The 2D $^1H$-$^{13}C$ HMQC NMR shows direct bonding between H and C atoms in the thymine ring structure. As expected it shows two signals, at the intersection of hydrogen atoms H6 and the hydrogens of the methyl group (f2: $^1H$ trace spectra) and carbon atoms C6 and the carbon atom of the methyl

group (f1: $^{13}$C trace spectra). The experiment confirms the identity of 9.99 ppm and 149.59 ppm peaks in 1D $^{13}$C NMR spectra as carbons of the methyl group and C6 respectively. **E**) The 2D $^{1}$H-$^{13}$C HMBC NMR shows signals that correspond to hydrogen and carbon atoms separated from each other by the distance of two or three chemical bonds in the thymine ring structure (blue arrows). The correct bond separation between atoms derived from the 2D $^{1}$H-$^{13}$C HMBC NMR confirms the identity of carbon peaks assigned in the 1D $^{13}$C NMR spectra. **F**) The 2D $^{1}$H-$^{15}$N HMBC NMR shows two bond separations between hydrogen attached to carbon and a nitrogen atom (blue arrows). The spectrum shows a single signal at the expected position. The signal corresponds to a separation of two chemical bonds between hydrogen H6 and nitrogen N1 in the thymine structure. The correct separations between atoms derived from the 2D $^{1}$H-$^{15}$N HMBC NMR confirm the identity of nitrogen peaks in the 1D $^{15}$N NMR spectra. The relationships between atoms derived from the 2D NMR, taken together with the 1D NMR data, further confirm peak assignments of the carbon, hydrogen and nitrogen 1D NMR spectra and support the hypothesis that the thymine ring remains unchanged and is stable in 98% w/w concentrated sulfuric acid.

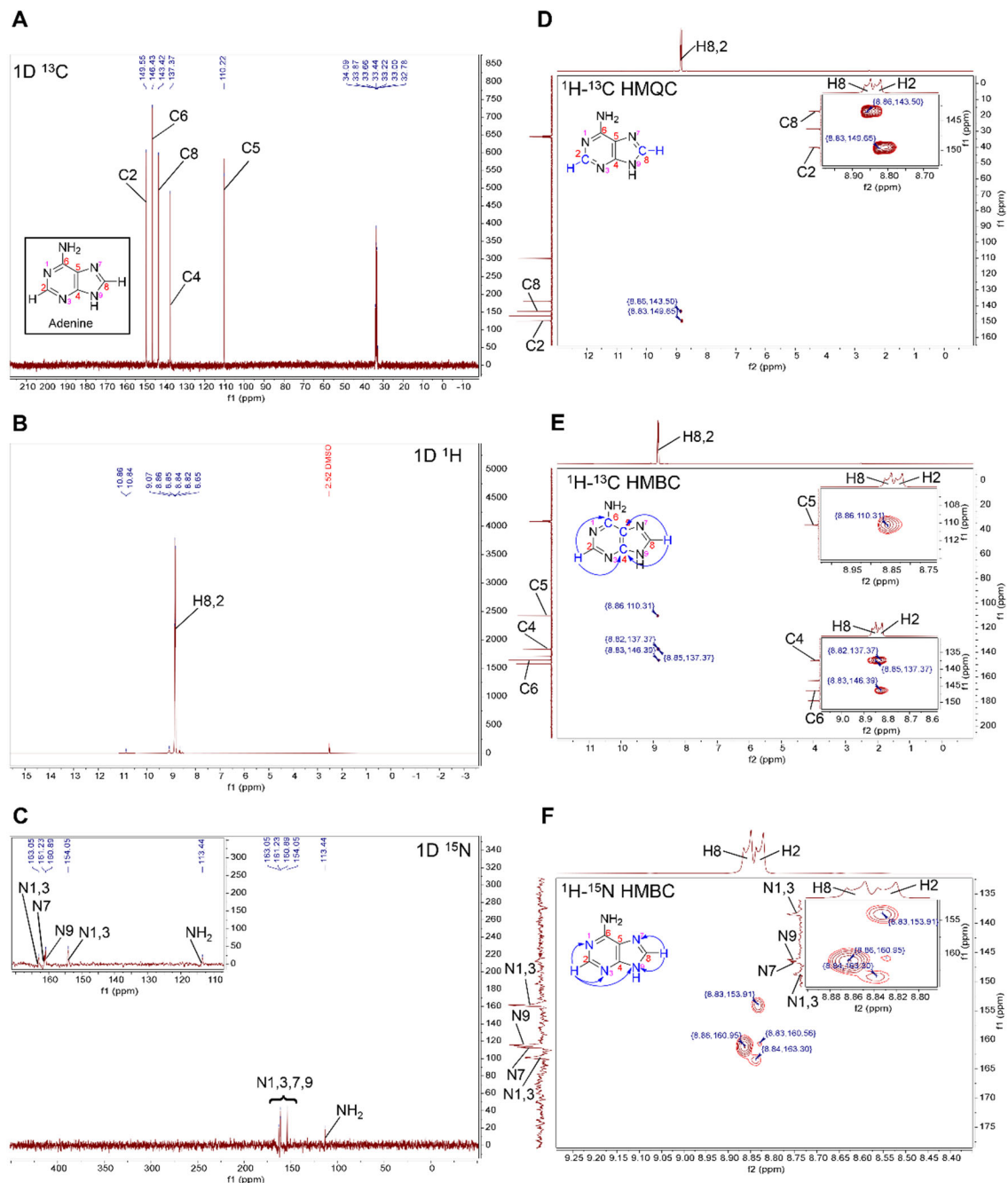

**Fig. S5.** NMR spectra for adenine concentrated sulfuric acid (98% $D_2SO_4$ and 2% $D_2O$, by weight, with reference DMSO-$d_6$) at room temperature. The NMR experiments confirm the stability of adenine in concentrated sulfuric acid. **A**) 1D $^{13}C$ NMR shows five peaks corresponding to five carbons in the adenine ring. The DMSO-$d_6$ reference peak shown at 33.44 ppm. **B**) 1D $^1H$ NMR shows overlapping peaks corresponding to hydrogen atoms H8 and H2 in the adenine ring. The 98% $D_2SO_4$ and 2% $D_2O$ solvent peak is suppressed for clarity. **C**) 1D $^{15}N$ NMR further reaffirms the integrity of the adenine ring in concentrated sulfuric acid by showing four peaks of the nitrogen atoms of the aromatic ring and one nitrogen belonging to the intact amino group attached to carbon C6. **D**) The 2D $^1H$-$^{13}C$ HMQC NMR shows direct bonding between H and C atoms in the adenine ring structure. As expected it shows two signals, at the intersection of hydrogen atoms H8 and H2 (f2: $^1H$ tracer spectra) and carbon atoms C8 and C2 (f1: $^{13}C$ tracer spectra). This confirms the identity of the 143.42 ppm and 149.55 ppm peaks in the 1D $^{13}C$ NMR

spectra as carbons C8 and C2 respectively. **E**) The 2D $^1$H-$^{13}$C HMBC NMR shows signals that correspond to hydrogen and carbon atoms separated from each other by 3 chemical bonds in the adenine ring structure (blue arrows). As expected for adenine, the spectrum shows three signals at the expected positions. The correct separations between atoms derived from the 2D $^1$H-$^{13}$C HMBC NMR confirm the identity of the carbon atom peaks in 1D $^{13}$C NMR spectra. **F**) The 2D $^1$H-$^{15}$N HMBC NMR shows bond distances between hydrogen atoms attached to carbon and nitrogen atoms (blue arrows). The spectrum shows signals at the expected positions allowing for the assignment of the N7 and N9 nitrogen atoms. The relationships between atoms derived from the 2D NMR, taken together with the 1D NMR data, further confirm peak assignments of the carbon, hydrogen and nitrogen 1D NMR spectra and support the hypothesis that the adenine ring remains unchanged and is stable in 98% w/w concentrated sulfuric acid.

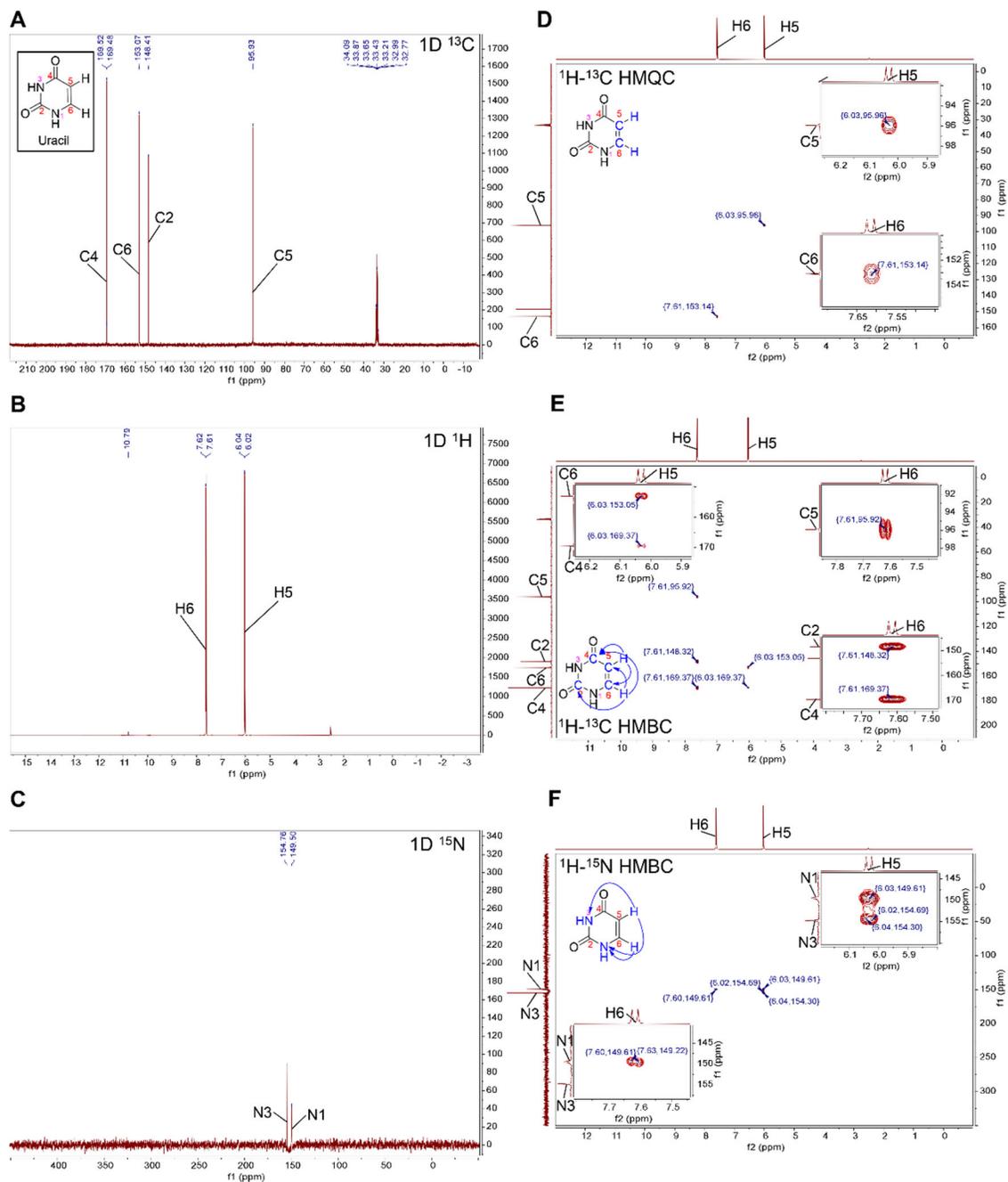

**Fig. S6.** NMR spectra for uracil in concentrated sulfuric acid (98% D$_2$SO$_4$ and 2% D$_2$O, by weight, with reference DMSO-d$_6$) at room temperature. The NMR experiments confirm the stability of uracil in concentrated sulfuric acid. **A**) 1D $^{13}$C NMR shows four peaks corresponding to four carbons in the uracil ring. DMSO-d$_6$ reference peak shown at 33.43 ppm. **B**) 1D $^1$H NMR shows two peaks corresponding to hydrogen atoms attached to carbons in the uracil ring. The Hs associated with Ns have not been detected. The solvent peak is suppressed for clarity. **C**) 1D $^{15}$N NMR further reaffirms the integrity of the uracil ring in concentrated sulfuric acid by showing two peaks corresponding to the nitrogen atoms of the aromatic ring. **D**) The 2D $^1$H-$^{13}$C HMQC NMR shows direct bonding between H and C atoms in the uracil ring structure. As expected it shows two signals, at the intersection of hydrogen atoms H5 and H6 (f2: $^1$H trace spectra) and carbon atoms C5 and C6 (f1: $^{13}$C trace spectra). The experiment confirms the identity of the 95.93 ppm and 153.07 ppm peaks in 1D $^{13}$C NMR spectra as carbons C5 and C6 respectively. **E**) The 2D

$^1$H-$^{13}$C HMBC NMR shows signals that correspond to hydrogen and carbon atoms separated from each other by two or three chemical bonds in the uracil ring structure (blue arrows). The correct separations between atoms derived from the 2D $^1$H-$^{13}$C HMBC NMR confirm the assignment of carbon peaks in 1D $^{13}$C NMR spectra. **F**) The 2D $^1$H-$^{15}$N HMBC NMR shows 2 or 3 bond distances between hydrogens attached to carbon and nitrogen atoms (blue arrows). The spectrum shows three signals at the expected positions. The three signals correspond to a distance of 2 or 3 chemical bonds between hydrogens H5 and H6 and nitrogens N1 and N3 in the uracil structure. The correct separations between atoms derived from the 2D $^1$H-$^{15}$N HMBC NMR confirm the identity of nitrogen peaks in the 1D $^{15}$N NMR spectra. The relationships between atoms derived from the 2D NMR, taken together with the 1D NMR data, further confirm peak assignments of the carbon, hydrogen and nitrogen 1D NMR spectra and support the hypothesis that the uracil ring remains unchanged and is stable in 98% w/w concentrated sulfuric acid.

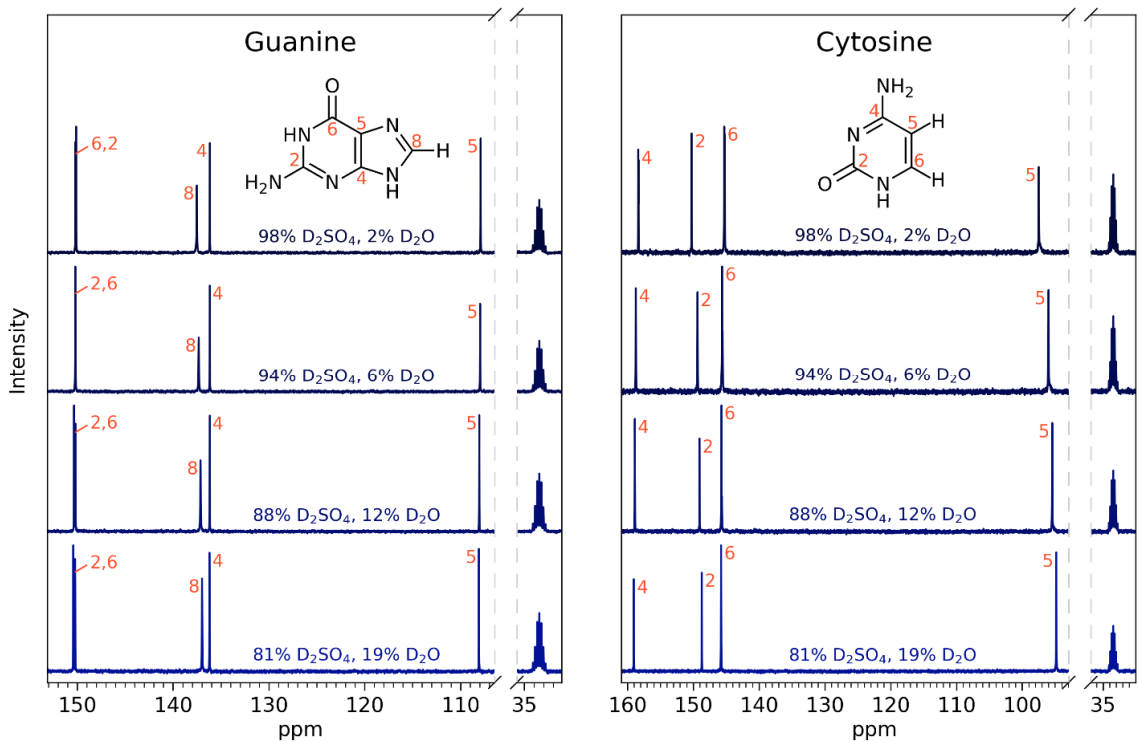

**Fig. S7**. 1D $^{13}$C NMR spectra for guanine (left) and cytosine (right) for a range of sulfuric acid concentrations found in the Venus clouds. From top to bottom are different concentrations (by weight) of sulfuric acid in water: 98% D$_2$SO$_4$/2% D$_2$O; 94% D$_2$SO$_4$/6% D$_2$O; 88% D$_2$SO$_4$/12% D$_2$O; 81% D$_2$SO$_4$/19% D$_2$O with DMSO-d$_6$ as a reference and at room temperature. The labeled NMR peaks show five peaks corresponding to five carbon atoms in the guanine ring, and 4 peaks corresponding to four carbons of the cytosine ring. All peaks are consistent with the molecules being stable and the structure not being affected by the concentrated sulfuric acid solvent. For a description of peak assignments, see Figure S1 and Figure S2.

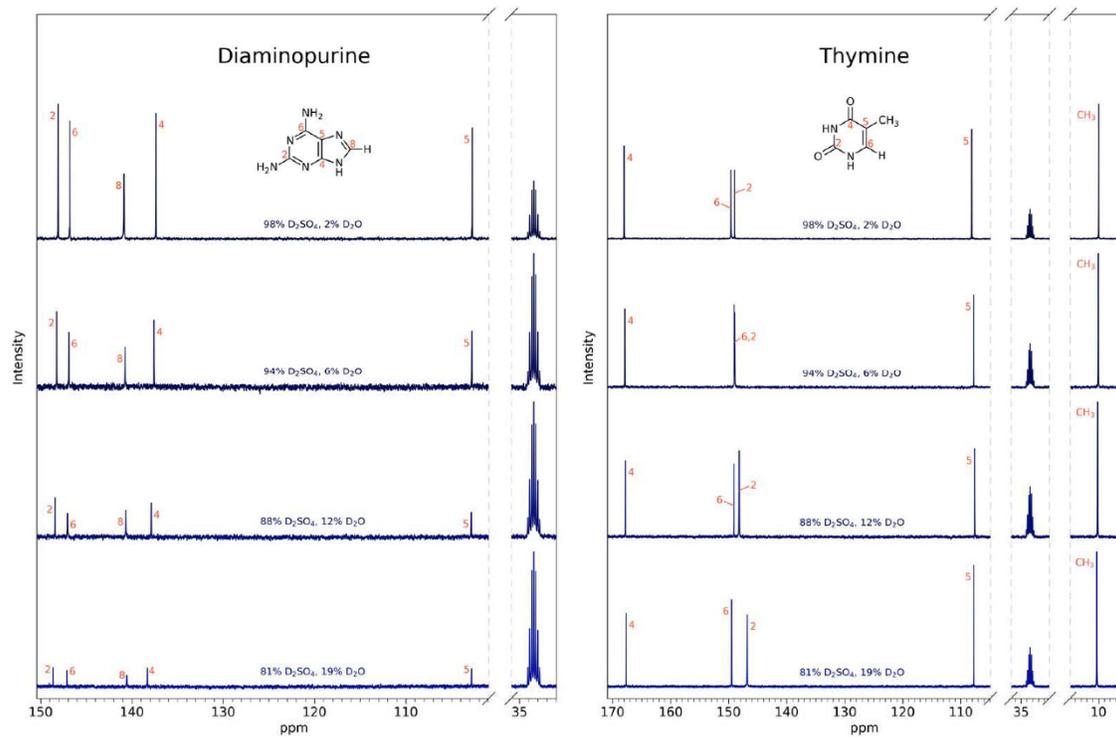

**Fig. S8.** 1D $^{13}$C NMR spectra for 2,6-diaminopurine (left) and thymine (right) for a range of sulfuric acid concentrations found in the Venus clouds. From top to bottom are different concentrations (by weight) of sulfuric acid in water: 98% $D_2SO_4$/2% $D_2O$; 94% $D_2SO_4$/6% $D_2O$; 88% $D_2SO_4$/12% $D_2O$; 81% $D_2SO_4$/19% $D_2O$ with DMSO-$d_6$ as a reference and at room temperature. The labeled NMR peaks show five peaks corresponding to five carbon atoms in the 2,6-diaminopurine ring, and 5 peaks corresponding to five carbons of the pyrimidine ring and a methyl group. All peaks are consistent with the molecules being stable and the structure not being affected by the concentrated sulfuric acid solvent. For a description of peak assignments, see Figure S3 and Figure S4.

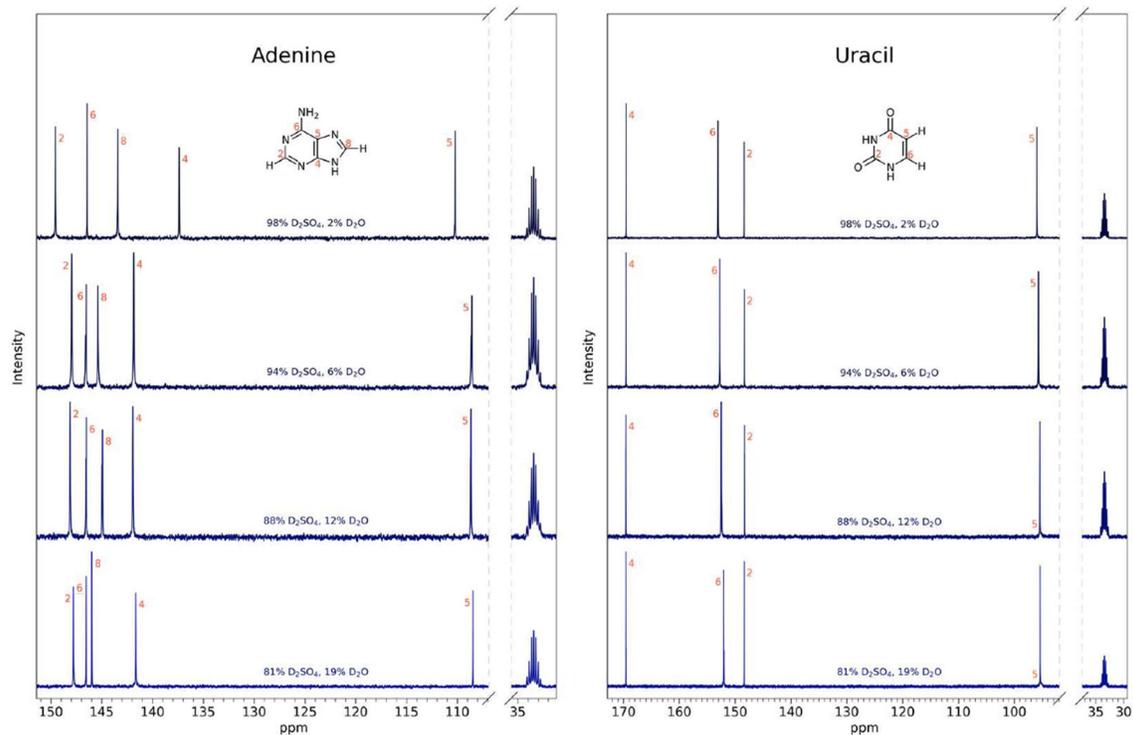

**Fig. S9.** 1D $^{13}$C NMR spectra for adenine (left) and uracil (right) for a range of sulfuric acid concentrations found in the Venus clouds. From top to bottom are different concentrations (by weight) of sulfuric acid in water: 98% $D_2SO_4$/2% $D_2O$; 94% $D_2SO_4$/6% $D_2O$; 88% $D_2SO_4$/12% $D_2O$; 81% $D_2SO_4$/19% $D_2O$ with DMSO-$d_6$ as a reference and at room temperature. The labeled NMR peaks show five peaks corresponding to five carbon atoms in the adenine ring, and 4 peaks corresponding to four carbons of the uracil ring. All peaks are consistent with the molecules being stable and the structure not being affected by the concentrated sulfuric acid solvent. For a description of peak assignments, see Figure S5 and Figure S6.

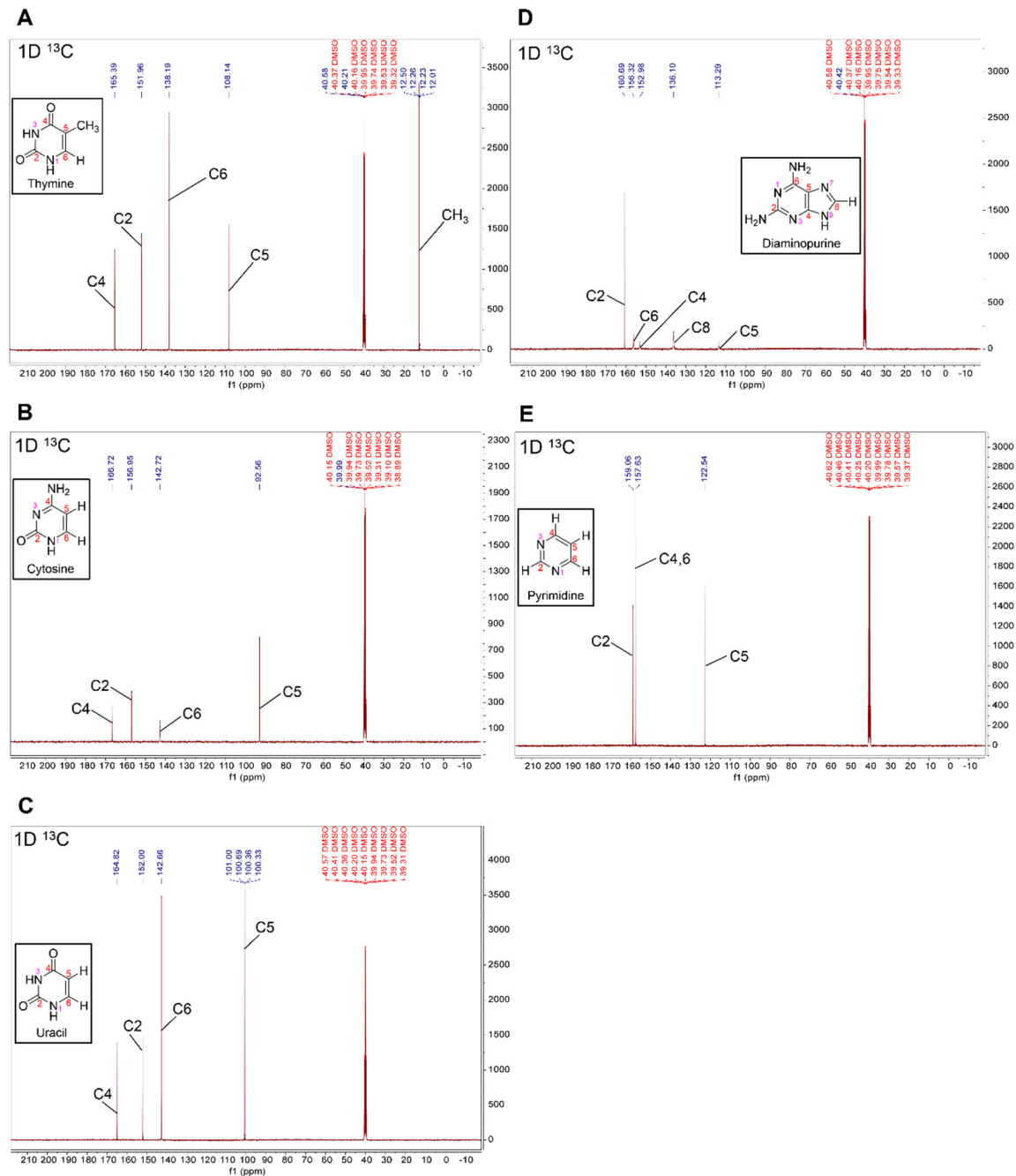

**Fig. S10.** 1D $^{13}$C NMR in DMSO-$d_6$ solvent for **A**) thymine, **B**) cytosine, **C**) uracil **D**) 2,6-diaminopurine, **E**) pyrimidine. The chemical shifts of the collected spectra are consistent with previously reported values (Table S2, Table S3, Table S4, Table S6 and Table S8). Spectra for adenine, purine and guanine could not be collected due to poor solubility of these compounds in DMSO-$d_6$ solvent.

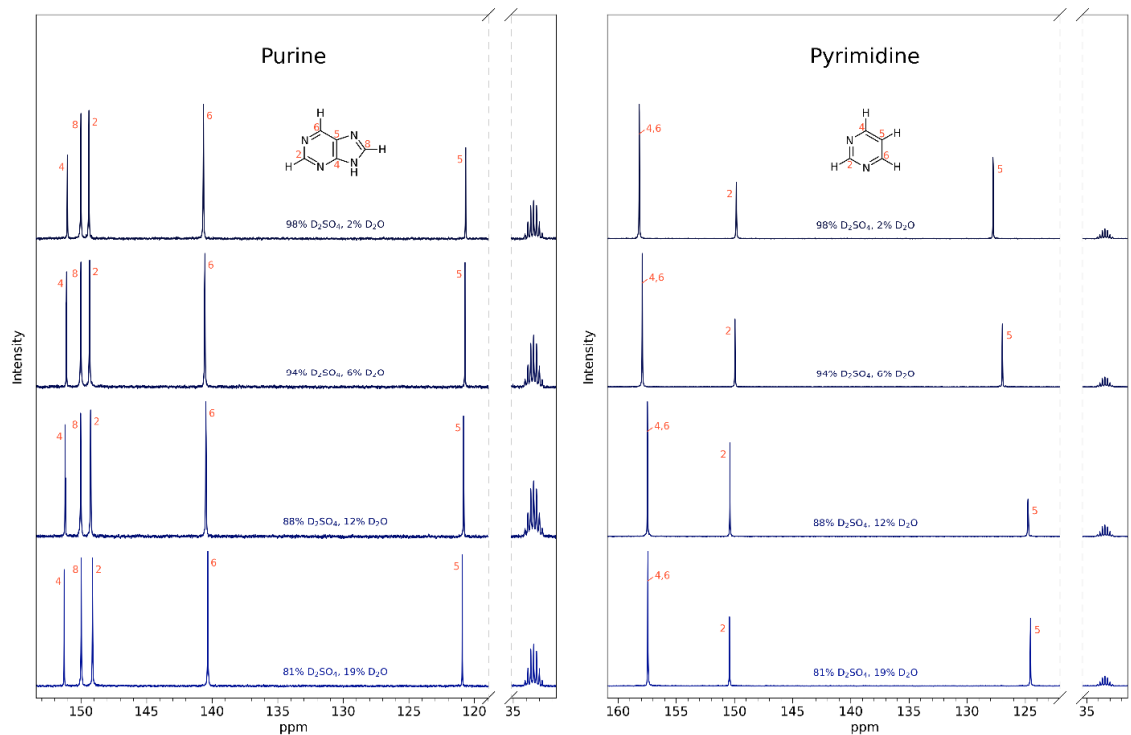

**Fig. S11.** 1D $^{13}$C NMR spectra for purine (left) and pyrimidine (right) for a range of sulfuric acid concentrations found in the Venus clouds. From top to bottom are different concentrations (by weight) of sulfuric acid in water: 98% $D_2SO_4$/2% $D_2O$; 94% $D_2SO_4$/6% $D_2O$; 88% $D_2SO_4$/12% $D_2O$; 81% $D_2SO_4$/19% $D_2O$ with DMSO-$d_6$ as a reference and at room temperature. The labeled NMR peaks show five peaks corresponding to five carbon atoms in the purine ring, and 4 peaks corresponding to four carbons of the pyrimidine ring. All peaks are consistent with the molecules being stable and the structure not being affected by the concentrated sulfuric acid solvent. For a description of peak assignments, see Figure 5 and Figure 6 of the main text.

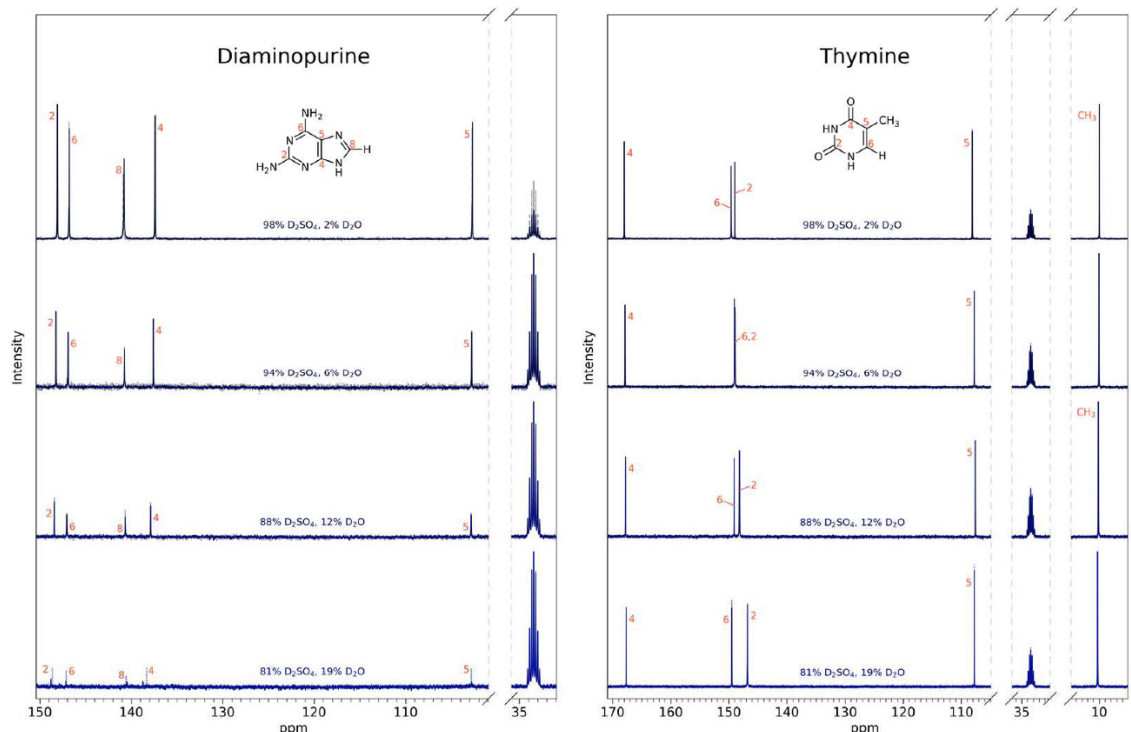

**Fig. S12.** 2,6-diaminopurine (left) and thymine (right) are stable for two weeks in a range of sulfuric acid concentrations found in the Venus clouds. We have incubated 10-80 mg of each compound in 81-98% w/w $D_2SO_4$ for two weeks. Due to low solubility of diaminopurine in concentrated sulfuric acid different amounts have been tested. After two-week incubation we have collected the 1D $^{13}C$ NMR spectra (solid line spectra), at each of the tested acid concentrations, and compared them to the original 1D $^{13}C$ NMR spectra collected after ~30-48 h (dashed line spectra and Figure S8). The two-week spectra and the ~30-48 h spectra look virtually identical for all tested concentrations, confirming long-term stability of the nucleic acid bases in concentrated sulfuric acid solvent. From top to bottom are different concentrations (by weight) of sulfuric acid in water: 98% $D_2SO_4$/2% $D_2O$; 94% $D_2SO_4$/6% $D_2O$; 88% $D_2SO_4$/12% $D_2O$; 81% $D_2SO_4$/19% $D_2O$ with DMSO-$d_6$ as a reference and at room temperature. All peaks are consistent with the molecules being stable and the structure not being affected by the concentrated sulfuric acid solvent.

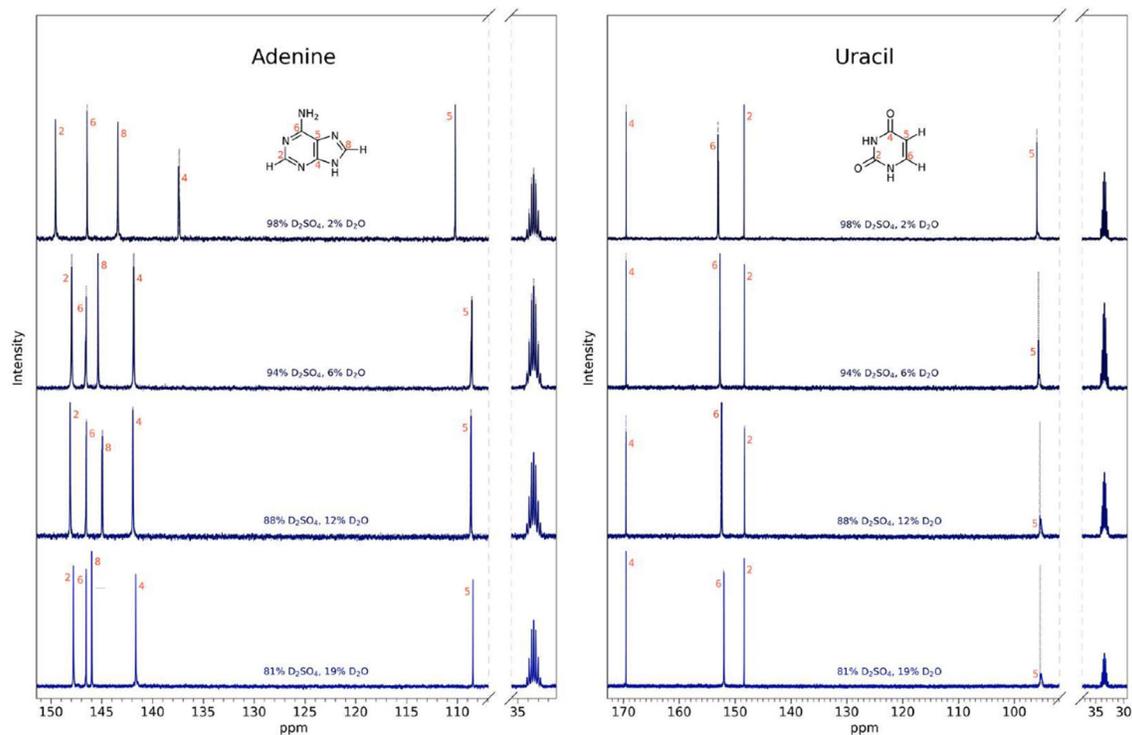

**Fig. S13.** Adenine (left) and uracil (right) are stable for 2 weeks in a range of sulfuric acid concentrations found in the Venus clouds. We have incubated 30 mg of each base in 81-98% w/w $D_2SO_4$ for two weeks. After two-week incubation we have collected the 1D $^{13}C$ NMR spectra (solid line spectra), at each of the tested acid concentrations, and compared them to the original 1D $^{13}C$ NMR spectra collected after ~30-48 h (dashed line spectra and Figure S9). The two-week spectra and the ~30-48 h spectra look virtually identical for all tested concentrations, confirming long-term stability of the nucleic acid bases in concentrated sulfuric acid solvent. From top to bottom are different concentrations (by weight) of sulfuric acid in water: 98% $D_2SO_4$/2% $D_2O$; 94% $D_2SO_4$/6% $D_2O$; 88% $D_2SO_4$/12% $D_2O$; 81% $D_2SO_4$/19% $D_2O$ with DMSO-$d_6$ as a reference and at room temperature. All peaks are consistent with the molecules being stable and the structure not being affected by the concentrated sulfuric acid solvent.

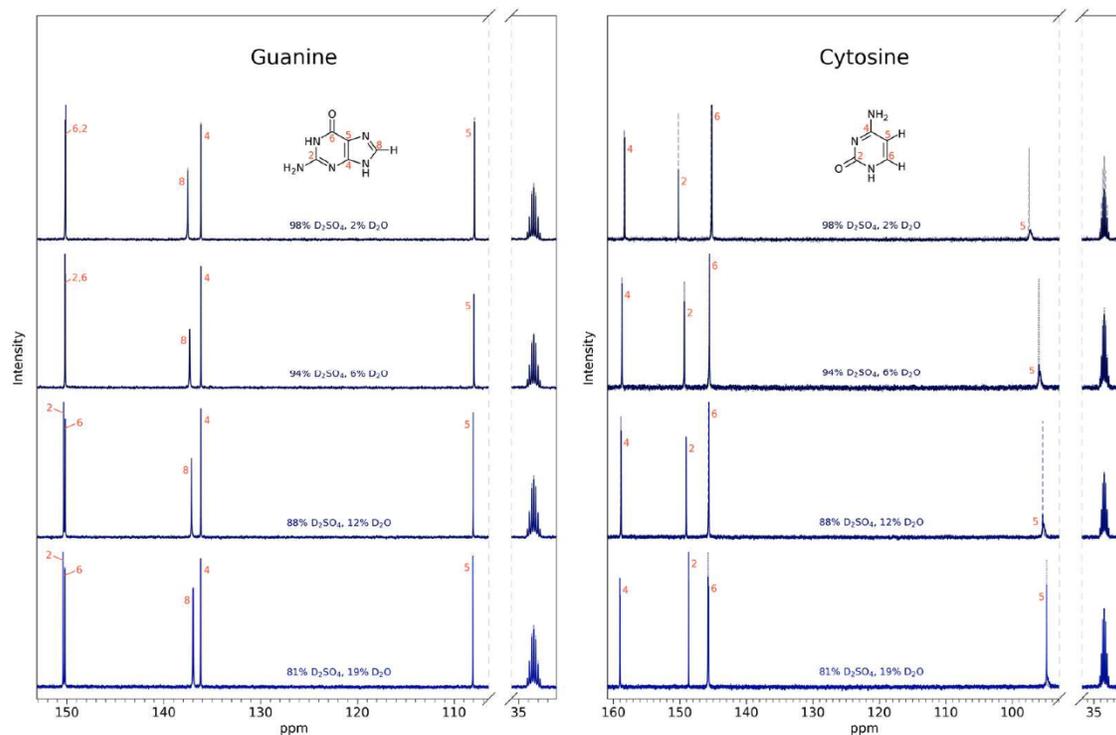

**Fig. S14.** Guanine (left) and cytosine (right) are stable for 2 weeks in a range of sulfuric acid concentrations found in the Venus clouds. We have incubated 30-40 mg of each base in 81-98% w/w $D_2SO_4$ for two weeks. After two-week incubation we have collected the 1D $^{13}C$ NMR spectra (solid line spectra), at each of the tested acid concentrations, and compared them to the original 1D $^{13}C$ NMR spectra collected after ~30-48 h (dashed line spectra and Figure 4). The two-week spectra and the ~30-48 h spectra look virtually identical for all tested concentrations, confirming long-term stability of the nucleic acid bases in concentrated sulfuric acid solvent. From top to bottom are different concentrations (by weight) of sulfuric acid in water: 98% $D_2SO_4$/2% $D_2O$; 94% $D_2SO_4$/6% $D_2O$; 88% $D_2SO_4$/12% $D_2O$; 81% $D_2SO_4$/19% $D_2O$ with DMSO-$d_6$ as a reference and at room temperature. All peaks are consistent with the molecules being stable and the structure not being affected by the concentrated sulfuric acid solvent.

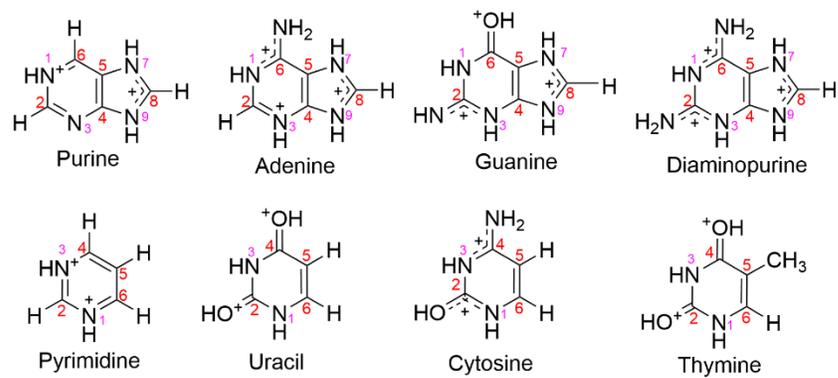

**Fig. S15.** Inferred protonation states of the nucleic acid bases studied in this paper in concentrated sulfuric acid. See text for details.

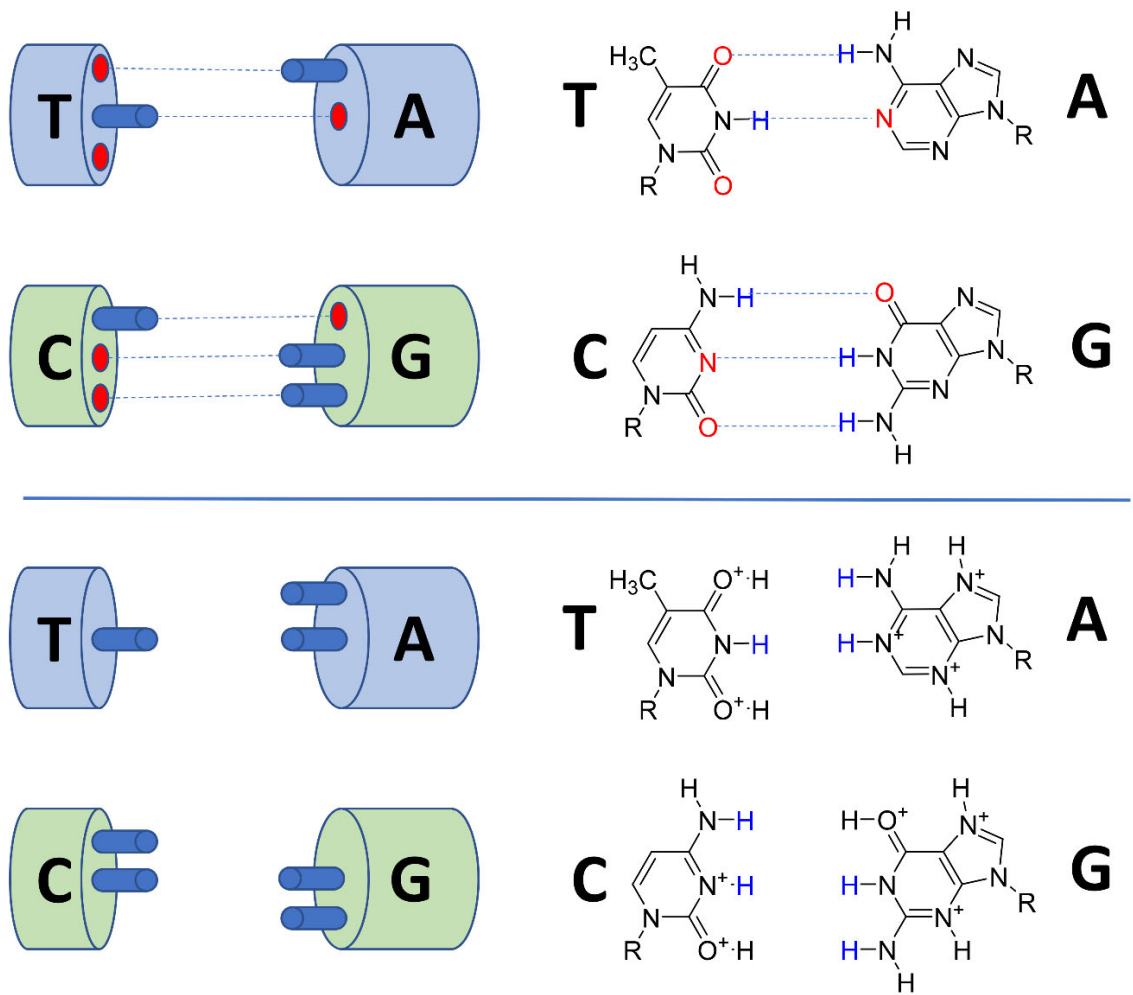

**Fig. S16.** Protonation and anticipated interference with nucleic acid base pairing. *Right:* Chemical structures of nucleic acid base pairs. *Top*: Formation of base pairs between adenine (A) and thymine (T), cytosine (C) and guanine (G) in water. *Bottom*: The efficient base-pairing of the canonical bases in concentrated sulfuric acid may be perturbed by the protonation of the nitrogen atoms which converts hydrogen bond acceptors (red circles) into hydrogen bond donors (blue rods).

Table S1. An overview and comparison of NMR chemical shifts of guanine from this work, obtained in 98% $D_2SO_4$, 2% $D_2O$ (by weight) to values in selected solvents reported in the literature. All listed literature NMR data have been obtained at, or close to, room temperature.

| Guanine | | | | | |
|---|---|---|---|---|---|
| $^{13}C$ | | | | | |
| Solvent (reference) | C2 (ppm) | C4 (ppm) | C5 (ppm) | C6 (ppm) | C8 (ppm) |
| $D_2O$ pH=13 (1) | 160.00 | 162.20 | 119.60 | 168.80 | 150.10 |
| $D_2O$/NaOH (7) | 160.60 | 162.80 | 120.30 | 169.30 | 150.50 |
| DMSO-$d_6$ (2)* | 153.80 | 150.45 | 108.42 | 155.74 | 137.87 |
| DMSO-$d_6$/HCl (8) | 156.00 | 150.00 | 108.00 | 154.00 | 138.00 |
| 98% $D_2SO_4$, 2% $D_2O$ | 150.13 | 136.17 | 107.90 | 150.19 | 137.53 |
| $^1H$ | | | | | |
| Solvent (reference) | H8 (ppm) | H9 (ppm) | H7 (ppm) | $NH_2$ (ppm) | |
| $D_2O$ (2)* | 8.75 | – | – | – | |
| 98% $D_2SO_4$, 2% $D_2O$ | 8.29 | 12.21 | 11.73 | 7.05 | |
| $^{15}N$ | | | | | |
| Solvent (reference) | N1 (ppm) | N3 (ppm) | N7 (ppm) | N9 (ppm) | $NH_2$ (ppm) |
| 98% $D_2SO_4$, 2% $D_2O$ | 102.79 | 142.94 | 154.76 | 159.34 | 85.92 |

(–) Available literature does not provide chemical shift values for the specified atoms or no peaks corresponding to hydrogens attached to nitrogen have been detected.
(*) Data for hydrochloride salt.

Table S2. An overview and comparison of NMR chemical shifts of cytosine from this work, obtained in 98% D₂SO₄, 2% D₂O (by weight) to values in selected solvents reported in the literature. All listed literature NMR data have been obtained at, or close to, room temperature.

| Cytosine | | | | |
|---|---|---|---|---|
| **$^{13}$C** | | | | |
| **Solvent (reference)** | **C2 (ppm)** | **C4 (ppm)** | **C5 (ppm)** | **C6 (ppm)** |
| D₂O (2) | 159.91 | 168.10 | 95.79 | 144.05 |
| D₂O (2)* | 149.74 | 160.76 | 94.61 | 146.92 |
| D₂O (9) | 166.40 | 167.20 | 94.20 | 155.30 |
| D₂O (10) | 160.20 | 168.30 | 95.90 | 143.90 |
| DMSO-d6 (2) | 157.77 | 167.49 | 93.35 | 143.47 |
| DMSO-d6 (11) | 156.90 | 166.60 | 92.50 | 142.70 |
| DMSO-d6 (12) | 156.80 | 166.50 | 92.40 | 142.70 |
| DMSO-d6** | 156.95 | 166.72 | 92.56 | 142.72 |
| conc. H₂SO₄ (5) | 150.19 | 161.03 | 95.80 | 147.60 |
| 98% D₂SO₄, 2% D₂O | 150.23 | 158.34 | 97.48 | 145.28 |
| **$^1$H** | | | | |
| **Solvent (reference)** | **H5 (ppm)** | **H6 (ppm)** | **H1 (ppm)** | **NH₂ (ppm)** |
| D₂O (2) | 5.97 | 7.50 | – | – |
| D₂O (9) | 5.78 | 7.62 | – | – |
| D₂O (13) | 5.94 | 7.50 | – | – |
| D₂O pH=3.35 (14) | 5.89 | 7.86 | – | – |
| DMSO-d6 (11) | 5.58 | 7.33 | – | – |
| DMSO-d6 (12) | 5.56 | 7.29 | 10.20 | 6.95 |
| DMSO-d6 (15, 16) | 5.56 | 7.32 | 10.39 | 7.03 |
| conc. H₂SO₄ (5) | 6.53 | 7.90 | see (5) | see (5) |
| 98% D₂SO₄, 2% D₂O | 5.88 | 7.17 | – | – |
| **$^{15}$N** | | | | |
| **Solvent (reference)** | **N1 (ppm)** | **N3 (ppm)** | **NH₂ (ppm)** | |
| 98% D₂SO₄, 2% D₂O | 136.44 | 138.61 | 102.59 | |

(–) Available literature does not provide chemical shift values for the specified atoms or no peaks corresponding to hydrogens attached to nitrogen have been detected.
(*) Data for hydrochloride salt.
(**) This study, see Figure S10 in the SI.

**Table S3.** An overview and comparison of NMR chemical shifts of 2,6-diaminopurine from this work, obtained in 98% $D_2SO_4$, 2% $D_2O$ (by weight) to values in different solvents reported in the literature.

| Diaminopurine | | | | | | |
|---|---|---|---|---|---|---|
| $^{13}C$ | | | | | | |
| Solvent (reference) | C2 (ppm) | C4 (ppm) | C5 (ppm) | C6 (ppm) | C8 (ppm) | |
| DMSO-$d_6$ (17) | 160.20 | 152.77 | 112.50 | 155.78 | 135.91 | |
| DMSO-$d_6$** | 160.69 | 152.98 | 113.29 | 156.32 | 136.10 | |
| 98% $D_2SO_4$, 2% $D_2O$ | 148.08 | 137.38 | 102.74 | 146.80 | 140.87 | |
| $^1H$ | | | | | | |
| Solvent (reference) | H8 (ppm) | H9 (ppm) | H7 (ppm) | $NH_2$ (ppm) | $NH_2$ (ppm) | |
| DMSO-$d_6$ (18, 19) | 7.76 | – | – | – | – | |
| DMSO-$d_6$ (20) | 8.48 | – | – | – | – | |
| 98% $D_2SO_4$, 2% $D_2O$ | 8.53 | – | – | – | – | |
| $^{15}N$ | | | | | | |
| Solvent (reference) | N1 (ppm) | N3 (ppm) | N7 (ppm) | N9 (ppm) | $NH_2$ (ppm) | $NH_2$ (ppm) |
| 98% $D_2SO_4$, 2% $D_2O$ | see text | see text | 156.96 | 160.12 | see text | see text |

(–) Available literature does not provide chemical shift values for the specified atoms or no peaks corresponding to hydrogens attached to nitrogen have been detected.
(**) This study, see Figure S10.

Table S4. An overview and comparison of NMR chemical shifts of thymine obtained in 98% $D_2SO_4$, 2% $D_2O$ (by weight) to values in selected solvents reported in the literature. All listed literature NMR data have been obtained at, or close to, room temperature.

| **Thymine** | | | | | |
|---|---|---|---|---|---|
| **$^{13}C$** | | | | | |
| Solvent (reference) | C2 (ppm) | C4 (ppm) | C5 (ppm) | C6 (ppm) | $CH_3$ (ppm) |
| $D_2O$ (2) | 153.90 | 168.29 | 110.88 | 139.79 | 12.06 |
| DMSO-$d_6$ (2) | 151.46 | 164.87 | 107.66 | 137.63 | 11.72 |
| DMSO-$d_6$ (21) | 151.49 | 164.93 | 107.68 | 137.72 | 11.79 |
| DMSO-$d_6$ (22) | 151.50 | 164.80 | 107.40 | 137.80 | 11.70 |
| DMSO-$d_6$ (23) | 151.50 | 164.90 | 107.60 | 137.80 | 11.80 |
| DMSO-$d_6$ (24) | 151.51 | 165.48 | 108.12 | 138.19 | 12.24 |
| DMSO-d6 (11) | 151.40 | 164.90 | 107.60 | 137.70 | – |
| DMSO-$d_6$** | 151.96 | 165.39 | 108.14 | 138.19 | 12.26 |
| conc. $H_2SO_4$ (5) | 150.86 | 169.60 | 109.72 | 150.57 | 11.96 |
| 98% $D_2SO_4$, 2% $D_2O$ | 148.95 | 167.98 | 108.09 | 149.59 | 9.99 |
| **$^1H$** | | | | | |
| Solvent (reference) | H6 (ppm) | H1 (ppm) | H3 (ppm) | | $CH_3$ (ppm) |
| $D_2O$/PBS (25) | 7.39 | – | – | | 1.87 |
| $D_2O$ (26) | 7.38 | – | – | | 1.89 |
| $D_2O$ (27) | 7.65 | – | – | | 1.90 |
| $D_2O$ (28, 29) | 7.41 | – | – | | 1.91 |
| $D_2O$ (30) | 7.38 | 10.60 | 10.96 | | – |
| $D_2O$ (31) | 7.38 | – | – | | 1.89 |
| DMSO-$d_6$ (2) | 7.28 | 10.60 | 11.00 | | 1.75 |
| DMSO-$d_6$ (21) | 7.24 | 10.98 | 10.57 | | 1.72 |
| DMSO-$d_6$ (22) | 6.80 | – | – | | 1.26 |
| DMSO-$d_6$ (23, 24) | 7.25 | – | – | | 1.72 |
| DMSO-d6 (11) | 7.24 | – | – | | – |
| conc. $H_2SO_4$ (5) | 7.82 | see (5) | see (5) | | 2.09 |
| 98% $D_2SO_4$, 2% $D_2O$ | 7.43 | – | – | | 1.58 |
| **$^{15}N$** | | | | | |
| Solvent (reference) | N1 (ppm) | N3 (ppm) | | | |
| 98% $D_2SO_4$, 2% $D_2O$ | 146.64 | 154.42 | | | |

(–) Available literature does not provide chemical shift values for the specified atoms or no peaks corresponding to hydrogens attached to nitrogen have been detected.
(**) This study, see Figure S10.

Table S5. An overview and comparison of NMR chemical shifts of adenine from this work, obtained in 98% $D_2SO_4$, 2% $D_2O$ (by weight) to values in selected solvents reported in the literature. All listed literature NMR data have been obtained at, or close to, room temperature.

| Adenine | | | | | |
|---|---|---|---|---|---|
| **[13]C** | | | | | |
| Solvent (reference) | C2 (ppm) | C4 (ppm) | C5 (ppm) | C6 (ppm) | C8 (ppm) |
| $D_2O$ pH=13 (1) | 150.5 | 160.4 | 121.0 | 155.0 | 153.7 |
| DMSO-$d_6$ (2) | 153.41 | 151.71 | 119.07 | 156.36 | 140.30 |
| DMSO-$d_6$ (32) | 152.40 | 151.30 | 117.50 | 155.30 | 139.30 |
| DMSO-$d_6$ (33) | 152.20 | 151.60 | 121.90 | 155.60 | 139.50 |
| DMSO-$d_6$ (34) | 152.35 | 151.15 | 117.54 | 155.41 | 139.63 |
| DMSO-$d_6$ (17, 35) | 152.37 | 151.30 | 117.61 | 155.30 | 139.29 |
| DMSO-$d_6$ (36) | 152.46 | 151.44 | 117.41 | 155.37 | 139.38 |
| DMSO-$d_6$/HCl (8) | 144.00 | 150.00 | 114.00 | 152.00 | 146.00 |
| 98% $D_2SO_4$, 2% $D_2O$ | 149.55 | 137.37 | 110.22 | 146.43 | 143.42 |
| **[1]H** | | | | | |
| Solvent (reference) | H2 (ppm) | H8 (ppm) | H9 (ppm) | H7 (ppm) | $NH_2$ (ppm) |
| $D_2O$ (3) | 8.62 | 8.57 | – | – | – |
| $D_2O$ (37) | 8.27 | 8.22 | – | – | – |
| $D_2O$ (38) | 8.05 | 8.09 | – | – | – |
| $D_2O$/NaOD (39) | 8.23 | 8.12 | – | – | – |
| $CDCl_3$ (2) | 8.14 | 8.11 | 12.80 | – | 7.09 |
| DMSO-$d_6$ (33) | 8.11 | 8.10 | 12.78 | – | 7.10 |
| DMSO-$d_6$ (36) | 8.12 | 8.11 | 12.75 | – | 7.10 |
| DMSO-$d_6$ (18) | – | 8.14 | – | – | – |
| 98% $D_2SO_4$, 2% $D_2O$ | 8.83 | 8.86 | – | – | – |
| **[15]N** | | | | | |
| Solvent (reference) | N1 (ppm) | N3 (ppm) | N7 (ppm) | N9 (ppm) | $NH_2$ (ppm) |
| DMSO-$d_6$ (40) | 236.00 | 230.20 | 241.40 | 159.10 | 80.50 |
| 98% $D_2SO_4$, 2% $D_2O$ | see text | see text | 161.23 | 160.89 | 113.44 |

(–) Available literature does not provide chemical shift values for the specified atoms or no peaks corresponding to hydrogens attached to nitrogen have been detected.

Table S6. An overview and comparison of NMR chemical shifts of uracil from this work, obtained in 98% $D_2SO_4$, 2% $D_2O$ (by weight) to values in selected solvents reported in the literature. All listed literature NMR data have been obtained at, or close to, room temperature.

| Uracil | | | | |
|---|---|---|---|---|
| **$^{13}$C** | | | | |
| **Solvent (reference)** | **C2 (ppm)** | **C4 (ppm)** | **C5 (ppm)** | **C6 (ppm)** |
| DMSO-$d_6$ (2) | 152.27 | 165.09 | 101.01 | 142.89 |
| DMSO-$d_6$ (21) | 151.39 | 164.20 | 100.11 | 142.07 |
| DMSO-$d_6$ (33) | 151.40 | 164.30 | 100.10 | 142.10 |
| DMSO-$d_6$ (22) | 151.40 | 164.20 | 100.10 | 142.10 |
| DMSO-$d_6$ (24) | 151.98 | 164.80 | 100.67 | 142.67 |
| DMSO-$d_6$ (41, 42) | 151.39 | 164.20 | 100.10 | 142.07 |
| DMSO-$d_6$ (42) | 151.45 | 164.26 | 100.28 | 142.13 |
| DMSO-$d_6$** | 152.00 | 164.82 | 100.69 | 142.66 |
| conc. $H_2SO_4$ (5) | 150.30 | 171.60 | 97.30 | 153.90 |
| 98% $D_2SO_4$, 2% $D_2O$ | 148.41 | 169.48 | 95.93 | 153.07 |
| **$^1$H** | | | | |
| **Solvent (reference)** | **H5 (ppm)** | **H6 (ppm)** | **H1 (ppm)** | **H3 (ppm)** |
| $D_2O$ (37) | 5.85 | 7.57 | – | – |
| $D_2O$ (31) | 5.85 | 7.51 | – | – |
| $D_2O$/PBS (25) | 5.78 | 7.49 | – | – |
| DMSO-$d_6$ (2) | 5.47 | 7.41 | 10.82 | 11.02 |
| DMSO-$d_6$ (21) | 5.45 | 7.39 | 11.00 | 10.60 |
| DMSO-$d_6$ (33) | 5.45 | 7.39 | 10.93 | 10.93 |
| DMSO-$d_6$ (43) | 5.37 | 7.29 | – | – |
| DMSO-$d_6$ (22) | 5.44 | 7.38 | – | – |
| DMSO-$d_6$ (24) | 5.45 | 7.39 | 10.85 | 10.85 |
| DMSO-$d_6$ (41, 42) | 5.45 | 7.39 | 10.80 | 11.00 |
| DMSO-$d_6$ (42) | 5.82 | 7.74 | 11.31 | 11.31 |
| conc. $H_2SO_4$ (5) | 6.70 | 8.30 | see (5) | see (5) |
| 98% $D_2SO_4$, 2% $D_2O$ | 6.02 | 7.61 | – | – |
| **$^{15}$N** | | | | |
| **Solvent (reference)** | **N1 (ppm)** | **N3 (ppm)** | | |
| DMSO-$d_6$ (43) | 135.00 | 162.5 | | |
| 98% $D_2SO_4$, 2% $D_2O$ | 149.50 | 154.76 | | |

(–) Available literature does not provide chemical shift values for the specified atoms or no peaks corresponding to hydrogens attached to nitrogen have been detected.
(**) This study, see Figure S10.

**Table S7.** An overview and comparison of NMR chemical shifts of purine from this work, obtained in 98% $D_2SO_4$, 2% $D_2O$ (by weight) to values in selected solvents reported in the literature. All listed literature NMR data have been obtained at, or close to, room temperature.

| Purine | | | | | |
|---|---|---|---|---|---|
| **$^{13}C$** | | | | | |
| Solvent (reference) | C2 (ppm) | C4 (ppm) | C5 (ppm) | C6 (ppm) | C8 (ppm) |
| $D_2O$ (2) | 152.39 | 155.74 | 129.04 | 145.36 | 148.04 |
| DMSO-$d_6$ (2) | 152.03 | 154.54 | 130.31 | 145.53 | 146.07 |
| DMSO-$d_6$ (32) | 152.10 | 154.70 | 130.40 | 145.50 | 146.10 |
| DMSO-$d_6$ (17) | 152.10 | 154.77 | 130.46 | 145.50 | 146.09 |
| DMSO-$d_6$ (44) | 151.80 | 154.50 | 130.20 | 145.30 | 145.90 |
| DMSO-$d_6$ (45) | 152.21 | 152.50 | 133.05 | 145.80 | 147.12 |
| 98% $D_2SO_4$, 2% $D_2O$ | 149.39 | 151.04 | 120.66 | 140.66 | 150.00 |
| **$^{1}H$** | | | | | |
| Solvent (reference) | H2 (ppm) | H6 (ppm) | H8 (ppm) | H9 (ppm) | H7 (ppm) |
| $D_2O$ (2) | 8.72 | 8.83 | 8.50 | – | – |
| $D_2O$ (37) | 9.00 | 9.19 | 8.64 | – | – |
| DMSO-$d_6$ (2) | 8.99 | 9.21 | 8.70 | 13.50 | – |
| DMSO-$d_6$ (32) | 8.85 | 9.05 | 8.54 | – | – |
| DMSO-$d_6$ (46) | 8.90 | 9.10 | 8.60 | – | – |
| DMSO-$d_6$ (18, 19) | – | – | 8.68 | – | – |
| DMSO-$d_6$ (44) | 8.88 | 9.09 | – | 12.80 | – |
| DMSO-$d_6$ (45) | 8.91 | 9.12 | 8.61 | 13.45 | – |
| 98% $D_2SO_4$, 2% $D_2O$ | 9.01 | 9.20 | 9.20 | – | – |
| **$^{15}N$** | | | | | |
| Solvent (reference) | N1 (ppm) | N3 (ppm) | N7 (ppm) | N9 (ppm) | |
| $H_2O$ (40) | 267.60 | 252.50 | 195.70 | 191.60 | |
| DMSO-$d_6$ (40) | 278.90 | 261.30 | 210.50 | 190.00 | |
| 90% $D_2SO_4$, 10% $D_2O$ (4)* | 186.03 | 260.03 | 158.33 | 163.43 | |
| 98% $D_2SO_4$, 2% $D_2O$ | 185.99 | 262.08 | 158.11 | 163.02 | |

(–) Available literature does not provide chemical shift values for the specified atoms or no peaks corresponding to hydrogens attached to nitrogen have been detected.
(*) chemical shift values converted from $\sigma$ in the original source (4) to $\delta$ for consistency (47).

Table S8. An overview and comparison of NMR chemical shifts of pyrimidine from this work, obtained in 98% $D_2SO_4$, 2% $D_2O$ (by weight) to values in selected solvents reported in the literature. All listed literature NMR data have been obtained at, or close to, room temperature.

| Pyrimidine | | | |
|---|---|---|---|
| $^{13}C$ | | | |
| Solvent (reference) | C2 (ppm) | C4,6 (ppm) | C5 (ppm) |
| $D_2O$ (48) | 157.1 | 157.0 | 122.3 |
| $D_2O/H_2SO_4$ (49, 50) | 152.2 | 158.8 | 125.1 |
| $CDCl_3$ (2) | 159.08 | 156.92 | 121.61 |
| DMSO-$d_6$ (51) | 158.39 | 156.90 | 121.86 |
| DMSO-$d_6$ (52) | 158.60 | 157.12 | 122.04 |
| DMSO-$d_6$** | 159.06 | 157.63 | 122.54 |
| 98% $D_2SO_4$, 2% $D_2O$ | 149.85 | 158.18 | 127.77 |
| $^1H$ | | | |
| Solvent | H2 (ppm) | H4,6 (ppm) | H5 (ppm) |
| $D_2O$ (48) | 8.98 | 8.67 | 7.47 |
| $D_2O$ (37) | 9.19 | 8.86 | 7.65 |
| $D_2O$/DCl (49) | 9.74 | 9.45 | 8.34 |
| $CDCl_3$ (2) | 9.27 | 8.78 | 7.38 |
| DMSO-$d_6$ (52) | 9.24 | 8.86 | 7.58 |
| $CH_2Cl_2$ (53) | 9.15 | 8.69 | 7.29 |
| 98% $D_2SO_4$, 2% $D_2O$ | 9.68 | 9.20 | 8.35 |
| $^{15}N$ | | | |
| Solvent (reference) | N1,3 (ppm) | | |
| $CH_2Cl_2$ (53) | 294.40 | | |
| DMSO-$d_6$ (54) | 295.3 | | |
| 98% $D_2SO_4$, 2% $D_2O$ | 201.21 | | |

(**) This study, see Figure S10.

**Table S9.** Concentrations and peak maxima for UV spectroscopy in this study.

| Compound | Concentration (μM) | UV Maxima (nm) |
| --- | --- | --- |
| Adenine | 7 | 197.0, 258.0 |
| Cytosine | 100 | 214.0, 266.0 |
| 2,6-Diaminopurine | 50 | 194.0, 217.0, 276.0 |
| Guanine | 60 | 233.0, 255.0 |
| Purine | 40 | 258.0 |
| Pyrimidine | 0.6 | 246.0 |
| Thymine | 80 | 194.0, 284.0 |
| Uracil | 70 | 193.0, 276.0 |

**Dataset S1.** Dataset S1 contains original UV-Vis data.

The folder "UV-VIS_plots-data-code-tables DATASET S1.zip" containing the original UV-Vis data can be downloaded from Zenodo at https://zenodo.org/

**Dataset S2.** Dataset S2 contains original NMR data.

The folder "ORIGINAL NMR DATA_DATASET S2.zip" containing the original NMR data can be downloaded from Zenodo at https://zenodo.org/ and contains the following data:

a) all 2 week stability data – that contains all the 1D-$^{13}$C NMR data collected after 2 week incubation in concentrated sulfuric acid.

b) DMSO solvent bases data – 1D-$^{13}$C NMR data for the selected bases measured directly in DMSO-$d_6$ solvent.

The 1D-$^{13}$C NMR measurements in sulfuric acid at different concentrations follows the following naming convention. Different concentrations (by weight) of sulfuric acid in water are denoted by different letters from H-K: 98% $D_2SO_4$/2% $D_2O$; 94% $D_2SO_4$/6% $D_2O$; 88% $D_2SO_4$/12% $D_2O$; 81% $D_2SO_4$/19% $D_2O$, all with DMSO-$d_6$ as a reference and at room temperature.

H - 81% $D_2SO_4$/19% $D_2O$ with DMSO-$d_6$ as a reference and at room temperature
I - 88% $D_2SO_4$/12% $D_2O$ with DMSO-$d_6$ as a reference and at room temperature
J - 94% $D_2SO_4$/6% $D_2O$ with DMSO-$d_6$ as a reference and at room temperature
K - 98% $D_2SO_4$/2% $D_2O$ with DMSO-$d_6$ as a reference and at room temperature

c) Cytosine NMR folder with the following files:

1D-$^{13}$C H-K - SZssea-Cytosine-H-DMSO-101322_1.zip, SZssea-Cytosine-I-DMSO-101322_1.zip, SZssea-Cytosine-J-DMSO-101322_1.zip, SZssea-Cytosine-K-DMSO-101322_1.zip (the same as SZssea-Cytosine-K-13C-DMSO-101322_1.zip)

1D-$^{13}$C - SZssea-Cytosine-K-13C-DMSO-101322_1.zip

1D-$^{1}$H - SZssea-1H-Cytosine-K-DMSO-101322_1.zip

1D-$^{15}$N - SZssea-Cytosine-K-15N-102022_1.zip

2D-$^{1}$H-$^{13}$C-HMQC - SZssea-Cytosine-K-HC-HMQC-101822_1.zip

2D-$^{1}$H-$^{13}$C-HMBC - SZssea-Cytosine-K-HC-HMBC-101822_1.zip

2D-1H-15N-HMBC - SZssea-Cytosine-K-HN-HMBC-101822_1.zip

d) Guanine NMR folder with the following files:

1D-$^{13}$C H-K - SZssea-Guanine-H-DMSO-101122_1.zip, SZssea-Guanine-I-DMSO-101122_1.zip, SZssea-Guanine-J-DMSO-101122_1.zip, SZssea-Guanine-K-DMSO-101122_1.zip (the same as SZssea-Guanine-K-13C DMSO-101122_1.zip)

1D-$^{13}$C - SZssea-Guanine-K-13C DMSO-101122_1.zip

1D-$^{1}$H - SZssea-Guanine-K-1H-DMSO-101122_1.zip

1D-$^{15}$N - SZssea-Guanine-K-15N-DMSO-101722_1.zip

2D-$^1$H-$^{13}$C-HMQC - SZssea-Guanine-K-HC-HMQC-DMSO-101722_1 (2).zip

2D-$^1$H-$^{13}$C-HMBC - SZssea-Guanine-K-HC-HMBC-DMSO-101722_1 (2).zip

2D-$^1$H-$^{15}$N-HMBC - SZssea-Guanine-K-HN-HMBC-DMSO-101722_1.zip

e) Adenine NMR folder with the following files:

1D-$^{13}$C H-K - SZssea-Adenine-H-DMSO-101322_1.zip, SZssea-Adenine-I-DMSO-101322_1.zip, SZssea-Adenine-J-DMSO-101322_1.zip, SZssea-Adenine-K-DMSO-101322_1.zip

1D-$^{13}$C - SZssea-Adenine-K-DMSO-101322_1.zip

1D-$^1$H - SZssea-1H-Adenine-K-DMSO-101322_1.zip

1D-$^{15}$N - SZssea-Adenine-K-15N-DMSO-110422_3.zip

2D-$^1$H-$^{13}$C-HMQC - SZssea-Adenine-K-HC-HMQC-102122_2.zip

2D-$^1$H-$^{13}$C-HMBC - SZssea-Adenine-K-HC-HMBC-102122_2.zip

2D-1H-$^{15}$N-HMBC - SZssea-Adenine-K-HN-HMBC-102122_2.zip

f) Diaminopurine NMR folder with the following files:

1D-$^{13}$C H-K - SZssea-Diap-H-DMSO-101422_1.zip, SZssea-Diap-H-DMSO-101422_2.zip, Szssea-Diap-H-DMSO-101422_4.zip, SZssea-Diap-I-DMSO-101422_1.zip, SZssea-Diap-I-DMSO-101422_2.zip, SZssea-Diap-J-101422_1.zip, SZssea-Diap-K-101422_1.zip

1D-$^{13}$C - SZssea-Diap-K-101422_1.zip

1D-$^1$H - SZssea-1H-Diap-K-101422_1.zip

1D-$^{15}$N - SZssea-Diapurine-K-15N-102122_1.zip

2D-$^1$H-$^{13}$C-HMQC - SZssea-Diaminopurine-K-HC-HMQC-102122_2.zip

2D-1H-$^{13}$C-HMBC - SZssea-Diaminopurine-K-HC-HMBC-102122_2.zip

2D-1H-$^{15}$N-HMBC - SZssea-Diaminopurine-K-HN-HMBC-102122_2.zip

g) Purine NMR folder with the following files:

1D-$^{13}$C H-K - SZssea-Purine-H-DMSO-101122_1.zip, SZssea-Purine-I-DMSO-101122_1.zip, SZssea-Purine-J-DMSO-101122_1.zip, SZssea-Purine-K-DMSO-101122_1.zip

1D-$^{13}$C - SZssea-Purine-K-DMSO-101122_1.zip

1D-$^1$H - SZssea-Purine-1H-K-DMSO-101722_2.zip

1D-$^{15}$N - SZssea-Purine-K-15N-102122_1.zip

2D-$^1$H-$^{13}$C-HMQC - SZssea-Purine-K-HC-HMQC-101722_1.zip

2D-$^1$H-$^{13}$C-HMBC - SZssea-Purine-K-HC-HMBC-101722_1.zip

2D-$^1$H-$^{15}$N-HMBC - SZssea-Purine-K-HN-HMBC-101722_1.zip

h) Pyrimidine NMR folder with the following files:

1D-$^{13}$C H-K - SZssea-Pyr-H-DMSO-101322_1.zip, SZssea-Pyr-I-DMSO-101322_1.zip, SZssea-Pyr-J-DMSO-101322_1.zip, SZssea-Pyr-K-DMSO-101322_1.zip

1D-$^{13}$C - SZssea-Pyr-K-DMSO-101322_1.zip

1D-$^1$H - SZssea-1H-Pyr-K-DMSO-101322_1.zip

1D-$^{15}$N - SZssea-Pyrimidine-K-15N_101922_1.zip

2D-$^1$H-$^{13}$C-HMQC - SZssea-Pyrimidine-K-HC-HMQC-102022_2.zip

2D-$^1$H-$^{13}$C-HMBC - SZssea-Pyrimidine-K-HC-HMBC_101922_1.zip

2D-$^1$H-$^{15}$N-HMBC - SZssea-Pyrimidine-K-HN-HMBC_101922_1.zip

i) Thymine NMR folder with the following files:

1D-$^{13}$C H-K - SZssea-Thymine-H-DMSO-101322_1.zip, SZssea-Thymine-I-DMSO-101322_1.zip, SZssea-Thymine-J-DMSO-101322_1.zip, SZssea-Thymine-K-DMSO-101322_1.zip

1D-$^{13}$C - SZssea-Thymine-K-DMSO-101322_1.zip

1D-$^1$H - SZssea-1H-Thymine-K-DMSO-101322_1.zip

1D-$^{15}$N - SZssea-Thymine-K-15N-102122_1.zip

2D-$^1$H-$^{13}$C-HMQC - SZssea-Thymine-K-HC-HMQC-102022_1.zip

2D-$^1$H-$^{13}$C-HMBC - SZssea-Thymine-K-HC-HMBC-102022_1.zip

2D-$^1$H-$^{15}$N-HMBC - SZssea-Thymine-K-HN-HMBC-102022_1.zip

j) Uracil NMR folder with the following files:

1D-$^{13}$C H-K - SZssea-Uracil-H-DMSO-101322_1.zip, SZssea-Uracil-I-DMSO-101322_1.zip, SZssea-Uracil-J-DMSO-101322_1.zip, SZssea-Uracil-K-DMSO-101322_1.zip

1D-$^{13}$C - SZssea-Uracil-K-DMSO-101322_1.zip

1D-$^1$H - SZssea-1H-Uracil-K-DMSO-101322_1.zip

1D-$^{15}$N - SZssea-Uracil-K-15N-102122_1.zip

2D-$^1$H-$^{13}$C-HMQC - SZssea-Uracil-K-HC-HMQC-101822_1.zip

2D-$^1$H-$^{13}$C-HMBC - SZssea-Uracil-K-HC-HMBC-101822_1.zip

2D-$^1$H-$^{15}$N-HMBC - SZssea-Uracil-K-HN-HMBC-101822_1.zip

**SI References**

1. L. G. Purnell, D. J. Hodgson, Carbon-13 nmr studies of purines and 8-azapurines in basic aqueous medium. *Org. Magn. Reson.* **10**, 1–4 (1977).
2. T. Saito, *et al.*, Spectral database for organic compounds (sdbs). *Natl. Inst. Adv. Ind. Sci. Technol.* (2006).
3. R. Wagner, W. von Philipsborn, Protonierung von Purin, Adenin und Guanin NMR.-Spektren und Strukturen der Mono-, Di-und Tri-Kationen. *Helv. Chim. Acta* **54**, 1543–1558 (1971).
4. M. Schumacher, H. Günther, Beiträge zur 15N-NMR-Spektroskopie Protonierung und Tautomerie in Purinen: Purin und 7-und 9-Methylpurin. *Chem. Ber.* **116**, 2001–2014 (1983).
5. R. L. Benoit, M. Frechette, 1H and 13C nuclear magnetic resonance and ultraviolet studies of the protonation of cytosine, uracil, thymine, and related compounds. *Can. J. Chem.* **64**, 2348–2352 (1986).
6. R. Wagner, W. von Philipsborn, Protonierung von Amino-und Hydroxypyrimidinen NMR-Spektren und Strukturen der Mono-und Dikationen. *Helv. Chim. Acta* **53**, 299–320 (1970).
7. M. D. Oza, R. Meena, K. Prasad, P. Paul, A. K. Siddhanta, Functional modification of agarose: A facile synthesis of a fluorescent agarose–guanine derivative. *Carbohydr. Polym.* **81**, 878–884 (2010).
8. T. Kozluk, I. D. Spenser, Carbon-13 NMR spectroscopy as a biosynthetic probe. The biosynthesis of purines in yeast. *J. Am. Chem. Soc.* **109**, 4698–4702 (1987).
9. S. Kopf, *et al.*, Base-Mediated Remote Deuteration of N-Heteroarenes–Broad Scope and Mechanism. *European J. Org. Chem.* **2022**, e202200204 (2022).
10. A. A. Shaw, M. D. Shetlar, 3-Ureidoacrylonitriles: novel products from the photoisomerization of cytosine, 5-methylcytosine, and related compounds. *J. Am. Chem. Soc.* **112**, 7736–7742 (1990).
11. Y. Shalom, J. Blum, R. G. Harvey, Adducts of phenanthrene 9, 10-imine and of benz [a] anthracene 5, 6-imine to some nitrogen heterocycles. *J. Heterocycl. Chem.* **33**, 681–686 (1996).
12. J. H. Clark, E. M. Goodman, The cytosine—fluoride interaction. *Spectrochim. Acta Part A Mol. Spectrosc.* **42**, 457–460 (1986).
13. S. Shirotake, Complexes between Nucleic Acid Bases and Bivalent Metal Ions. III. Syntheses and Spectral Analyses of Cytosine-Calcium Chloride Complexes. *Chem. Pharm. Bull.* **28**, 956–963 (1980).
14. A. Bourafai-Aziez, *et al.*, Development, Validation, and Use of 1H-NMR Spectroscopy for Evaluating the Quality of Acerola-Based Food Supplements and Quantifying Ascorbic Acid. *Molecules* **27**, 5614 (2022).
15. S. P. Samijlenko, *et al.*, Structural peculiarities of 6-azacytosine and its derivatives imply intramolecular H-bonds. *J. Mol. Struct.* **484**, 31–38 (1999).
16. S. P. Samijlenko, *et al.*, 1H NMR investigation on 6-azacytidine and its derivatives. *Spectrochim. Acta Part A Mol. Biomol. Spectrosc.* **55**, 1133–1141 (1999).
17. M. C. Thorpe, W. C. Coburn Jr, J. A. Montgomery, The 13C nuclear magnetic resonance spectra of some 2-, 6-, and 2, 6-substituted purines. *J. Magn. Reson.* **15**, 98–112 (1974).
18. W. C. J. Coburn, M. C. Thorpe, J. A. Montgomery, K. Hewson, Correlation of the Proton Magnetic Resonance Chemical Shifts of Substituted Purines with Reactivity Parameters. II. 6-Substituted Purines. *J. Org. Chem.* **30**, 1114–1117 (1965).
19. W. C. J. Coburn, M. C. Thorpe, J. A. Montgomery, K. Hewson, Correlation of the Proton Magnetic Resonance Chemical Shifts of Substituted Purines with Reactivity Parameters. I. 2,6-Disubstituted Purines. *J. Org. Chem.* **30**, 1110–1113 (1965).
20. L.-L. Xu, *et al.*, Chiroptical Activity from an Achiral Biological Metal–Organic Framework. *J. Am. Chem. Soc.* **140**, 11569–11572 (2018).
21. D. Kubica, S. Molchanov, A. Gryff-Keller, Solvation of uracil and its derivatives by DMSO:

A DFT-supported 1H NMR and 13C NMR study. *J. Phys. Chem. A* **121**, 1841–1848 (2017).

22. S. Kan, *et al.*, Chemical constituents from the roots of Xanthium sibiricum. *Nat. Prod. Res.* **25**, 1243–1249 (2011).
23. A. W. Newaz, K. Yong, W. Yi, B. Wu, Z. Zhang, Antimicrobial metabolites from the Indonesian mangrove sediment-derived fungus Penicillium chrysogenum sp. ZZ1151. *Nat. Prod. Res.*, 1–7 (2022).
24. J. E. Okokon, *et al.*, In vivo antihyperglycaemic and antihyperlipidemic activities and chemical constituents of Solanum anomalum. *Biomed. Pharmacother.* **151**, 113153 (2022).
25. S. Lamichhane, *et al.*, Strategy for nuclear-magnetic-resonance-based metabolomics of human feces. *Anal. Chem.* **87**, 5930–5937 (2015).
26. A. Sułkowska, Effect of temperature on the stability of association of pyrimidine bases with serum albumin: Proton NMR study. *Appl. Spectrosc.* **51**, 428–432 (1997).
27. M. Martini, J. Termini, Peroxy radical oxidation of thymidine. *Chem. Res. Toxicol.* **10**, 234–241 (1997).
28. P. J. W. Pouwels, R. Kaptein, R. F. Hartman, S. D. Rose, Photo-CIDNP study of pyrimidine dimer splitting I: reactions involving pyrimidine radical cation intermediates. *Photochem. Photobiol.* **61**, 563–574 (1995).
29. P. J. W. Pouwels, R. Kaptein, R. F. Hartman, S. D. Rose, Photo-CIDNP study of pyrimidine dimer splitting II: reactions involving pyrimidine radical anion intermediates. *Photochem. Photobiol.* **61**, 575–583 (1995).
30. H. Asanuma, T. Hishiya, M. Komiyama, Direct evidences for the hydrogen bonding in water by polymeric receptors carrying diaminotriazine. *Chem. Lett.* **27**, 1087–1088 (1998).
31. A. Sulkowska, A. Michnik, Proton NMR studies on the interaction of alkyl derivatives of pyrimidine bases, their nucleosides and nucleotides with bovine serum albumin. *J. Mol. Struct.* **348**, 73–76 (1995).
32. M. T. Chenon, R. J. Pugmire, D. M. Grant, R. P. Panzica, L. B. Townsend, Carbon-13 magnetic resonance. XXVI. Quantitative determination of the tautomeric populations of certain purines. *J. Am. Chem. Soc.* **97**, 4636–4642 (1975).
33. L. RunHui, *et al.*, N-containing compounds from the traditional Chinese medicine ChanSu. *Chem. Nat. Compd.* **45**, 599–600 (2009).
34. K. Dybiec, S. Molchanov, A. Gryff-Keller, Adenine and some of its analogues in DMSO-d6 solution: an NMR and GIAO-DFT study. *Pol. J. Chem.* **83**, 857–868 (2009).
35. P. Chittepu, V. R. Sirivolu, F. Seela, Nucleosides and oligonucleotides containing 1, 2, 3-triazole residues with nucleobase tethers: Synthesis via the azide-alkyne 'click'reaction. *Bioorg. Med. Chem.* **16**, 8427–8439 (2008).
36. H. Moriyama, T. Iizuka, M. Nagai, K. Hoshi, Adenine, an inhibitor of platelet aggregation, from the leaves of Cassia alata. *Biol. Pharm. Bull.* **26**, 1361–1364 (2003).
37. H. J. Schneider, *et al.*, Complexation of nucleosides, nucleotides, and analogs in an azoniacyclophane. Van der Waals and electrostatic binding increments and NMR shielding effects. *J. Am. Chem. Soc.* **114**, 7704–7708 (1992).
38. A. Sułkowska, Temperature effect on the stability of the complexes between purine derivatives and serum albumin: proton NMR study. *J. Mol. Struct.* **410**, 23–25 (1997).
39. M. Mano, T. Seo, K. Imai, Synthesis of 6-methylaminopurine by thermal cyclization of 4, 6-bis (methylamino)-5-phenylazopyrimidine. *Chem. Pharm. Bull.* **31**, 3454–3459 (1983).
40. R. Marek, V. Sklenar, NMR Studies of Purines. *Annu. reports NMR Spectrosc.* **54**, 201–242 (2005).
41. E. Bednarek, *et al.*, Theoretical and experimental 1H, 13C, 15N, and 17O NMR spectra of 5-nitro, 5-amino, and 5-carboxy uracils. *J. Mol. Struct.* **482**, 333–337 (1999).
42. E. Bednarek, *et al.*, Theoretical and experimental 1H, 13C, 15N, and 17O NMR chemical shifts for 5-halogenouracils. *J. Mol. Struct.* **554**, 233–243 (2000).
43. J. H. Clark, J. S. Taylor, A. J. Goodwin, Multinuclear NMR studies of the fluoride-uracil complex. *Spectrochim. Acta Part A Mol. Spectrosc.* **38**, 1101–1104 (1982).
44. A. Unciti-Broceta, M. J. Pineda de las Infantas, M. A. Gallo, A. Espinosa, Reduction of Different Electron-Poor N-Heteroarylhydrazines in Strong Basic Conditions. *Chem. Eur. J.*

**13**, 1754–1762 (2007).
45. M. Česnek, *et al.*, Synthesis and properties of 2-guanidinopurines. *Collect. Czechoslov. Chem. Commun.* **71**, 1303–1319 (2006).
46. T. H. Graham, W. Liu, D.-M. Shen, A Method for the Reductive Scission of Heterocyclic Thioethers. *Org. Lett.* **13**, 6232–6235 (2011).
47. R. K. Harris, E. D. Becker, S. M. Cabral de Menezes, R. Goodfellow, P. Granger, NMR Nomenclature: Nuclear Spin Properties and Conventions for Chemical Shifts. IUPAC Recommendations 2001. *Solid State Nucl. Magn. Reson.* **22**, 458–483 (2002).
48. K. Goel, S. Bera, M. Singh, D. Mondal, Synthesis of dual functional pyrimidinium ionic liquids as reaction media and antimicrobial agents. *RSC Adv.* **6**, 106806–106820 (2016).
49. J. Clark, G. Hitiris, Covalent hydration of 5-substituted pyrimidines. *Spectrochim. Acta Part A Mol. Spectrosc.* **40**, 75–79 (1984).
50. R. J. Pugmire, D. M. Grant, Carbon-13 magnetic resonance. X. Six-membered nitrogen heterocycles and their cations. *J. Am. Chem. Soc.* **90**, 697–706 (1968).
51. J. Riand, M. T. Chenon, N. Lumbroso-Bader, Etude par rmn du carbone-13 des effets de substituants dans le noyau de la pyrimidine. *Tetrahedron Lett.* **15**, 3123–3126 (1974).
52. A. Y. Denisov, V. I. Mamatyuk, O. P. Shkurko, Additivity of 13C-1H and 1H-1H spin-spin coupling constants in six-membered aromatic nitrogen-containing heterocycles. *Chem. Heterocycl. Compd.* **21**, 821–825 (1985).
53. K. J. Sheehy, L. M. Bateman, N. T. Flosbach, M. Breugst, P. A. Byrne, Identification of N- or O-Alkylation of Aromatic Nitrogen Heterocycles and N-Oxides Using 1H–15N HMBC NMR Spectroscopy. *European J. Org. Chem.* **2020**, 3270–3281 (2020).
54. A. Dokalik, H. Kalchhauser, W. Mikenda, G. Schweng, NMR spectra of nitrogen-containing compounds. Correlations between experimental and GIAO calculated data. *Magn. Reson. Chem.* **37**, 895–902 (1999).